\documentclass[longauth]{aa} 

\usepackage{amsmath} 
\usepackage{graphicx}
\usepackage{txfonts}
\usepackage{xcolor}
\usepackage{multirow,bigdelim} 

\usepackage{booktabs,siunitx}
\newcommand{\au}{au}
\def\ms{m\,s$^{-1}$}

\def\prot{169.3^{+3.7}_{-3.6}}

\newcommand{\juliet}{\texttt{juliet}}
\newcommand{\serval}{\texttt{serval}}
\newcommand{\raccoon}{\texttt{raccoon}}
\newcommand{\exostriker}{\texttt{Exo-Striker}}
\newcommand{\aliasfinder}{\texttt{AliasFinder}}
\newcommand{\tess}{TESS}
\newcommand{\gaia}{\textit{Gaia}}
\newcommand{\jwst}{JWST}
\newcommand{\wolf}{Wolf~1069}

\usepackage[breaklinks, colorlinks, citecolor=blue, linkcolor=blue]{hyperref}
\makeatletter
\def\instrefs#1{{\def\scsep{\def\scsep{,}}\@for\w:=#1\do{\scsep\ref{inst:\w}}}}
\renewcommand{\inst}[1]{\unskip$^{\instrefs{#1}}$}

\begin{document}

\title{The CARMENES search for exoplanets around M dwarfs}
\subtitle{\wolf~b: Earth-mass planet in the \\ habitable zone of a nearby, very low-mass star\thanks{RVs and additional data (i.e., stellar activity indicators as shown in Fig.~\ref{fig:glsperiodogramwolf1069activity} and long-term photometry as in
Fig.~\ref{fig:phottimeseries}) are available in electronic form at the CDS via anonymous ftp to \url{cdsarc.u-strasbg.fr} (\url{TBD}).}}
\titlerunning{\wolf~b: Earth-mass planet in the habitable zone of an M-dwarf}

\author{
D.~Kossakowski\inst{mpia} 
\and M.~K\"urster\inst{mpia} 
\and T.~Trifonov\inst{mpia,sofia} 
\and Th.~Henning\inst{mpia} 
\and J.~Kemmer\inst{lsw} 
\and J.\,A.~Caballero\inst{cabesac} 
\and R.~Burn\inst{mpia} 
\and S.~Sabotta\inst{lsw} 
\and J.\,S.~Crouse\inst{gsfc,seec,maryland,cress} 
\and T.\,J.~Fauchez\inst{gsfc,au,seec} 
\and E.~Nagel\inst{hamburg,tls} 
\and A.~Kaminski\inst{lsw} 
\and E.~Herrero\inst{ice,ieec} 
\and E.~Rodr\'iguez\inst{iaa} 
\and E.~Gonz\'alez-\'Alvarez\inst{cabinta} 
\and A.~Quirrenbach\inst{lsw} 
\and P.\,J.~Amado\inst{iaa} 
\and I.~Ribas\inst{ice,ieec} 
\and A.~Reiners\inst{iag} 
\and J.~Aceituno\inst{caha,iaa} 
\and V.\,J.\,S.~B\'ejar\inst{iac,ull} 
\and D.~Baroch\inst{ice,ieec} 
\and S.\,T.~Bastelberger\inst{maryland,gsfc,cress,seec} 
\and P.~Chaturvedi\inst{tls} 
\and C.~Cifuentes\inst{cabesac} 
\and S.~Dreizler\inst{iag} 
\and S.\,V.~Jeffers\inst{mig} 
\and R.~Kopparapu\inst{gsfc,seec} 
\and M.~Lafarga\inst{warwick,ice,ieec} 
\and M.\,J.~L\'opez-Gonz\'alez\inst{iaa} 
\and S.~Mart\'in-Ruiz\inst{iaa} 
\and D.~Montes\inst{ucm} 
\and J.\,C.~Morales\inst{ice,ieec} 
\and E.~Pall\'e\inst{iac,ull} 
\and A.~Pavlov\inst{mpia} 
\and S.~Pedraz\inst{caha} 
\and V.~Perdelwitz\inst{ariel,hamburg} 
\and M.~P\'erez-Torres\inst{iaa,cyprus} 
\and M.~Perger\inst{ice,ieec} 
\and S.~Reffert\inst{lsw}
\and C.~Rodr\'iguez~L\'opez\inst{iaa} 
\and M.~Schlecker\inst{arizona,mpia} 
\and P.~Sch\"ofer\inst{iaa,iag} 
\and A.~Schweitzer\inst{hamburg} 
\and Y.~Shan\inst{oslo,iag} 
\and A.~Shields\inst{irvine} 
\and S.~Stock\inst{lsw} 
\and E.~Wolf\inst{boulder,gsfc,nexss} 
\and M.\,R.~Zapatero~Osorio\inst{cabinta}
\and M.~Zechmeister\inst{iag}
}

\institute{
\label{inst:mpia}Max-Planck-Institut f\"{u}r Astronomie, K\"{o}nigstuhl  17, 69117 Heidelberg, Germany
\email{astrodianakossakowski@gmail.com}
\and \label{inst:sofia} Department of Astronomy, Sofia University ``St Kliment Ohridski'', 5 James Bourchier Blvd, BG-1164 Sofia, Bulgaria
\and \label{inst:lsw}Landessternwarte, Zentrum f\"ur Astronomie der Universit\"at Heidelberg, K\"onigstuhl 12, 69117 Heidelberg, Germany
\and \label{inst:cabesac}Centro de Astrobiolog\'ia (CSIC-INTA), ESAC, Camino bajo del castillo s/n, 28692 Villanueva de la Ca\~nada, Madrid, Spain
\and \label{inst:gsfc}NASA Goddard Space Flight Center, 8800 Greenbelt Road, Greenbelt, MD 20771, USA
\and \label{inst:seec}NASA GSFC Sellers Exoplanet Environments Collaboration
\and \label{inst:maryland}Department of Astronomy, University of Maryland, College Park, MD 20742, United States of America
\and \label{inst:cress}Center for Research and Exploration in Space Science and Technology, NASA/GSFC, Greenbelt, MD 20771, United States of America
\and \label{inst:au}American University, College of Arts and Sciences, Washington DC, United States of America
\and \label{inst:hamburg}Hamburger Sternwarte, Gojenbergsweg 112, 21029 Hamburg, Germany
\and \label{inst:tls}Th\"uringer Landessternwarte Tautenburg, Sternwarte 5, 07778 Tautenburg, Germany
\and \label{inst:ice}Institut de Ci\`encies de l'Espai (ICE, CSIC), Campus UAB, C/ de Can Magrans s/n, 08193 Cerdanyola del Vall\`es, Spain
\and \label{inst:ieec}Institut d'Estudis Espacials de Catalunya (IEEC), C/ Gran Capit\`a 2-4, 08034 Barcelona, Spain
\and \label{inst:iaa}Instituto de Astrof\'isica de Andaluc\'ia (CSIC), Glorieta de la Astronom\'ia s/n, 18008 Granada, Spain
\and \label{inst:cabinta}Centro de Astrobiolog\'ia (CSIC-INTA), Carretera de Ajalvir km 4, 28850 Torrej\'on de Ardoz, Madrid, Spain
\and \label{inst:iag}Institut f\"ur Astrophysik und Geophysik, Georg-August-Universit\"at, Friedrich-Hund-Platz 1, 37077 G\"ottingen, Germany
\and \label{inst:caha}Centro Astron\'onomico Hispano en Andaluc\'ia, Observatorio de Calar Alto, Sierra de los Filabres, E-04550 G\'ergal, Spain
\and \label{inst:iac}Instituto de Astrof\'isica de Canarias (IAC), 38205 La Laguna, Tenerife, Spain
\and \label{inst:ull}Departamento de Astrof\'isica, Universidad de La Laguna, 38206 La Laguna, Tenerife, Spain
\and \label{inst:mig}Max-Planck-Institut f\"ur Sonnensystemforschung, Justus-von-Liebig Weg 3, 37077 G\"ottingen, Germany
\and \label{inst:warwick}Department of Physics, University of Warwick, Gibbet Hill Road, Coventry CV4 7AL, United Kingdom
\and \label{inst:ucm}Departamento de F\'isica de la Tierra y Astrof\'isica \& IPARCOS-UCM (Instituto de F\'isica de Part\'iculas y del Cosmos de la UCM), Facultad de Ciencias F\'isicas, Universidad Complutense de Madrid, E-28040 Madrid, Spain
\and \label{inst:ariel}Department of Physics, Ariel University, Ariel 40700, Israel
\and \label{inst:cyprus}{School of Sciences, European University Cyprus, Diogenes street, Engomi, 1516 Nicosia, Cyprus}
\and \label{inst:arizona}Department of Astronomy/Steward Observatory, The University of Arizona, 933 North Cherry Avenue, Tucson, AZ 85721, United States of America
\and \label{inst:oslo}Centre for Earth Evolution and Dynamics, Department of Geosciences, Universitetet i Oslo, Sem S\ae{}lands vei 2b, 0315 Oslo, Norway
\and \label{inst:irvine}Department of Physics \& Astronomy, University of California, Irvine, CA 92697, Irvine
\and \label{inst:boulder}University of Colorado, Boulder Laboratory for Atmospheric and Space Physics, Department of Atmospheric and Oceanic Sciences, Boulder, CO 80303, United States of America
\and \label{inst:nexss}NASA NExSS Virtual Planetary Laboratory, Seattle, WA
}

\date{Received 28 October 2022 / accepted 21 December 2022}

\abstract{We present the discovery of an Earth-mass planet ($M_\textnormal{b}\sin i = 1.26\pm0.21\,M_\oplus$) on a \num{15.6}\,d orbit of a relatively nearby ($d \sim$ 9.6\,pc) and low-mass ($0.167 \pm 0.011\,M_\odot$) M5.0\,V star, \wolf. Sitting at a separation of $0.0672\pm0.0014$\,\au\ away from the host star puts \wolf~b in the habitable zone (HZ), receiving an incident flux of $S = 0.652\pm 0.029\,S_\oplus$.
The planetary signal was detected using telluric-corrected radial-velocity (RV) data from the CARMENES spectrograph, amounting to a total of 262 spectroscopic observations covering almost four years. There are additional long-period signals in the RVs, one of which we attribute to the stellar rotation period. This is possible thanks to our photometric analysis including new, well-sampled monitoring campaigns undergone with the OSN and TJO facilities that supplement archival photometry (i.e., from MEarth and SuperWASP), and this yielded an updated rotational period range of $P_\textnormal{rot} = 150-170$\,d, with a likely value at $\prot$\,d. The stellar activity indicators provided by the CARMENES spectra likewise demonstrate evidence for the slow rotation period, though not as accurately due to possible factors such as signal aliasing or spot evolution. 
Our detectability limits indicate that additional planets more massive than one Earth mass with orbital periods of less than 10 days can be ruled out, suggesting that perhaps \wolf~b had a violent formation history. This planet is also the sixth closest Earth-mass planet situated in the conservative HZ, after \object{Proxima Centauri}~b, \object{GJ~1061}~d, \object{Teegarden's Star}~c, and \object{GJ~1002}~b and~c. Despite not transiting, \wolf~b is nonetheless a very promising target for future three-dimensional climate models to investigate various habitability cases as well as for sub-\ms\ RV campaigns to search for potential inner sub-Earth-mass planets in order to test planet formation theories.
} 


   \maketitle
%
\section{Introduction}

An impressive 5000 exoplanets and counting have been detected thus far\footnote{\url{https://exoplanetarchive.ipac.caltech.edu/}, accessed on 16 September 2022.}, largely thanks to the past and ongoing radial-velocity (RV) and transit surveys. On the hunt for an Earth analog, out of these thousands of planets, only $\sim$50 have been found to sit in the so-called habitable zone (HZ) of their stellar host\footnote{The Habitable Exoplanet Catalog:  \url{http://phl.upr.edu/projects/habitable-exoplanets-catalog}, last updated on 6 December 2021 and further discussed in Appendix~\ref{appendix:planetshz}.}, which is defined to be the circumstellar region in which liquid water could potentially exist on the surface of the planet \citep{Kasting1993,Kopparapu2013}. Only 20 of these are considered to be Earth-sized, defined by radii of 0.8\,$R_\oplus<R_\textnormal{p}<1.6\,R_\oplus$ or by masses of 0.5\,$M_\oplus<M\sin{i}_\textnormal{p}<3\,M_\oplus$. Moreover, a majority of them have been discovered around M-dwarf stellar hosts, most likely due to the ease in detectability considering the higher planet-to-star mass and radius ratios \citep[see e.g.,][]{ZechmeisterKuerster2009,Seager2010,Bonfils2013,Shields2016,Perryman2018}. 

The definition of the HZ is not a definite implication for a life-hosting world, but rather acts as a good indicator for a planet to show further promising potential for markers of surface habitability. There are in fact a variety of factors that affect its habitability potential, such as the X-ray/UV emission or age in regards to the stellar host, or processes due to the planet itself including its atmospheric composition or its ability to retain certain elements in its atmosphere \citep[e.g.,][]{Dong2018,Kopparapu2019}. For this reason, it is not only crucial to first gather planets that are situated within the HZ, but also, it is necessary to build a better understanding of the stellar effects on the planet's habitability \citep[e.g.,][]{Segura2003,Segura2010,Hilton2011,Cohen2014,Chadney2016}, and to also characterize its atmosphere observability, with instruments such as the James Webb Space Telescope \citep[\jwst;][]{jwst}.  

Even though most HZ planets around low-mass M dwarfs are RV-only detections that do not transit, we can nonetheless generate useful indicators to investigate their habitability further. Three-dimensional (3D) general circulation model (GCM) climate simulations \citep[e.g.,][]{Way2017,Wolf2022} can produce predictions to investigate various atmosphere compositions to test how durably habitable the planet is. These analyses, along with a push for improvements in thermal emission and reflected light phase curve observations, are crucial given the rise of nontransiting planets found in the HZ of stellar hosts within the solar neighborhood \citep[e.g.,][]{Anglada-Escude2016a,Dreizler2020_gj1061}.  

In this paper, we turn our attention to the discovery of \wolf~b, an Earth-mass planetary companion orbiting a mid-type M dwarf within the conservative HZ limits, as defined by \cite{Kopparapu2013}. This planet could very well possess the key factors in making it indeed a habitable world according to preliminary 3D GCM models. Also, in contrast to other habitable worlds in the conservative HZ with similar host stars (i.e., Kepler-1649, Proxima Centauri, GJ~1061, Teegarden's~Star, and GJ~1002), \wolf~b is the only one within the conservative HZ without an inner planet, based on our current detection limits. This notion is supported by the works of \cite{Burn2021}, \cite{Mulders2021}, and \cite{Schlecker2021b}, where we expect a lower planet occurrence rate for stars with $M_\star<0.2\,M_\odot$ than for stars with $0.2\,M_\odot<M_\star<0.5\,M_\odot$ for both the pebble and core accretion scenarios. Granted, these are theoretical predictions as more observation-based evidence is required to confirm this, and \wolf~b could still be accompanied by closer-in and outer planets. Nevertheless, the concept that only one planet survives is predicted by formation models if there were at least one giant impact at the late stage. 
This would enhance the chance of having a massive moon similar to the Earth and might also stir up the interior of the planet to prevent stratification and sustain a magnetic field \citep[e.g.,][]{Jacobson2017}. As remote as this appears, the search for exo-moons is no longer so far-fetched in recent times \citep[e.g.,][]{MartinezRodiguez2019,Dobos2022}. 

The paper is outlined as follows.
Section~\ref{sec:data} first presents the comprehensive spectroscopic and photometric data collected for this work.
Then, the host star and its properties are introduced in Sect.~\ref{sec:stellar}, in which we determine and update its rotational period using newly taken photometry from our facilities. The various signals in the RVs for this system are investigated and modeled in Sect.~\ref{sec:analysis}, where the results and prospects for \wolf~b are then discussed in Sect.~\ref{sec:discussion}. We finally display our conclusions in Sect.~\ref{sec:conclusion}.

\section{Observational data} \label{sec:data}

\subsection{CARMENES high-resolution spectroscopy} \label{sec:carmenes}

\begin{figure*}
\centering
\begin{minipage}{0.92\textwidth}
  \centering
  \includegraphics[width=1\linewidth]{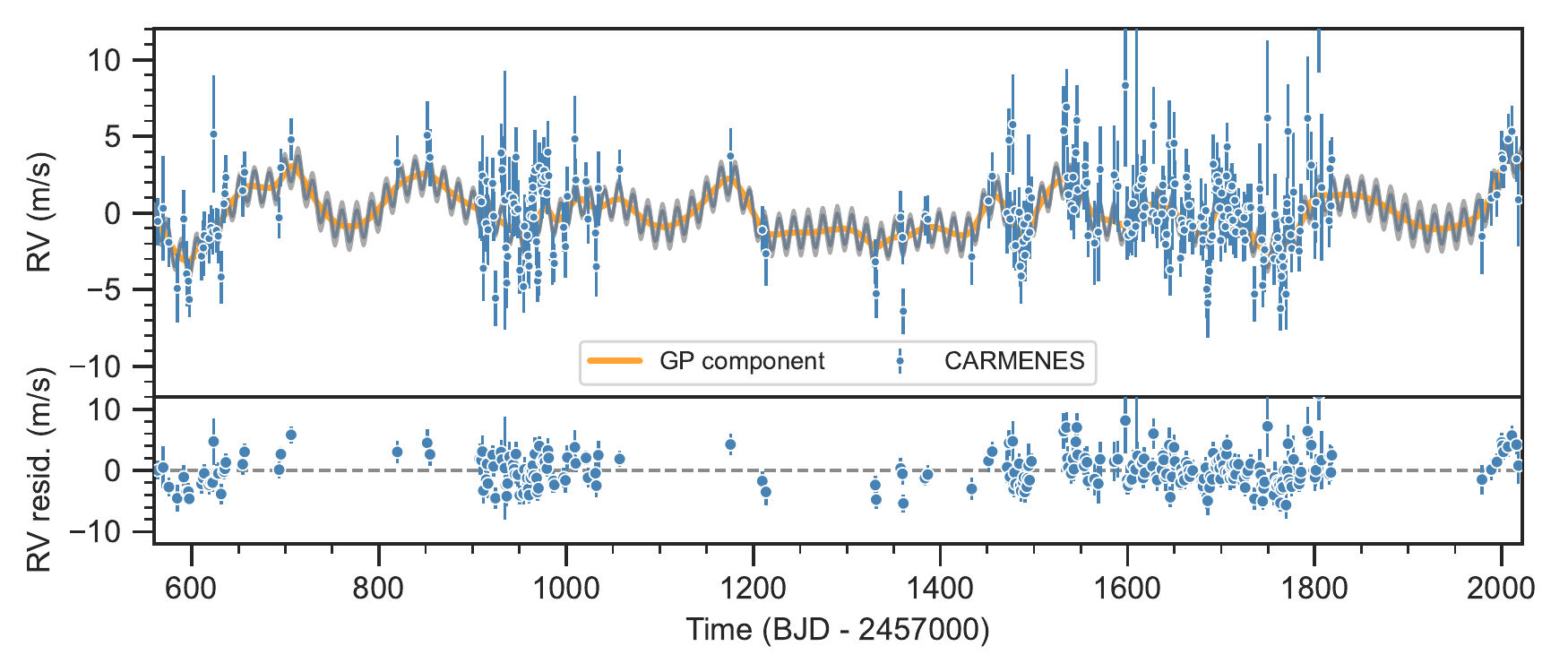}
\end{minipage}
\begin{minipage}{.45\textwidth}
  \centering
  \includegraphics[width=1\linewidth]{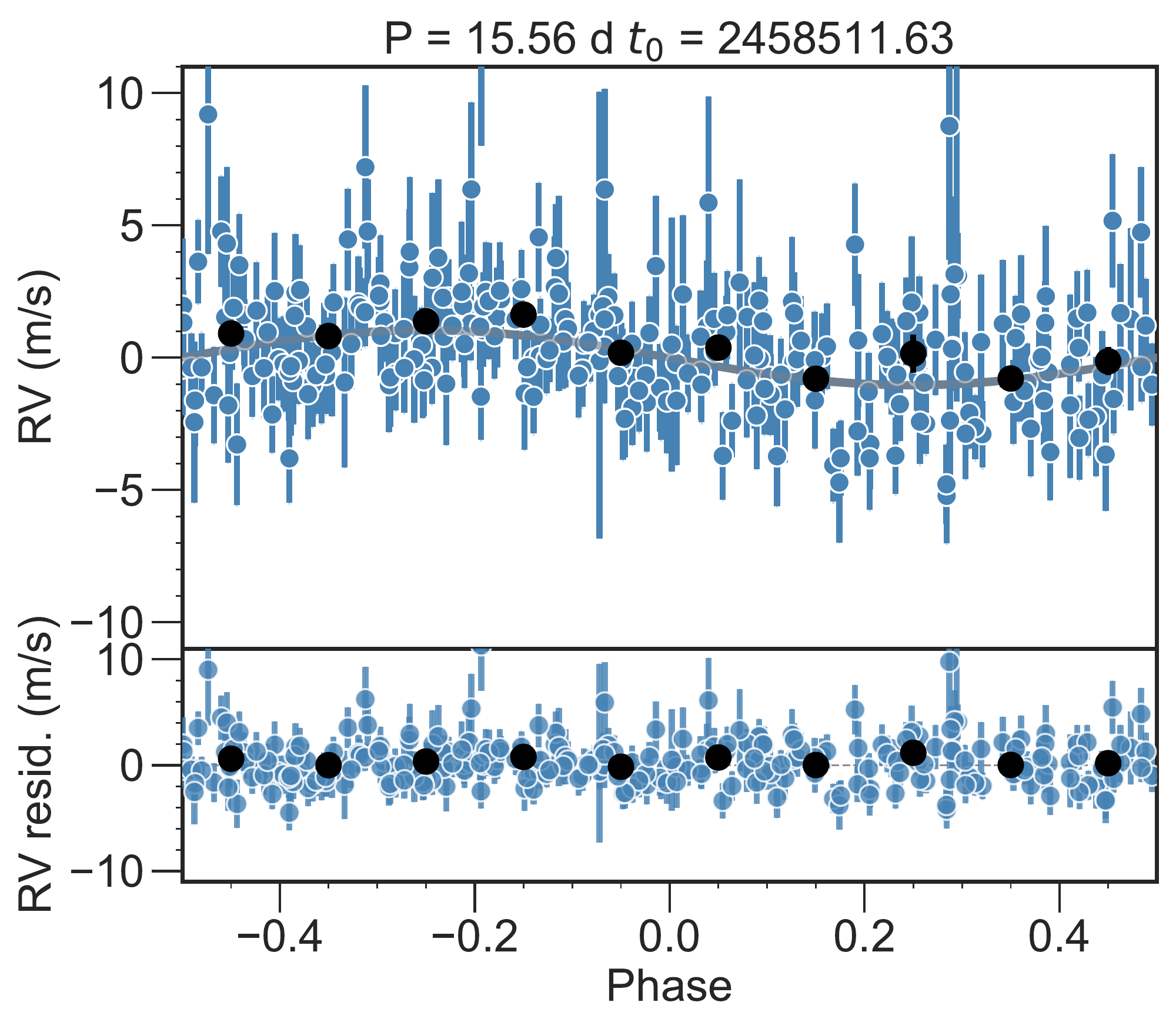}
\end{minipage}
\caption[RV time series for and phase-folded RVs for \wolf~b]{
RV time series and phase-folded plots for \wolf~b. 
\textit{Top panel:} CARMENES VIS RV measurements for \wolf\, along with the best-fit model (dark gray line) and the stellar rotation period modeled by a dSHO-GP (orange). The light gray band indicates the 68\% confidence interval of the model. \textit{Bottom panel:} RVs phase-folded to the period of \wolf~b at 15.6\,d ($K_{\rm b} = 1.07\pm0.17$\,\ms) and with the GP component subtracted out. The black circles represent the data points binned to 0.1 in phase space for visualization purposes. The bottom panel for each plot represents the residuals after subtracting out the model. There are two data points that did not fit within the boundary for visual reasons.}
\label{fig:rv}
\end{figure*}

The CARMENES\footnote{Calar Alto high-Resolution search for
M dwarfs with Exo-earths with Near-infrared and optical \'Echelle Spectrographs, \url{http://carmenes.caha.es}} instrument is located at the 3.5\,m telescope at the Calar Alto Observatory in Spain and consists of two separate spectrographs residing in two channels: the visual (VIS), which covers the spectral range 520--960\,nm (spectral resolution of $\mathcal{R}$ = 94\,600), and the near-infrared (near-IR), which covers the 960--1710\,nm range ($\mathcal{R}$ = 80\,400) \citep{CARMENES, CARMENES18}. \wolf\ (Karmn J20260+585) was one of the $\sim$300 stars initially chosen as part of the CARMENES Guaranteed Time Observation (GTO) program \citep{Reiners2018}, and 276 observations were since accumulated spanning \num{1450} days (June 2016 -- June 2020).
There were six measurements that were missing a drift correction as well as an additional eight that had low signal-to-noise ratio, and were, for this reason, discarded, resulting in 262 usable RV measurements. Furthermore, the spectra were notably affected by telluric absorption \citep[e.g.,][]{Reiners2018}. We corrected them by employing the template division telluric modeling methodology, a technique to remove telluric absorption lines from stellar spectra with numerous intrinsic lines \citep{Nagel2022}. This technique is suitable for separating telluric and spectral components based on the Earth's barycentric motion throughout the year. The telluric-free spectra of Wolf~1069 are then produced by fitting a synthetic transmission model of the Earth's atmosphere to each individually extracted telluric spectrum with \texttt{molecfit} \citep{Smette2015}. The weighted root mean square (wrms) and the median uncertainty of the remaining 262 data points are 2.66\,\ms\ and 1.67\,\ms, respectively.
The simultaneously taken near-IR measurements were not considered as part of the analysis given the notably higher wrms of 7.0\,\ms\ along with the significantly higher mean uncertainty of each observation. Therefore, we continue the analysis with the 262 telluric-absorption-corrected VIS spectra. The CARMENES RV data with their uncertainties are displayed in the top panel of Fig.~\ref{fig:rv}. 

The raw data are first pipelined through the standard guaranteed time observations via \texttt{caracal} \citep{Caballero2016_SPIE}. Then, the RVs are determined using \serval\footnote{\url{https://github.com/mzechmeister/serval}} \citep{serval}, where the spectra are corrected for barycentric motion, secular acceleration, and instrumental drift, and then nightly zero-points were calculated and applied \citep{Trifonov2020}. In addition, \serval\ produces various stellar activity indicators such as the chromatic index (CRX), the differential line width (dLW), the H$\alpha$ index, the Ca~\textsc{ii} IR triplet (IRT) lines, and the Na~\textsc{i}~D doublet lines. 
We also obtained the photospheric absorption band indices TiO~$\lambda\ 7050\,\AA$, TiO~$\lambda\ 8430\,\AA$, and TiO~$\lambda\ 8860\,\AA$ from the nontelluric-corrected spectra using spectral ranges without notable telluric contamination, as defined by \cite{Schoefer2019}. Lastly, we computed the cross-correlation function (CCF) and its full-width half-maximum (FWHM), contrast (CTR), and bisector velocity span (BVS) values computed with the \raccoon\footnote{\url{https://github.com/mlafarga/raccoon}} code, adopting the approach of using binary masks as explained in \cite{Lafarga2020}. These indicators were investigated for the rotation period of the stellar host (Sect.~\ref{sec:rotperiod}).

\subsection{Our photometric campaigns} \label{sec:ourownphot}
We carried out simultaneous, continuous photometric follow-up of \wolf\ from 2017 to 2020 with the photometric facilities as listed below. A summary of the various photometric data sets is found in Table~\ref{tab:photlogbook}. They are also displayed as a time series in Fig.~\ref{fig:phottimeseries}. 

\paragraph{Observatorio de Sierra Nevada (OSN).}
The Observatorio de Sierra Nevada (OSN)\footnote{\url{https://www.osn.iaa.csic.es/en}}, currently maintained by the Instituto de Astrof\'isica de Andaluc\'ia (IAA) and situated at Loma de Dilar in Granda, Spain, hosts two Nasmyth optical telescopes with apertures of 90\,cm (T90) and 150\,cm (T150). Both telescopes are equipped with similar VersArray 2k $\times$ 2k CCD cameras, which deliver images with fields of view (FOV) of 13.2\,arcmin $\times$ 13.2\,arcmin \citep[T90;][]{Amado2021} or 7.92\,arcmin $\times$ 7.92\,arcmin \citep[T150;][]{Rodriguez2010}. Each camera is based on a high quantum efficiency back-illuminated CCD chip, type Marconi-EEV CCD42-4, with optimized response in the ultraviolet. We monitored \wolf\ using both telescopes and various observing runs between the years 2017 and 2020. The observations with the T90 telescope were collected using both Johnson $V$ and $R$ filters during three observing runs as tabulated in Table~\ref{tab:photlogbook}. Each epoch typically consisted of 20 exposures of 80\,s in each filter per night. The resulting light curves were obtained by the method of synthetic aperture photometry. Each CCD frame was corrected in a standard way for bias and flat-fielding. Different aperture sizes were tested in order to choose the best one for our observations. A number of nearby and relatively bright stars within the frames were selected as reference stars to produce differential photometry of \wolf. Outliers due to poor observing conditions or very high airmass were removed. 

The observations with the T150 telescope were collected during a single observing run and were reduced in the same way. In this case, each epoch typically consisted of 30 exposures of 80\,s, 35\,s, and 10\,s in each $V$, $R$, and $I$ filter, respectively, per night. 

\paragraph{Telescopi Joan Or\'o (TJO).}
Simultaneous to the OSN photometry, observations for \wolf\ were carried out with the 80\,cm Telescopi Joan Or\'o (TJO)\footnote{\url{http://www.ieec.cat/content/206/what-s-the-oadm/}} at the Observatori del Montsec in Lleida, Spain. The images were obtained with an exposure time of 60 seconds using the Johnson $R$ filter of the LAIA imager, a 4k $\times$ 4k CCD with a FOV of 30\,arcmin and a plate scale of 4\,arcsec/pixel. The images were calibrated with darks, bias, and flat fields with the ICAT pipeline \citep{Colome2006} of the TJO. The differential photometry was extracted with \texttt{AstroImageJ} \citep{Collins2017} using the aperture size that minimized the rms of the resulting relative fluxes, and a selection of the 20 brightest comparison stars in the field, which did not show variability. Then, data points with a low S/N were filtered from the light curve, and any points with relative fluxes greater than 1.08 or less than 0.96 were removed before nightly binning.

\subsection{Photometric monitoring surveys} \label{sec:photsurveys}
\begin{figure*}
\centering
\includegraphics[width=\textwidth]{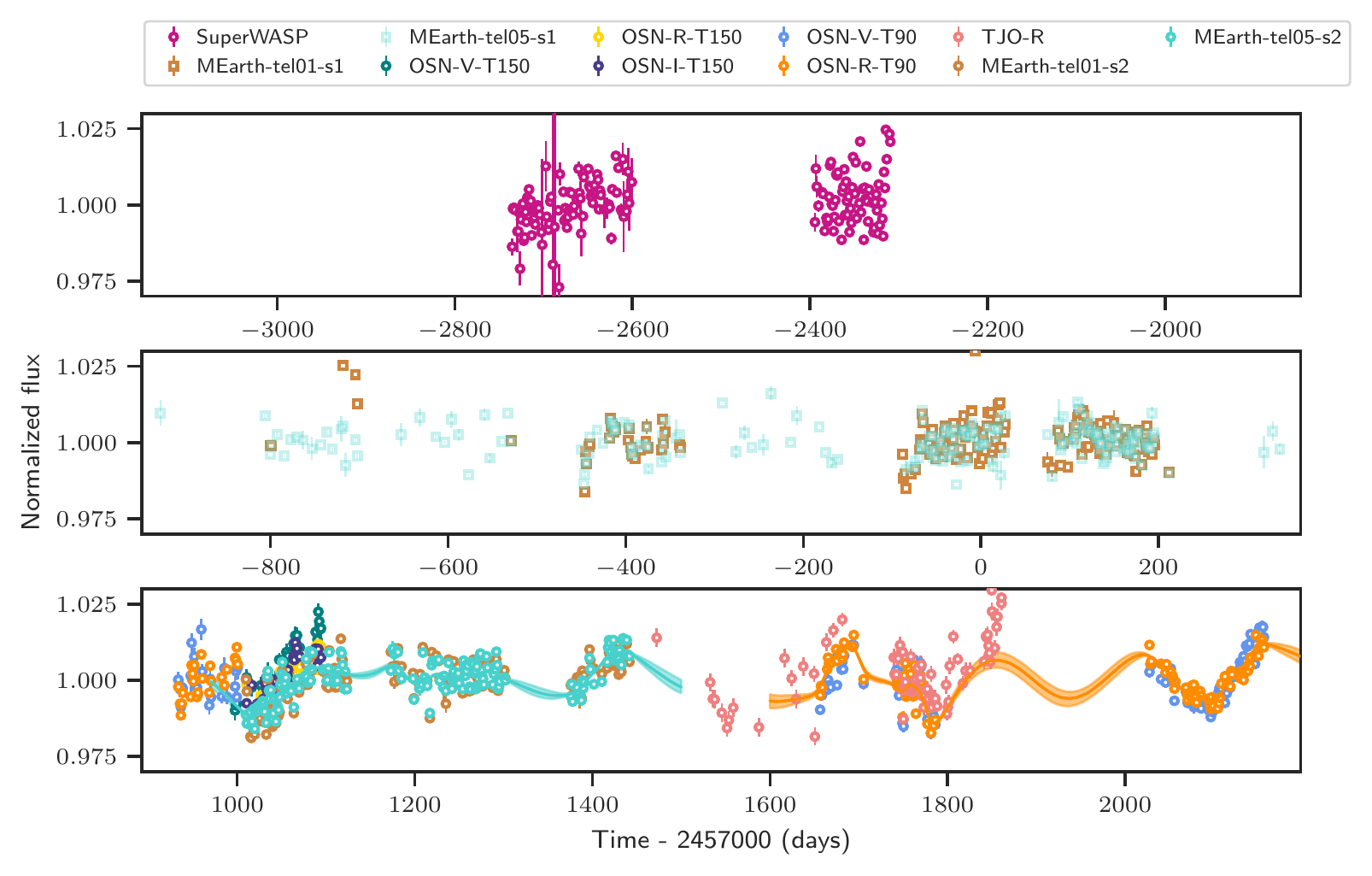}
\caption{Time series of the long-term photometry for \wolf\ color coded by instrument and filter. The time range for each panel is consistent among all panels. The MEarth-tel05-s1 data were not included in the final rotational period determination but are faintly shown for illustrative purposes. Given that the GP model is unique to each instrument with its own amplitude hyperparamter \citep[see for example Fig.~8 in][]{Kemmer2020_toi488}, the extrapolated GP models of two instruments (MEarth-tel-05 and OSN-R-T90) are overplotted with the same color as their respective data sets for illustrative purposes.}
\label{fig:phottimeseries}
\end{figure*}

Additionally, we compiled a collection of archival long-term photometric data taken by monitoring surveys as described below. They are also listed in Table~\ref{tab:photlogbook} and illustrated in Fig.~\ref{fig:phottimeseries}.

\paragraph{SuperWASP.}
The SuperWASP\footnote{Super-Wide Angle Search for Planets.} project is led by a chiefly UK-based consortium. Using two arrays of robotic telescopes operating in the northern and southern hemispheres, the survey obtains light curves for millions of objects at high cadence to look for transiting planets and study other astrophysical phenomena across the entire sky \citep{Pollacco2006,Butters10}. SuperWASP-North is located at the Roque de los Muchachos Observatory in La Palma, whereas SuperWASP-South is at the South African Astronomical Observatory near Sutherland, South Africa. Each observatory consists of eight wide-angle cameras with Canon 200\,mm, f/1.8 lenses that feed into 2048 $\times$ 2048 CCDs. The pixel scale is 13.7\,arcsec. 

\wolf\ was monitored for four seasons with SuperWASP-North from 2007 to 2010, though only sparsely in the last two seasons. The usable data spans from June 2007 to August 2008 with a $\sim$6 month gap. We received the complete light curve, corrected for instrumental and atmospheric systematics, from the SuperWASP team. The detrending procedure is nonaggressive and expected to preserve true astrophysical signals (including rotational modulation), as documented by \citet{Tamuz05}. After clipping the final two seasons and iteratively rejecting outliers commensurate with the size of the data set in each season, we binned the data nightly such that
the weighted mean and error of all the data points that go into each bin constitutes the flux and error of that bin. 



\paragraph{MEarth-North.}
\wolf\ was observed by the MEarth\footnote{\url{https://www.cfa.harvard.edu/MEarth/Welcome.html}} project \citep{Irwin2015_MEarth}, specifically with the MEarth-North array composed of eight 40\,cm telescopes, each equipped with a 25.6\,arcmin $\times$ 25.6\,arcmin FOV Apogee U42 camera, and located at the Fred Lawrence Whipple Observatory on Mount Hopkins nearby Tucson, Arizona, USA. Using the light curves from the latest data release\footnote{DR10: \url{https://lweb.cfa.harvard.edu/MEarth/DR10/north2011-2020/}}, the target was observed for more than six years with telescope one (``MEarth-tel01'') and telescope five (``MEarth-tel05''). The MEarth project generally uses a RG715 long-pass filter, except for the 2010–2011 season when an I$_{715-895}$ interference filter was chosen. In the case of \wolf, with observations collected from both telescopes later than October 2011, the RG715 filter was always used, and thus, we consider each photometric light curve as its own. The data were nightly binned following the same procedure as for the SuperWASP data. Particularly for first season with MEarth-tel05, we excluded certain nightly measurements where only one observation was taken to ensure accurate data quality. This constituted $\sim$15 nights out of the final 228 (Table~\ref{tab:photlogbook}), which were in the end not considered for the final rotational period determination due to large noisiness (see Sect.~\ref{sec:rotperiod}).







\paragraph{\tess.} 
\wolf\ was thus far observed in three of the northern sectors (15 -- camera \#2 CCD \#4; 16 -- camera \#2 CCD \#3; 17 -- camera \#3 CCD \#4, 15 August -- 2 November 2019) during the nominal mission of the Transiting Exoplanet Survey Satellite \citep[\tess;][]{Ricker2015} with the short 2-minute cadence photometry, as well as in three sectors (41 -- camera \#2 CCD \#1; 23 July 2021 -- 20 August 2021, 55 -- camera \#3 CCD \#3; 05 August 2022 -- 01 September 2022; 56 -- camera\#3 CCD \#3; 01 September 2022 -- 30 September 2022) during the extended mission with 2-minute and 20-second cadence\footnote{\url{https://heasarc.gsfc.nasa.gov/cgi-bin/tess/webtess/wtv.py}}. The target is being currently observed  in one sector (57 -- camera \#3 30 September 2022 -- 29 October 2022). The publicly available data from all sectors were downloaded from the Mikulski Archive for Space Telescopes\footnote{\url{https://mast.stsci.edu}}. Following the typical procedure, these data are corrected for artifacts and systematic trends \citep[Presearch Data Conditioning, \texttt{PDC\_SAP} flux --][]{Smith2012,Stumpe2012,Stumpe2014}, provided by the Science Processing Operations Center \citep[SPOC;][]{Jenkins2016}. We use these data for our analysis.

\begin{figure*}
    \centering
    \includegraphics[width=1\linewidth]{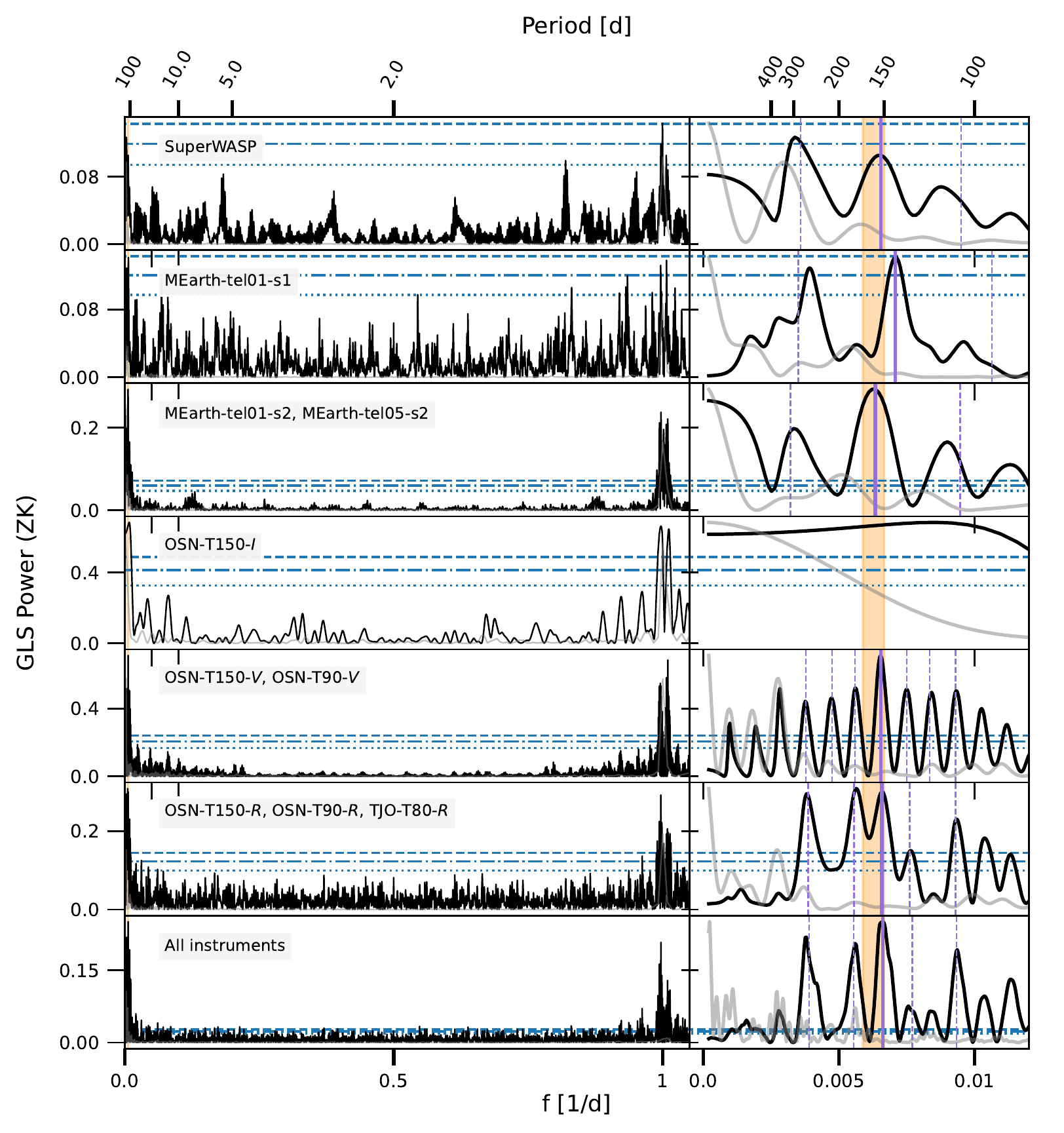}
    \caption{
    GLS periodograms for the long-term photometric data of \wolf. The right panels are zoomed in to the longer-period regime. Each horizontal panel represents an effective instrument that was considered for the determination of the stellar rotation period (Sect.~\ref{sec:rotperiod}). The normalized GLS power of the sampling of the data for each row is shown in gray. The range for the photometric rotation period of 150--170\,d is shaded in orange. Some significant alias signals due to the sampling of each respective data set are illustrated with a vertical dashed line, whereas the true signal is marked with a solid line.}
    \label{fig:glsperiodogramphot}
\end{figure*}

\begin{table*}
\caption{Observing log of ground-based long-term photometric observations acquired for \wolf.}
\label{tab:photlogbook}
\centering
\begin{tabular}{lcccccccc}
\hline \hline
\noalign{\smallskip}
Instrument  & \multicolumn{2}{c}{Date} & Filter  & $\Delta$t\tablefootmark{a} & \multicolumn{1}{c}{$N_\textnormal{obs}$} & \multicolumn{1}{c}{$N_\textnormal{nights}$\tablefootmark{b}}   &  \multicolumn{1}{c}{rms\tablefootmark{c}} \\
& Begin & End & &   (d)  &  &  & \multicolumn{1}{c}{(ppt)}  \\
\noalign{\smallskip}
\hline 
\noalign{\smallskip}
SuperWASP       & 13 June 2007   & 12 August 2008  & 400--700\,nm     &  426 & 10\,436 & 172 & 10.05 \\
\noalign{\smallskip}
\ldelim\{{2}{22mm}[MEarth-tel01 \  ]    & 29 September 2012   & 08 July 2015  & \rdelim\}{4}{17.5mm}[\ \  RG715]  & 1\,012  & 2\,595  & 183 &  6.02 \\
  & 13 September 2017   & 19 November 2018  &   & 431  & 6\,237  & 177 & 6.85  \\
\noalign{\smallskip}
\ldelim\{{2}{22mm}[MEarth-tel05 \  ]    & 28 May 2012{$^\dagger$}      & 10 November 2015{$^\dagger$}  &  & 1\,260  & 2\,716  & 228 & 5.44  \\ 
& 13 September 2017      & 19 November 2018  &  & 431  & 5\,665  & 175 &  6.52 \\ 
\noalign{\smallskip}
\ldelim\{{3}{17.5mm}[OSN-T150\ \ \ \ \ \ ]     & 31 August 2017  & 06 December 2017  & $V$     &  97 & 1\,322 & 41 &  8.65 \\
        & 31 August 2017  & 06 December 2017  & $R$     &  97 & 1\,276 & 39 &  5.90 \\
        & 14 September 2017  & 06 December 2017  & $I$     &  84 & 1\,078 & 37 &  6.38 \\
\noalign{\smallskip}
\ldelim\{{6}{20mm}[OSN-T90\ \ \ \ \ \ \ \ ]         & 29 June 2017      & 04 September 2017 &    \rdelim\}{3}{17.5mm}[\ \ \ \ \ \ $V$]    &  67 & 439 & 24 &  6.07 \\
        & 21 June 2019      & 29 October 2019 &        &  130 & 719 & 45 &  7.42 \\
        & 26 June 2020      & 02 November 2020 &       &  129 & 1187 & 64 &  8.73 \\
\noalign{\smallskip}
         & 29 June 2017      & 04 September 2017 & \rdelim\}{3}{17.5mm}[\ \ \ \ \ \ $R$]       &  67 & 442 & 24 &  5.68 \\
        & 21 June 2019      & 29 October 2019 &       &  130 & 716 & 45 &  8.10 \\
        & 26 June 2020      & 02 November 2020 &        &  129 & 1207 & 64 &  6.40 \\
\noalign{\smallskip}
TJO-T80         & 19 December 2018      & 12 January 2020 & $R$  &  388 & 1\,345 & 79 & 10.02  \\ 
\noalign{\smallskip}
\hline 
\end{tabular}
\tablefoot{
\tablefoottext{a}{Time span of the observation.}
\tablefoottext{b}{Number of nightly binned observations.}
\tablefoottext{c}{Root mean square in parts-per-thousand.}

Data sets that were not used for the photometric rotational period determination are indicated by a dagger {$^\dagger$}.
}
\end{table*}
\section{Stellar properties} \label{sec:stellar}

\subsection{Basic astrophysical properties}

\wolf\ (GJ~1253, Karmn J20260+585) is a slowly rotating, high-proper motion M5.0\,V star at less than 10\,pc discovered by \cite{Wolf1920}.
For decades, it was the subject only of astro-photometric analysis of stars in the solar neighborhood \citep{Luyten1955,GlieseJahreiss1979,Probst1983,Weis1984}, with the first spectral analysis by \cite{Bidelman1985}.
Since the end of the 20th century, \wolf\ was more investigated in X-rays \citep{Wood1994,Stelzer2013}, with high-resolution imaging \citep{Dieterich2012,Jodar2013,Janson2014,Lamman2020}, and especially for determining its astrophysical stellar parameters \citep{RojasAyala2012,Mann2015,Rajpurohit2018,Marfil2021}.

We summarize the stellar properties of \wolf\ in Table~\ref{tab:stellarparams}.
Following \cite{Cifuentes2020}, we integrated the spectral energy distribution and computed the bolometric luminosity using broadband photometry and the latest parallactic distance from \gaia\ Data Release 3 \citep[DR3;][]{GaiaDR3}.
From this value and the effective temperature of the star, determined spectroscopically by \cite{Passegger2019}, we derived the radius from the Stefan-Boltzmann law, and the mass from the mass-radius relation by \cite{Schweitzer2019}.
All tabulated parameters agree within uncertainties with published values \cite[e.g.,][]{Jenkins2009,Terrien2015,Lafarga2020}, except for the rotation period, which we concluded to be 150--170\,d. Our $P_{\rm rot}$ determination is explained in detail and compared with the literature in Sect.~\ref{sec:rotperiod}.
Although \wolf\ kinematically belongs to the Galactic thin disk (i.e., younger ages of $\sim$1-5\,Gyr) according to the galactocentric velocities calculated as in \cite{Montes2001} with a custom made python code (Cort\'es-Contreras et al. in prep.), the very long $P_{\rm rot}$ determined by us points toward older ages \citep[$\sim$7-11\,Gyr;][]{Newton2016}. 
The low activity indicators (H$\alpha$ and X-ray emission) and slow rotational velocity, $v \sin i$, are also in line with a long $P_{\rm rot}$. As a result, \wolf\ is a very weakly active star, which facilitates the identification of RV signals with very low semi-amplitudes.

\begin{table}
\centering 
\small
\begin{center}
\caption{Stellar parameters of \wolf.}
\label{tab:stellarparams}
\centering
\begin{tabular}{lcl}
\hline \hline
\noalign{\smallskip}
Parameter &  Value & Reference \\
\hline
\noalign{\smallskip}
\multicolumn{3}{c}{\textit{Basic identifiers and data}}\\
\noalign{\smallskip}
~~~Name                 & \wolf\             & Wolf1920 \\
~~~GJ                   & 1253                  & Gli57       \\
~~~Karmn                & J20260+585            & Cab16        \\
~~~\gaia\ DR3    & 2188318517720321664   &\gaia\ DR3  \\  
~~~$G$ (mag)        & $12.352 \pm 0.003$     & \gaia\ DR3\tablefootmark{a}\\
\noalign{\smallskip}
\multicolumn{3}{c}{\textit{Coordinates and spectral classification}}\\
\noalign{\smallskip}
~~~$\alpha$ (ICRS) & 20:26:05.80 & \gaia\ DR3  \\ 
~~~$\delta$ (ICRS) & +58:34:31.4 & \gaia\ DR3\\  
~~~Sp. type        & M5.0\,V     & Reid95 \\
\noalign{\smallskip}
\multicolumn{3}{c}{\textit{Parallax and kinematics}}\\
\noalign{\smallskip}
~~~$\pi$ (mas)  & 104.441 $\pm$ 0.026 & \gaia\ DR3     \\ 
~~~$d$ (pc)   & $9.5747\pm0.0024 $  & \gaia\ DR3      \\
~~~$\mu_\alpha \cos \delta$ (mas yr$^{-1}$)     & 261.038 $\pm$ 0.032    &   \gaia\ DR3   \\ 
~~~$\mu_\delta$ (mas yr$^{-1}$)                 & 542.906 $\pm$ 0.031   & \gaia\ DR3     \\ 
~~~$\gamma$ (k\ms) & --60.047 $\pm$ 0.020  & Lafa20 \\
~~~$U$ (k\ms)  & --23.1335 $\pm$ 0.0057 & This work \\
~~~$V$ (k\ms) & --61.2071 $\pm$ 0.0200 & This work \\
~~~$W$ (k\ms) & --8.4689 $\pm$ 0.0060 & This work \\
~~~Galactic population & Thin disk & This work \\
\noalign{\smallskip}
\multicolumn{3}{c}{\textit{Photospheric parameters}}\\
\noalign{\smallskip}
~~~$T_\textnormal{eff}$ (K)         & $3158 \pm 54$      & Pass19 \\
~~~$\log g_\star$ (cgs)                 & $4.93 \pm 0.06$   & Pass19\\
~~~[Fe/H] (dex)                     & $0.07 \pm 0.19$     & Pass19\\
~~~$v \sin{i}_\star$ (k\ms)              & <2 & Rein18 \\ 
~~~pEW(H$\alpha$)              & $-0.045 \pm 0.082$ & Fuhr22 \\
~~~$\left< B \right>$ (G) & <620 & Rein22\\
~~~$\log{F_X}$ (mW m$^{-2}$) & $-13.27$ & Stel13\\ 
~~~$P_\textnormal{rot}$ (d) &  150--170 & This work\tablefootmark{b}\\
\noalign{\smallskip}
\multicolumn{3}{c}{\textit{Physical parameters}}\\
\noalign{\smallskip}
~~~$L_\star$ ($10^{-4}\ L_\odot$)  & $29.44 \pm 0.28$  &  This work \\
~~~$R_\star$ ($R_\odot$)           &  $0.1813 \pm 0.0063$  & This work \\ 
~~~$M_\star$ ($M_\odot$)           & $0.167 \pm 0.011$   & This work \\
~~~Conservative HZ (\au)\tablefootmark{c} & [0.056,0.111] & This work\\
\noalign{\smallskip}
\hline
\end{tabular}
\tablefoot{
\tablefoottext{a}{See Table~\ref{tab:multibandphotometry} for multiband photometry different from \gaia\ DR3 $G$ band.}
\tablefoottext{b}{See Sect.~\ref{sec:rotperiod} for the $P_\textnormal{rot}$ determination.}
\tablefoottext{c}{For planets with $M_p = 1 M_\oplus$}
}
\tablebib{
    Cab16: \cite{Cab16}; 
    Cif20: \cite{Cifuentes2020};
    \gaia\ DR3: \cite{GaiaDR3}; 
    Fuhr20: \cite{Fuhrmeister2020};
    Fuhr22: \cite{Fuhrmeister2022};
    Gli57: \cite{GL57};
    Lafa20: \cite{Lafarga2020};
    Mar21: \cite{Marfil2021}.
    Pas19: \cite{Passegger2019};
    Reid95: \cite{Reid1995};
    Rein22: \cite{Reiners2022};
    Schw19: \cite{Schweitzer2019}; 
    Stel13: \cite{Stelzer2013};
    Wolf1920: \cite{Wolf1920}.
} 

\end{center}
\end{table}

\subsection{Stellar rotation period} \label{sec:rotperiod}

Using MEarth light curves from 2011 to 2014, \citet{DiezAlonso2019} determined a photometric rotation period of 57.7\,d for \wolf. However, the same data were analyzed earlier by \citet{Newton2016}, who reported only a tentative signal at 142.1\,d and noted that it did not meet their criteria for a confident detection. To further establish the stellar rotation period, we examined all available photometric measurements, including the newer-taken observations from the OSN and TJO telescopes, along with various stellar activity indicators available from the CARMENES spectra.

\paragraph{Long-term photometry.}

Considering each photometric data set alone, namely, OSN, TJO, SuperWASP, and MEarth, all indicate hints toward a prominent peak of $\sim$150--165\,d when taking a look at the generalized Lomb-Scargle \citep[GLS;][]{Zechmeister2009_GLS} periodograms, along with strong peaks present at the respective alias frequencies (Fig.~\ref{fig:glsperiodogramphot}). 
Interestingly, while this periodicity does present itself significantly in most photometric data sets, the most prominent peak in some is rather at a longer period ($\sim$200--300\,d), which can also be connected to an alias signal of the presumed rotational period due to the sampling for each of the respective data sets. Focusing on the first season of the photometry from MEarth, solely from tel-01 as the data from tel-05 are rather noisy, the highest peak is around 140\,d, similar to \cite{Newton2016}. On the other hand, the second observing block of the MEarth data, considering now data from both the tel-01 and tel-05 instruments, peaks clearly at $\sim$158\,d, with present alias signals (i.e., $\sim$110\,d and $\sim$300\,d) due to the sampling of 320\,d. 

To determine the rotational period of \wolf, we performed a fit with \juliet\footnote{\url{https://juliet.readthedocs.io/en/latest/index.html}} \citep{juliet} using the sum of two stochastically driven, damped harmonic oscillator (SHO) terms, or a double SHO Gaussian process (dSHO-GP), as done so in previous works such as, \cite{David2019}, \cite{Gillen2020}, and \cite{Kossakowski2021_toi1201} first considering all the data sets. A summary of the priors used for the analysis is found in Table~\ref{tab:protphotpriors}.
To set this up, we treated the OSN data as effectively different instruments arranged together only if the telescope size (i.e., T90 and T150) and filter (i.e., $V$, $R$, and $I$) were the same across multiple observational seasons. Each telescope of the MEarth data were separated into two temporal subsets, namely MEarth-tel05-s1 and MEarth-tel05-s2, and the same for MEarth-tel01 though without tel05, to account for possible changes in stellar activity behavior over long periods of time (i.e., $\sim$800\,d). The SuperWASP data were kept as is. We imposed a log-uniform prior for the rotational period, $P_\textnormal{rot}$, shared among all instruments ranging from 10\,d to 200\,d to avoid samples from populating near the longer-period alias signal. Likewise, the quality factor for the secondary oscillation, $Q_0$, and the difference between the quality factors of the first and second oscillations, $\delta Q$, were also shared. As for the fractional amplitude of the secondary oscillation with respect to the primary one, $f$, this parameter was shared among instruments with the same filter, that is, wavelength. The amplitude of the dSHO-GP, $\sigma_\textnormal{GP}$, followed suit as the dSHO-GP is trying to model the underlying physical behavior of the star that is wavelength dependent. To then account for any individual instrumental systematics, each respective instrument had its own offset value and jitter term\footnote{The jitter term is an additional noise count that is added in quadrature to the nominal uncertainty values.} \citep{Baluev2009}. 

From the posterior results of this fit, we obtain a rotational period of $\prot$\,d, compatible with the peaks in the periodograms and their widths (Fig.~\ref{fig:glsperiodogramphot}). 
To further experiment, we also tested out the same model setup, but this time solely on the photometry contemporaneously taken with the RVs, meaning those data comprising the last panel of Fig.~\ref{fig:phottimeseries}. The reason for considering just these data is that, it could be possible that the star underwent some changes in activity on the long scale due to spot migration or differential rotation. The results of this run give a rotational period of $170^{+15}_{-11}$\,d, which is 1-$\sigma$ well in agreement when considering all photometry, ensuring that the behavior of the star is still applicable to today. 
Therefore, we determine the photometric stellar rotation period to be $\prot$\,d. 

\paragraph{Spectroscopy.} 



We furthermore explored the various stellar activity indicators provided by the CARMENES spectra as well as the RVs themselves to look for agreement with the photometric rotational period.
\wolf\ is an H$\alpha$-inactive star \citep[pEW'(H$\alpha)\,\AA > -0.3$;][]{Schoefer2019}.

While there are no strong or moderate correlations found between the RVs and the stellar activity indicators using the Pearson's p-probability, the GLS periodograms of the indicators nonetheless consist of a variety of prominent peaks (see Fig.~\ref{fig:glsperiodogramwolf1069activity}). While some peaks found in the periodograms do coincide with the rotation period derived from the photometry, that is, $\sim$169\,d, there are nonetheless other existing signals with even higher significance ranging in periods from $\sim$200--260\,d, or even in the longer-period regime of up to a few hundred days. Particularly, these signals are present in the CRX, dLW, H$\alpha$ index, TiO~$\lambda\ 7050\,\AA$, TiO~$\lambda\ 8430\,\AA$, TiO~$\lambda\ 8860\,\AA$, CTR, and FWHM. 
There are still instances where the significant peaks coincide with the expected alias frequencies, however, the power at the rotational period is sometimes consistent with zero, as for instance in the Ca~\textsc{ii}~IRT3 indicator. This could be pure coincidental, we cannot exclude the possibility of it being a chance alignment. There are also peaks at even longer-period regimes (i.e., >1000\,d), which could be due to a stellar magnetic cycle not related to the stellar rotation itself.
This behavior appears similar to \object{EV~Lac} \citep[see][for a detailed analysis of the various periodicites]{Jeffers2022}, where it was shown that different indicators can respond with a phase lag, or nonuniformly with the rotational variation of the star.
In this work, we focus only on signals pertaining to the stellar rotation period, as well as those connected to the other RV signals (see Sect.~\ref{sec:signaldetection}). 

\wolf\ is considered to be a low-activity, low-mass star following the categorizations in \cite{Lafarga2021}. Hence, applying the findings made by \cite{Lafarga2021}, the most effective indicators for identifying the stellar rotation period comprise the dLW, CTR, and FWHM, which are tracing the varying width of the absorption lines in the spectra, as well as the indices of the chromospheric lines, H$\alpha$ and Ca~\textsc{ii} IRT, and sometimes the RVs themselves. The CRX and BVS in this instance are not as beneficial. 
Taking a closer look at these, we see that in fact, the dLW does peak at the expected rotation period and fits quite well to it. Likewise, when fitting for 169\,d, the CTR and FWHM also match quite well in the phase-folded plots (not shown). 
The H$\alpha$ index and Ca~\textsc{ii} IRT, however, do not agree with this periodicity, but rather show long-term trends, that may be related to a longer magnetic cycle as these indicators are also associated with magnetic cycles.
Another significant signal to mention in the dLW and the CTR appears at $\sim$29.5\,d, which is close to twice the orbital period of the planetary signal of interest (i.e., 15.6\,d, see Sect.~\ref{sec:modelcomparison}). However, this periodicity is most likely due to instrumental effects, namely, the contamination from the light of the Moon.
Within the RVs, there is a peak at 165\,d, though not significant, after subtracting out the other more prominent peaks (see Fig.~\ref{fig:rvperiodogram}~d). When performing the final fit on the RVs (see Sect.~\ref{sec:modelcomparison}), we apply a dSHO-GP on this signal arriving at a value of $P_\textnormal{rot,\ RV} = 165.6^{+3.3}_{-3.4}$\,d.

To bring all this evidence together, the photometric data point to a rotational period of $\sim$169\,d, the RVs to $\sim$165\,d, and a portion of stellar activity indicators similarly hint toward this value, though sometimes exhibit higher peaks at longer periods. 
Based on the photometry, it is evident that the variability of the star is changing from season to season (see Fig.~\ref{fig:phottimeseries}), so the periodicities detected in the indices skewing away from the predicted rotational period may be exhibiting the evolution of the spot configuration.
For example, some activity indices are evidently still consistent (i.e., same period or aliasing due to the sampling), although others are not as consistent, which could be a result of other issues, stellar cycles, spot evolution, data precision and sampling, that make the interpretation nontrivial. 
We therefore adopt the rotational period of \wolf\ to be within the range of 150$-$170\,d, with indications toward $P_\textnormal{rot} =\prot$\,d, though values outside this constrained range cannot be easily excluded. This is consistent with the predictions for a low-mass, inactive M dwarf \cite[][their Fig.~5]{Newton2017}.

\section{Analysis and results} \label{sec:analysis}
\subsection{Signal detection} \label{sec:signaldetection}
\begin{table*}[]
    \centering
    \begin{tabular}{c|l|SSSS}
\hline
\hline
\noalign{\smallskip}
    Sampling periods (d) & Alias order & \multicolumn{4}{c}{Present periods (d)} \\
\noalign{\smallskip}
\hline
\noalign{\smallskip}
& & {15.6} & {90.3} & {165} & {388}\\
& & {($f=0.06410\,\textnormal{d}^{-1}$)} & {($f=0.01107\,\textnormal{d}^{-1}$)} & {($f=0.00602\,\textnormal{d}^{-1}$)} & {($f=0.00258\,\textnormal{d}^{-1}$)}\\
\noalign{\smallskip}
\hline
\noalign{\smallskip}
 & $m=1$ & 14.8{$^\dagger$} & 67.7{$^\dagger$} & 102.8 & $\mathbf{159.2}$ {$^*$} \\
270& $m=-1$  & 16.6 & 135.7 & $\mathbf{431.0}$ {$^*$} & 887.8 \\

($f_s=0.00370\,\textnormal{d}^{-1}$)& $m=2$ & 14.0{$^\dagger$} & 54.1{$^\dagger$} & 74.5 & 100.2 \\
 & $m=-2$ & 17.6 & 272.7 & 722.9 & 207.0 \\
\noalign{\smallskip}
\hline
\noalign{\smallskip}
 & $m=1$ & $\mathbf{15.0}$ & 72.4{$^\dagger$} & 114.1{$^\dagger$} & $\mathbf{188.1}$ \\
365& $m=-1$ &  $\mathbf{16.3}$ & 120.0 & 304.5 {$^\dagger$} & 6157.4 \\
($f_s=0.00274\,\textnormal{d}^{-1}$)& $m=2$ & 14.4 {$^{\dagger *}$}  & 60.4{$^*$} & 86.9 & 124.1 \\
 & $m=-2$ & 17.1 & 178.7 & 1836.1 & 344.6 \\
\noalign{\smallskip}
\hline
\noalign{\smallskip}
 & $m=1$ & 15.3{$^\dagger$} & 82.1{$^\dagger$} & 140.1 & 271.0 \\
899& $m=-1$ & 15.9{$^\dagger$} & 100.4 & 203.6 & 682.6 \\

($f_s=0.00111\,\textnormal{d}^{-1}$)& $m=2$ & 15.1 & 75.2{$^{\dagger *}$} & 121.2 & 208.3 \\
 & $m=-2$ & 16.2 & 113.0 & 263.2 & 2835.9 \\
\noalign{\smallskip}
\hline
\noalign{\smallskip}
    \end{tabular}
    \caption{Aliases of the significant peaks found in the RVs given the sampling period tabulated. The periods of the aliases are computed using $P_\textnormal{alias} = 1/f_\textnormal{alias}$, where $f_\textnormal{alias} = f + m \cdot f_s$ following $f$ as the frequency of the true signal, $m$ as the alias order, and $f_s$ as the sampling frequency. Aliases that are significant (FAP > 10\%) are bolded, whereas those that are present but not significant are indicated with a dagger $^\dagger$. Ones marked with an asterisk signify peaks that are close to, but not exactly, the center of the expected alias.}
    \label{tab:rvaliases}
\end{table*}

\begin{figure*}
    \centering
    \includegraphics[width=1\textwidth]{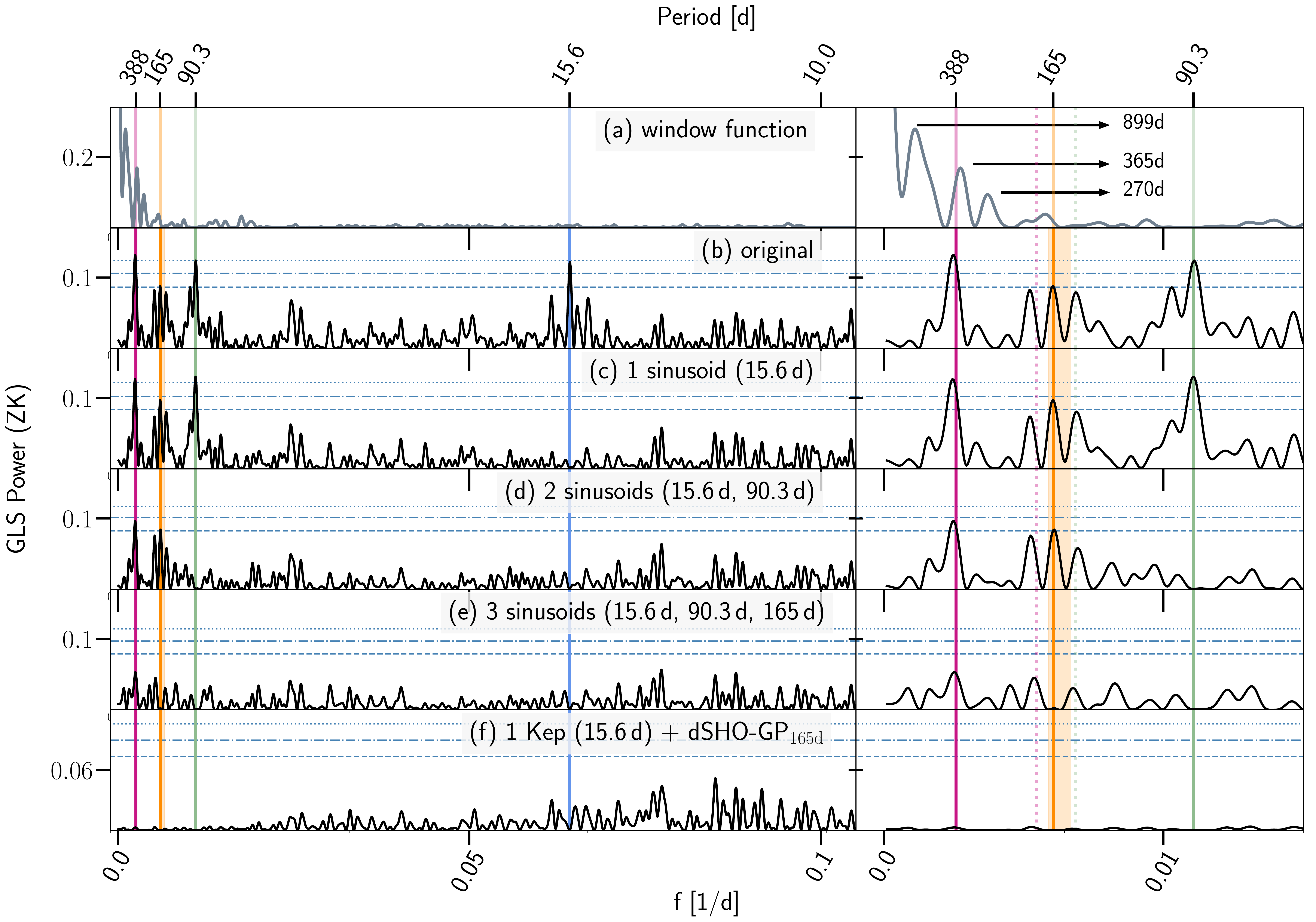}
    \caption{GLS periodograms for the RVs of \wolf\ after sequentially subtracting out the most prominent signals. The horizontal dashed, dot-dashed, and dotted lines represent the 10\,\%, 1\,\%, and 0.1\,\% FAP levels (from bottom to top). There were no significant signals with periods shorter than 10\,d other than the aliases due to the daily sampling. The right panel is a closer zoom-in of the left panel to highlight the longer-period signals. 
    The 15.6\,d planetary signal is illustrated with a vertical blue solid line, and the 90.3\,d signal and its alias due to the 270-d sampling with a green solid and dashed line, respectively. The range for the photometric rotation period is shaded in orange, where the stellar rotation period within the RVs is marked with an orange solid line.
    The component of residual telluric contamination at 388\,d and its alias due to the 365-d sampling period are also represented with a vertical magenta solid and dashed line, respectively.  
    Panel (a): the window function of the data set. Panel (b): no signal fitted, solely the original RVs with an offset and jitter term. Panel (c): residuals after subtracting the 15.6\,d signal. Panel (d): residuals after subtracting a simultaneous model fit of two sinusoids at 15.6\,d and 90.3\,d. Panel (e): residuals after subtracting a simultaneous model fit of three sinusoids at 15.6\,d, 90.3\,d, and 165\,d. Panel (f): residuals after subtracting the final model choice including 1 Keplerian at 15.6\,d (further described in Sect.~\ref{sec:modelcomparison}). 
    }
    \label{fig:rvperiodogram}
\end{figure*}

\begin{figure*}
    \centering
    \includegraphics[height=0.9\textheight]{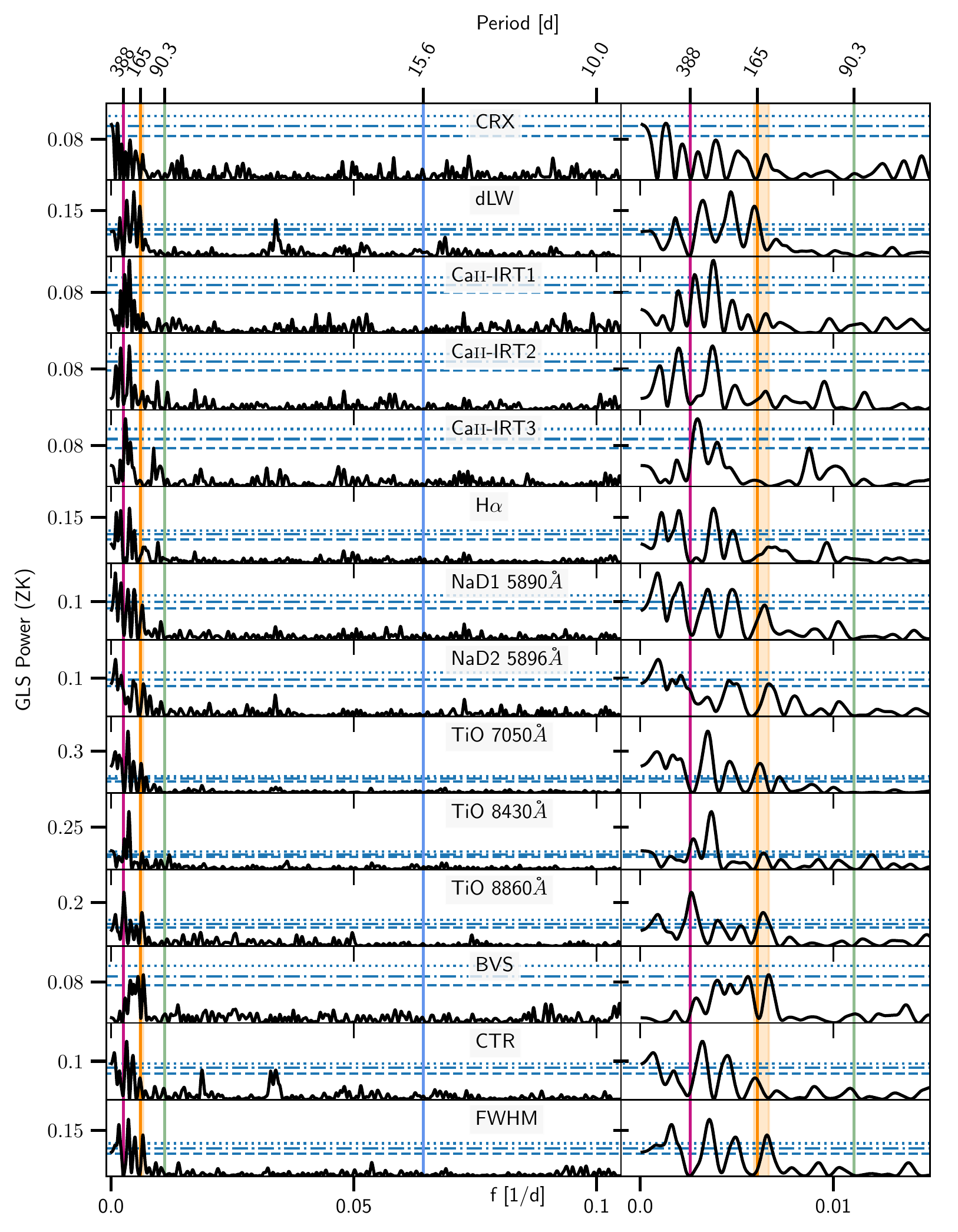}
    \caption{GLS periodograms for various known stellar activity indicators from the CARMENES spectroscopy. The window function of the data sampling can be found in Fig.~\ref{fig:rvperiodogram}~a. For consistency, the colored vertical lines and the frequency width of the panels on the left correspond exactly to those in the RV GLS periodograms (Fig.~\ref{fig:rvperiodogram}). The horizontal dashed, dot-dashed, and dotted blue lines represent the 10\,\%, 1\,\%, and 0.1\,\% FAP levels (from bottom to top).}
    \label{fig:glsperiodogramwolf1069activity}
\end{figure*}

We first computed a GLS periodogram on the RVs using the \exostriker\footnote{\url{https://github.com/3fon3fonov/exostriker}} tool \citep{exostriker} to initially identify potentially interesting signals. 
A series of GLS periodograms after sequentially subtracting out the most prominent sinusoidal signals, along with the discrete Fourier transform of the window function, can be found in Fig.~\ref{fig:rvperiodogram}. The sampling of the data showed three notable peaks at 899\,d, 365\,d (the yearly sampling), and 270\,d.
To quantify the significance of a signal's peak, we computed the false alarm probability (FAP) levels of 10, 1, and 0.1\,\% using 10\,000 randomizations of the input data where the frequency ranged from $1/t_\textnormal{baseline}$ to 1, with a frequency resolution of $1/t_\textnormal{baseline}$ and oversampling factor of ten. 

Already, in the original RV data set there are three prominent peaks at 15.6\,d, $\sim$90--100\,d and $\sim$400\,d, all hovering the 0.1\,\% FAP level (Fig.~\ref{fig:rvperiodogram}~b). To also mention, there are equally prominent peaks at 0.94\,d and 1.066\,d, both aliases of the 15.6\,d signal due to the daily sampling. Furthermore, there are strong yearly aliases for the 15.6\,d signal, particularly at 14.9\,d and 16.2\,d. Performing tests with the \aliasfinder\ \citep{Stock2020_aliasfinder}, we find that the true periodicity is the one at 15.6\,d (see Fig.~\ref{fig:15daliasfinder}), which we adopt when considering this signal. 
Continuing on, there are an additional three peaks around the 10\,\% FAP level, namely at 165\,d, $\sim$190\,d, and $\sim$145\,d, in order of significance, where the first one corresponding to the rotation period. The other two can be explained as aliases due to the 365-d and 270-d sampling of the $\sim$400\,d and $\sim$90--100\,d signal, respectively.
A full outline of all the aliases is tabulated in Table~\ref{tab:rvaliases}.

As it does not matter which of the peaks of equal significance is chosen first to subtract out for the prewhitening, we opt for the 15.6\,d considering its aliases are also those that are most prominent. 
After subtracting the 15.6\,d signal (Fig.~\ref{fig:rvperiodogram}~c), all corresponding aliases disappear, as expected.
Both the $\sim$90--100\,d and $\sim$400\,d signal reach above the 0.1\,\% FAP level, where the 165\,d, $\sim$190\,d, and $\sim$145\,d signals also become more prominent.
In the next succession of simultaneously modeling a two-sinusoidal model with the next highest peak (15.6\,d, 90.3\,d), there are no longer any signals that have an FAP level less than 0.1\,\%. The highest peak at $\sim$400\,d is above the 10\,\% FAP level, and its alias due to the yearly sampling (i.e., $\sim$190\,d) is apparent but no longer significant. The 165\,d is still above 10\,\% FAP level. 

The next signal to subtract for would be the one at $\sim$400\,d.
However, we have evidence that this signal is related to telluric contamination. We specifically compared the nontelluric-corrected RVs with those corrected for tellurics, finding that there is a prominent peak at 388\,d in the telluric component (Fig.~\ref{fig:redblue}, top panel). 
We conclude that even though we use the telluric-corrected spectra, the contamination still likely manifests as this process of telluric removal is nontrivial and, nonetheless, introduces residuals in the corrected spectra \citep{Nagel2022}.
Moreover, the peak in the RVs is more so at 404\,d, whereas the peak in the periodogram of the telluric component is actually at 388\,d. We presume that this could be due to the fact that the alias due to the 270-d sampling of the 165\,d signal is 420\,d, thus, adding some power there, in turn broadening the peak and veering the periodicity away from 388\,d to a value in between the two.
With this in mind, and considering the photometrically determined rotation period, we proceed with simultaneously subtracting the 165\,d signal next. Finally, there are no more signals above 10\,\% FAP after removing three simultaneous sinusoidal signals (15.6\,d, 90.3\,d, 165\,d). 

\subsection{RV model comparison}\label{sec:modelcomparison} 
 
Out of the three prominent signals, only the one at 165\,d does not reach our significance criterium (FAP < 0.1\,\%, though greater than 10\,\%), but is however related to stellar activity and thus might be important to model. 
We therefore continue the analysis by testing out ``three-signal models'' while properly considering the activity, with comparison to ``two-signal models'' by ignoring it. For reasons explained in Sect.~\ref{sec:90dsignal}, the origin of the 90.3\,d is ambiguous. Thus, we model it solely with a sinusoid when including this signal in a fit. For the 15.6\,d signal, there is no evidence against it to not be planetary, therefore, we apply circular or eccentric Keplerian orbits. For eccentric fits, the eccentricity was parametrized with $\mathcal{S}_1 = \sqrt{e}\sin\omega$ and $\mathcal{S}_2 = \sqrt{e}\cos\omega$ with uniform priors, $\mathcal{U}(-1,1)$ as recommended by \cite{Eastman2013}. 

The modeling and comparison of models is all performed with \juliet, a versatile python package for simultaneous transit and RV fitting, as already described in depth in other works such as \cite{Kemmer2020_toi488}, \cite{Stock2020_threesuperearths}, \cite{Bluhm2020_toi1235}, and \cite{Kossakowski2021_toi1201}. Following the same recipe as in the mentioned references, we use the computation of the Bayesian log evidence ($\ln \mathcal{Z}$) to compare models. Models with a $\Delta \ln \mathcal{Z}$ $\gtrsim$ 2.5 in comparison to another one show moderate evidence in favor of the winning model \citep[e.g.,][]{Trotta2008,Feroz2011}. Any value greater than 5 signifies strong evidence toward the winning model and anything below 2.5 indicates that neither of the two models are favored. 

To begin, we first ran a flat model with a white-noise term (i.e., jitter term, $\sigma_\textnormal{CARMENES-VIS}$) to act as a basis. Likewise, we ran a red-noise model using a dSHO-GP centered around the stellar rotation period (see below for setup details).
For the 15.6\,d signal, we imposed an appropriately sized uniform prior for the period, $\mathcal{U}$(15.4\,d, 15.7\,d) in order to cover the width of the peak as in the periodogram and to ensure that this true signal would be picked up rather than its nearby aliases. The time-of-transit center, which is used in \juliet\ to parametrize the phase of the orbit, was chosen to be uniform and set to cover one cycle during the most-sampled time epoch of the RV data set, $\mathcal{U}$(2458502\,d, 2458515\,d). 
Likewise, for the $\sim$90\,d signal, the prior on the period was kept wide enough to capture the tails of the posterior sample distribution, $\mathcal{U}$(85\,d, 95\,d).
To account for the stellar rotation period, we experimented modeling it with a sinusoid and with a dSHO-GP.
The photometrically determined rotational period of 150--170\,d (Sect.~\ref{sec:rotperiod}) was taken into consideration for the prior setup on the period. Within the RV periodograms (Sect.~\ref{sec:signaldetection}), the periodicity shows up as closer to 165\,d and encounters various neighboring signals due to aliasing of the other signals. For this reason, the prior on the period was kept relatively narrow and uniform from 155\,d to 175\,d. 
Table~\ref{tab:priors} showcases a full overview of the employed priors, including those for the instrument. 

\paragraph{Results.}
A table showcasing the Bayesian log evidence for the assortment of the tested models tested can be found in Table~\ref{tab:modelcomparison}. The winning model comprises one Keplerian for the 15.6\,d signal alongside a dSHO-GP centered on 165\,d to describe the behavior of the stellar rotation period. 
We note that including an extra sinusoid term for the 90.3\,d signal to this model is equivalent in terms of the Bayesian log evidence, even though this signal is evidently prominent in the GLS periodogram (Fig.~\ref{fig:rvperiodogram}). Given the ambiguity of the nature of this signal (Sect.~\ref{sec:90dsignal}), we find it appropriate to omit it from the final model and allow the dSHO-GP to moderately absorb it for the time being. The crucial aspect is that the interpretation of this signal and, thus, how we consider it in our models, which we acknowledge may adjust in the future, does not drastically alter the planetary parameters of the 15.6\,d signal (Fig.~\ref{fig:modelchoice}).

For the 15.6\,d signal, an eccentric Keplerian orbit was consistent with a circular one. The distribution of the eccentricity was consistent with zero. Focusing on the stellar rotation period, the dSHO-GP was preferred over a sinusoid ($\Delta \ln \mathcal{Z} > 10$) as it is most likely better suited in describing the quasi-periodic behavior of the stellar activity. Furthermore, this corresponds well to the fact that there seems to be a high level of spot evolution as we have encountered in the stellar activity indicators (Sect.~\ref{sec:rotperiod}) such that this is also demonstrated as an effect in the RVs. 

To conclude, the final model consists of ten free parameters applied on the 262 RV data points. The 15.6\,d signal is best described by a circular Keplerian model and the 165\,d signal by a dSHO-GP to account for the stellar rotation period. 
The best RV model from the posteriors is shown in Fig.~\ref{fig:rv}, where values for the derived planetary parameters can be found in Table~\ref{tab:posteriors_planet}. The full posterior overview for all model parameters is located in Table~\ref{tab:posteriors_all}. A visual inspection of the posterior probability densities for key parameters are displayed in Figs.~\ref{fig:cornerplot_p1} and \ref{fig:cornerplot_gp}.

\begin{table}
\caption{Model comparison using the Bayesian log evidences on the RVs for \wolf.}
\label{tab:modelcomparison}
\centering
\footnotesize
\begin{tabular}{l  S[table-format=1] S}
\hline \hline
\noalign{\smallskip}
Model & \multicolumn{1}{c}{$\ln{\mathcal{Z}}$} & \multicolumn{1}{c}{$\Delta\ln{\mathcal{Z}}$} \\\\
\noalign{\smallskip}
\hline 
\noalign{\smallskip}
\multicolumn{3}{l}{\textit{Base models}}\\
\noalign{\smallskip}
Flat & -640.00 & -33.14 \\
dSHO-GP$_\textnormal{165\,d}$ & -618.51 & -11.65 \\
\noalign{\smallskip}
\multicolumn{3}{l}{\textit{One-signal model}}\\
\noalign{\smallskip}
1 Kep$_\textnormal{15.6\,d}$ & -631.49 & -24.63 \\
\noalign{\smallskip}
\multicolumn{3}{l}{\textit{Two-signal models}}\\
\noalign{\smallskip}
1 Kep$_\textnormal{15.6\,d}$ + 1 Sin$_\textnormal{90.3\,d}$ &   -620.62 & -13.76 \\
\textbf{1 Kep$_\textnormal{15.6\,d}$ + dSHO-GP$_\textnormal{165\,d}$} &  $\mathbf{-606.86}$ & $\mathbf{0.0}$ \\
\noalign{\smallskip}
\multicolumn{3}{l}{\textit{Three-signal models}}\\
\noalign{\smallskip}
1 Kep$_\textnormal{15.6\,d}$ + 2 Sin$_\textnormal{90.3\,d, 165\,d}$ &  -616.83 & -9.96 \\
1 Kep$_\textnormal{15.6\,d}$ + 1 Sin$_\textnormal{90.3\,d}$ + dSHO-GP$_\textnormal{165\,d}$ &  -606.94 & -0.08 \\
\noalign{\smallskip}
\hline
\end{tabular}
\tablefoot{The chosen model was the 1 Kep$_\textnormal{15.6\,d}$ + dSHO-GP$_\textnormal{165\,d}$, as marked as the bold-faced row. 
A better model would have a larger, more positive $\Delta\ln{\mathcal{Z}}$. Regarding the model names, ``Kep'' refers to a circular Keplerian orbit and ``Sin'' to a sinusoidal signal. The period values are quoted as the median of the posterior distribution and can vary slightly depending on the model choice. 
}
\end{table}

\begin{table}
\centering
\caption{Derived posterior parameters for \wolf~b.}
\label{tab:posteriors_planet}
\begin{tabular}{lcl}
\hline
\hline
\noalign{\smallskip}
Parameter name & \multicolumn{1}{c}{Posterior\tablefootmark{(a)}} & Unit \\
 & b & \\
\noalign{\smallskip}
\hline
\noalign{\smallskip}

$P_\textnormal{p}$ & $15.564^{+0.015}_{-0.015}$ & d\\[0.1 cm]
$t_{0,p}$ (BJD) & $2458511.63^{+0.45}_{-0.46}$ & d\\[0.1 cm]
$K_\textnormal{p}$ & $1.07^{+0.17}_{-0.17}$ & m\,s$^{-1}$\\[0.1 cm]
$S_{1,p} = \sqrt{e_p}\sin \omega_\textnormal{p}$ & $0.0$ (fixed) & $\dots$\\[0.1 cm]
$S_{2,p} = \sqrt{e_p}\cos \omega_\textnormal{p}$ & $0.0$ (fixed) & $\dots$\\[0.1 cm]
$M\sin i_\textnormal{p}$ & $1.26^{+0.21}_{-0.21}$ & $M_\oplus$\\[0.1 cm]
$a_\textnormal{p}$ & $0.0672^{+0.0014}_{-0.0014}$ & au\\[0.1 cm]
$T_\textnormal{eq}$\tablefootmark{(b)} & $250.1^{+6.6}_{-6.5}$ & K\\[0.1 cm]
$S_\textnormal{p}$ & $0.652^{+0.029}_{-0.027}$ & $S_\oplus$\\[0.1 cm]

\noalign{\smallskip}
\hline
\end{tabular}
\tablefoot{\tablefoottext{a}{Error bars denote the $68\%$ posterior credibility intervals.}
\tablefoottext{b}{The equilibrium temperature of the planet assuming zero Bond Albedo and one emissivity.}}
\end{table}

\subsection{Investigating the low-amplitude 90.3\,d signal} \label{sec:90dsignal}

The 90.3\,d signal is significant with an FAP level near 0.1\,\% in the GLS periodograms (Fig.~\ref{fig:rvperiodogram}), yet it does not statistically improve the fit when including it in the RV models including also the rotation period (Table~\ref{tab:modelcomparison}). This could be an artifact that the semi-amplitude hovers around the 1\,\ms\ limit ($K_\textnormal{90.3\,d}$ = 91 $\pm$ 38\,c\ms), thus, making it possibly difficult to justify including such a low-amplitude signal with a relatively large uncertainty in the models. Likewise, the dSHO-GP kernel used in the model is equipped to pick up the rotational period, $P_\textnormal{rot}$, and half of the rotational period, $P_\textnormal{rot}$/2. It could be that 90.3\,d is decently close to 83\,d  (i.e., $P_\textnormal{rot}$/2) and, thus, the dSHO-GP is performing its job of absorbing this signal. In fact, this is demonstrated in the GP component of the RV model shown in Fig.~\ref{fig:rv}. It is nonetheless evident that this periodicity is present in the RV data as can be seen in the residuals of the RV model. Its nature, however, appears to be quite dubious. Below, we explore the signal further to better understand its origin. 

\begin{figure}
    \centering
    \includegraphics[width=1\linewidth]{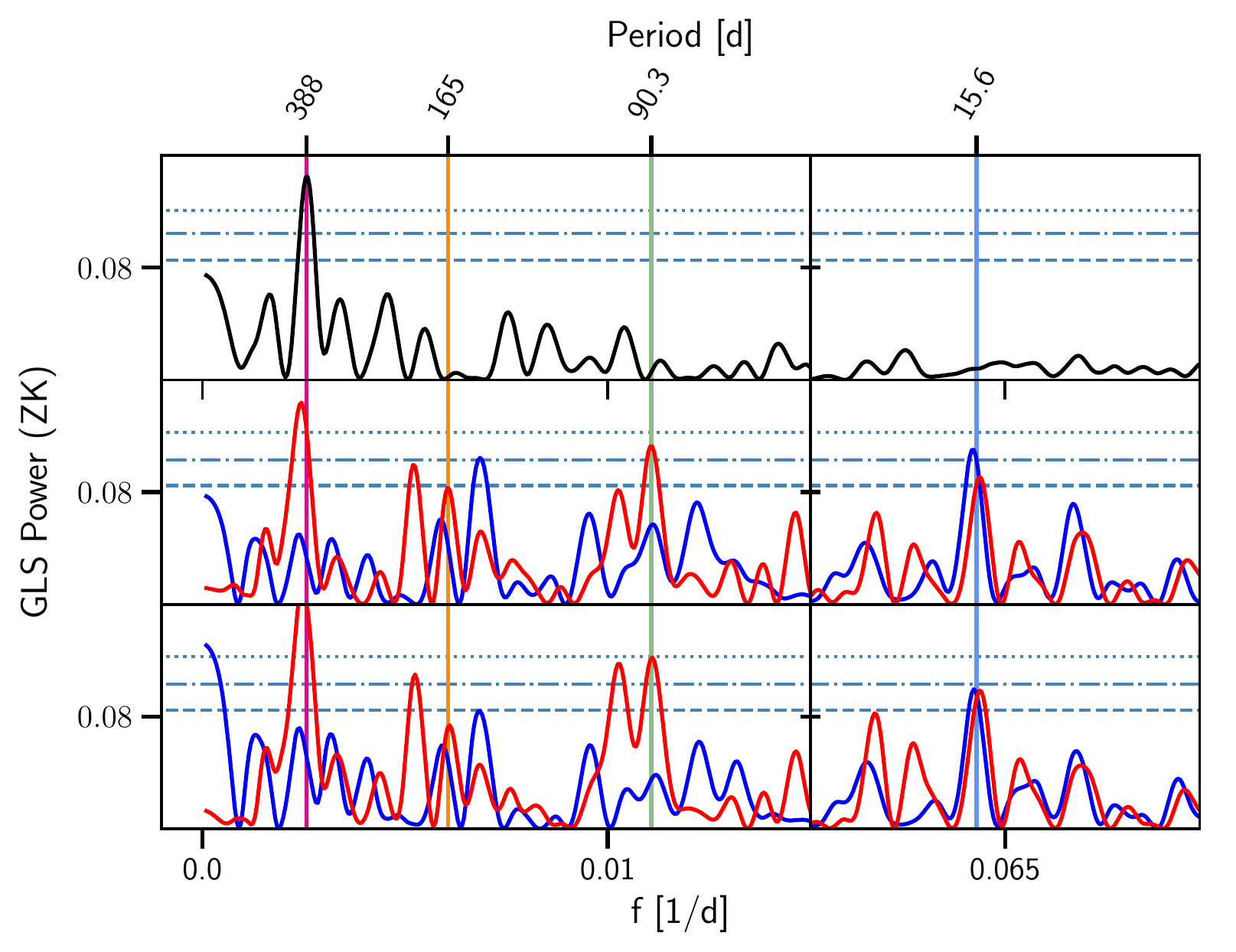}
    \caption{GLS periodograms of the telluric-only (nonTC subtracted from TC; \textit{top}), nonTC (\textit{middle}) and TC (\textit{bottom}) RVs. The red and blue colors represent the ``blue'' and ``red'' subsets within the VIS channel of the CARMENES instrument (Sect.~\ref{sec:90dsignal}). The vertical lines match those in Fig.~\ref{fig:rvperiodogram} for consistency and the horizontal lines are identical in the bottom panels but differ slightly from the top panel.
    }
    \label{fig:redblue}
\end{figure}

Its periodicity is near a quarter of one year. A suspicion could be that this signal should be present in the telluric-contamination-only component of the RVs, though, it is not (Fig.~\ref{fig:redblue}, top panel). We additionally tested breaking up both the  telluric-corrected (TC) and nontelluric-corrected (nonTC) RVs into ``blue'' and ``red'' subsets. To do this, we can recompute the RVs by selecting certain orders. The CARMENES VIS channel consists of 55 RV orders, 42 of which are used to compute the RV measurement via a weighted mean through \serval\ \citep{serval}.
For the blue and red subset, we considered the first 21 and last 21 orders, respectively. Thus, the blue and red subsets span roughly 570\,nm to 700\,nm and 700\,nm to 910\,nm, respectively. A GLS periodogram for both subsets and for both data sets is shown in Fig.~\ref{fig:redblue}. 
Taking a look at the nonTC spectra, the 388\,d (telluric-attributed) signal, its yearly alias at 188\,d, and the 90.3\,d, as well as a neighboring signal at 97\,d, are substantially stronger in the red than in the blue. This is in agreement that the telluric contamination is stronger in the red than in the blue, given that there are sharper, deeper telluric-absorption features in the redder part of the VIS channel \citep[Fig.~1 in][]{Reiners2018}. 
Meanwhile, the power of the 15.6\,d and 165\,d signals is consistent within one another in both subsets. 
After the telluric correction, some residual telluric contamination remains, though dampened, indicating that the correction for tellurics was indeed effective but left some residual effect as can be seen in the RV periodograms (Fig.~\ref{fig:rvperiodogram}). The power of the 15.6\,d and 165\,d signals is still compatible, which is most important. 
Therefore, even though the 90.3\,d has no appearance in the telluric-contamination-only component, it does exhibit chromaticity and behavior similar to other telluric signals, pointing less in favor for a planetary signal and potentially more in favor of telluric effects. It is, however, puzzling as to why the 90.3\,d signal does not peak in the telluric-only RVs (Fig.~\ref{fig:redblue}). 

\begin{figure*}
    \centering
    \includegraphics[width=1\linewidth]{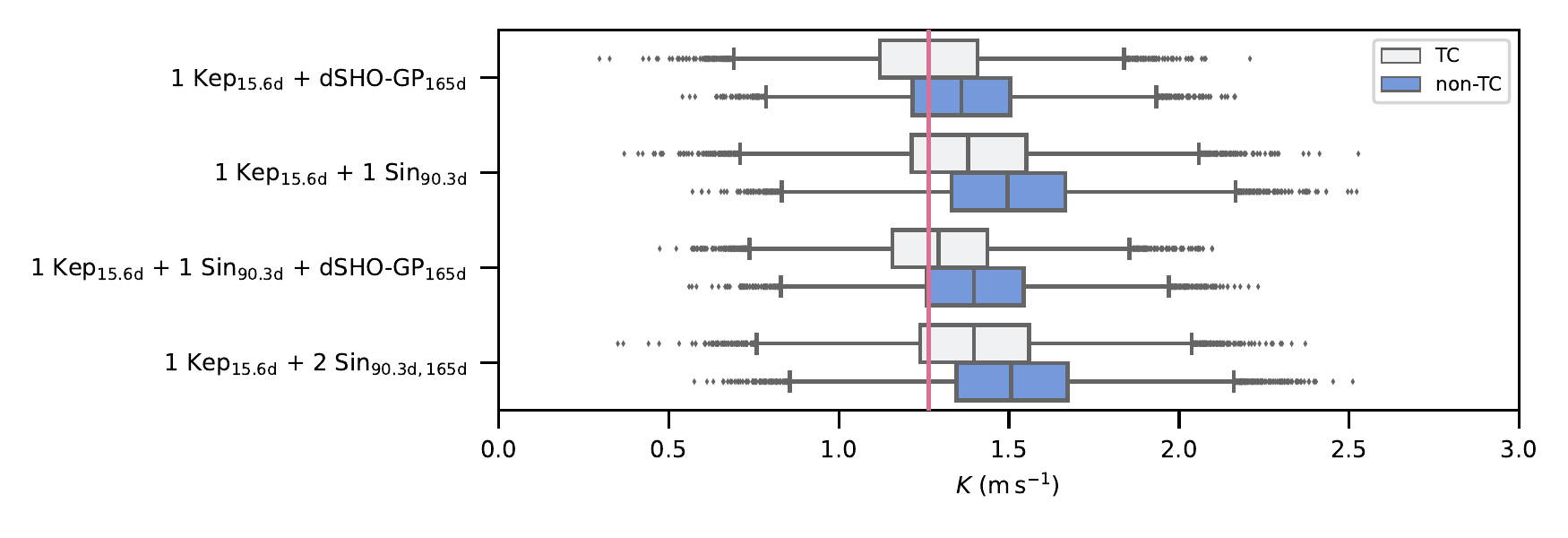}
    \caption{Box plot of the posteriors for the distributions of the minimum mass for the 15.6\,d signal based on the model choice. The gray and blue boxes represent the 25\,\% and 75\,\% quartiles of the posterior from the telluric-corrected (TC) and nontelluric-corrected (nonTC) RVs, respectively. The red vertical line represents the median value of the minimum mass of the 15.6\,d signal when applying the most favored model. The extending gray lines depict the rest of the distribution and the dots are deemed as ``outliers''. The models named here match those in Table~\ref{tab:modelcomparison}.}
    \label{fig:modelchoice}
\end{figure*}

To summarize, we are not able to distinctly decipher the origin of the 90.3\,d signal. Even if it were planetary, we currently do not have enough evidence to support this claim. As our main concern is the 15.6\,d signal and we presented that its planetary parameters are independent of whether the 90.3\,d signal is considered or not (Fig.~\ref{fig:modelchoice}, particularly between 1 Kep$_\textnormal{15.6\,d}$ + 1 dSHO-GP$_\textnormal{165\,d}$ and 1 Kep$_\textnormal{15.6\,d}$ + 1 Sin$_\textnormal{90.3\,d}$ + dSHO-GP$_\textnormal{165\,d}$), we choose not to include this signal in the final model as a precaution. Further investigation or RV monitoring may be beneficial, however, this lies beyond the scope of this paper.

\subsection{Transit search within \tess} \label{sec:transitsearchtess}

With a minimum mass of $1.26 \pm 0.21\,M_\oplus$ and assuming an Earth-like core-mass fraction of 0.26, we obtained a radius estimate of $1.08\,R_\oplus$ for \wolf~b using the mass-radius relation as given by \cite{Zeng2016}. This then translates to an expected transit depth of $\sim$3.6\,ppt, which should be easily detectable with \tess, though not with the other available photometric facilities (i.e., SuperWASP and MEarth). The transit probability is however rather low at 1.2\% ($p \approx R_\star/a_\textnormal{p}$). Nonetheless, propagating the orbital period and $t_0$ from the RV fit with their 1-$\sigma$ uncertainties (taken from Table~\ref{tab:posteriors_planet}), we unfortunately did not find any hint of a possible transit. We additionally checked to confirm that the transits could not have happened during the data gaps, where only one would fall within an observational gap. Likewise, we checked with the transit-least-squares\footnote{\url{https://github.com/hippke/tls}} method \citep{Hippke2019}, though no interesting signals popped up. 
Given this information, we were able to obtain a maximum inclination for \wolf~b of $i_{\mathrm max} = \arccos\left(R_\star/a_\textnormal{p}\right) = 89.35$\,deg.


\section{Discussion} \label{sec:discussion}

\subsection{On the promising habitability of \wolf~b} \label{sec:hz}

Plugging in the stellar luminosity and effective temperature (Table~\ref{tab:stellarparams}), \wolf~b, with a distance of $0.0672 \pm 0.0014$\,\au\ to the star, sits comfortably within the conservative HZ limits, namely, 0.056\,\au\ to 0.111\,\au\ given the runaway-greenhouse and maximum-greenhouse limits, respectively \citep{Kopparapu2013,Kopparapu2014}. 
Even more so, it is very likely that \wolf~b is indeed an Earth-like planet with Earth-like composition (32.5\% iron mass fraction and 67.5\% silicates) and radius around one Earth radii \citep[following Fig.~1 in][]{LuquePalle2022}, as we also estimated in Sect.~\ref{sec:transitsearchtess}.
Figure~\ref{fig:hz} puts \wolf~b in context with other planets around M-dwarf stars that are most likely to have a rocky composition and maintain surface liquid water as listed in the Habitable Exoplanet Catalog\footnote{\url{https://personal.ems.psu.edu/~ruk15/planets/}} with some modifications (Appendix~\ref{appendix:planetshz}). To this effect, \wolf~b resembles best \object{Proxima Centauri}~b, \object{GJ~1061}~d, \object{Teegarden's Star}~c, \object{Kepler-1649}~c, and \object{GJ~1002}~b and~c. With the exception of \object{Kepler-1649}~c, all are RV-only detections. Furthermore, all 14 planetary systems illustrated contain more than one planetary companion, excluding \wolf, \object{Ross~128}, and \object{Kepler-1229}, as discussed in Sect.~\ref{sec:wolfplanetarysimulations}. 
When considering the occurrence rate for planets with 1--10 Earth masses on periods shorter than 10\,d around later-type M dwarfs, this value lies between $\sim$0.56--1.75 planets per star \citep{Ribas2022, Hardegree-Ullman2019}. 
Regarding the proximity of these systems, \cite{DressingCharbonneau2015} estimated that the nearest nontransiting HZ planet could be $2.6 \pm 0.4$\,pc away, and is within 3.5\,pc with $95\%$ confidence for potentially habitable 1--1.5\,$R_\oplus$ planets. Soon after, \object{Proxima Centauri}~b was discovered at a distance of 1.30\,pc \citep{Anglada-Escude2016a}, \object{GJ~1061}~d at 3.67\,pc \citep{Dreizler2020_gj1061}, Teegarden's~Star~c at 3.83\,pc \citep{Zechmeister2019_Teegarden}, and \object{GJ~1002}~b and~c at 4.85\,pc \citep{Mascareno2022}. \wolf~b is located at a distance of 9.57\,pc, making it the sixth closest, conservative HZ Earth-mass planet to us. Other closer contenders included \object{Ross~128}~b ($d=3.38$\,pc) and \object{GJ~273}~b ($d=5.83$\,pc), though these planets lie in the optimistic HZ. 
 
\begin{figure}
    \centering
    \includegraphics[width=1\linewidth]{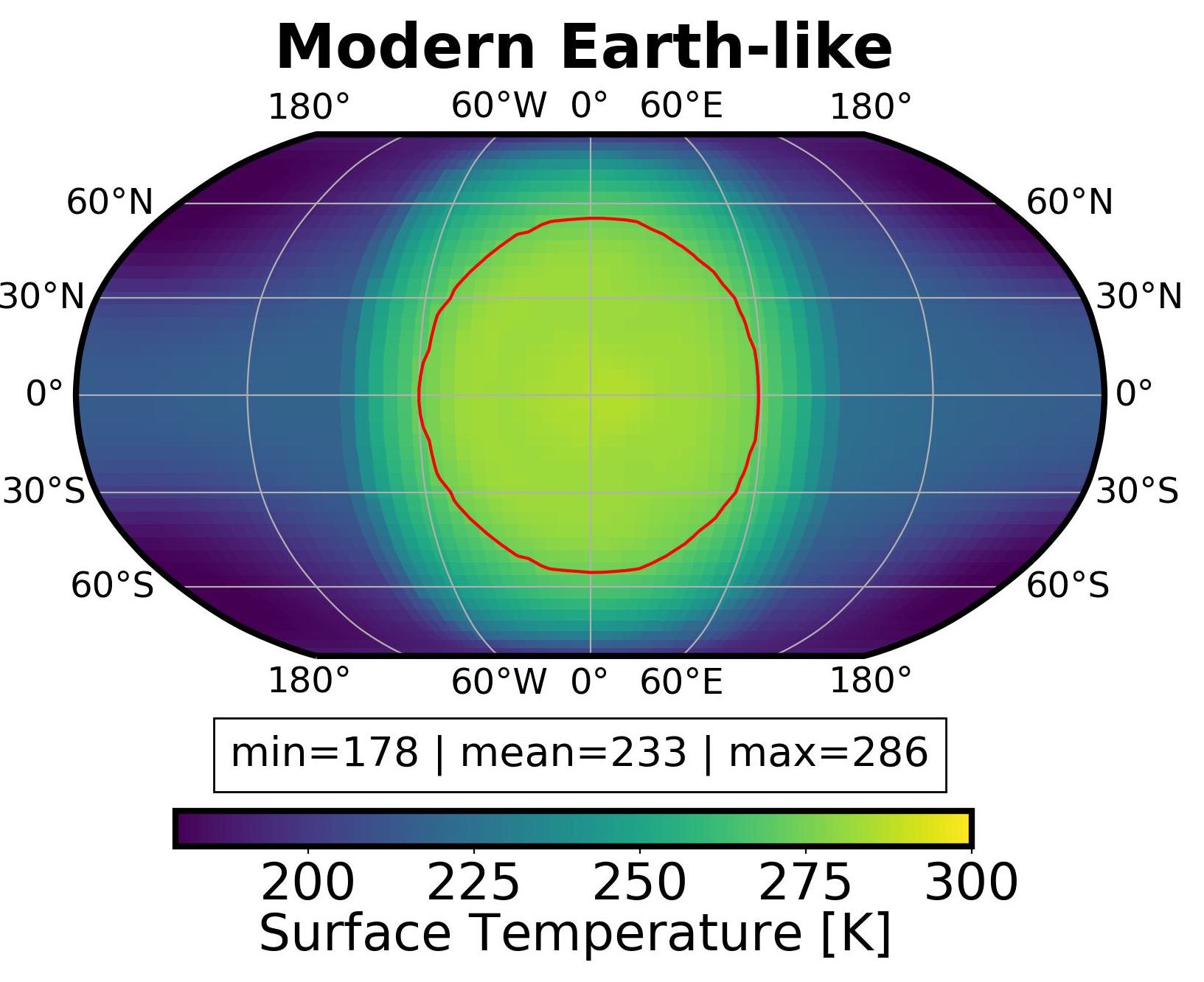}
    \caption{Surface temperature map of Wolf 1069~b produced by the ExoCAM GCM, assuming a Modern Earth-like atmosphere. The map is centered at the substellar point and the red line delimits the area where water is at the liquid phase on the surface.}
    \label{fig:ExoCAM}
\end{figure}

\begin{figure}
    \centering
    \includegraphics{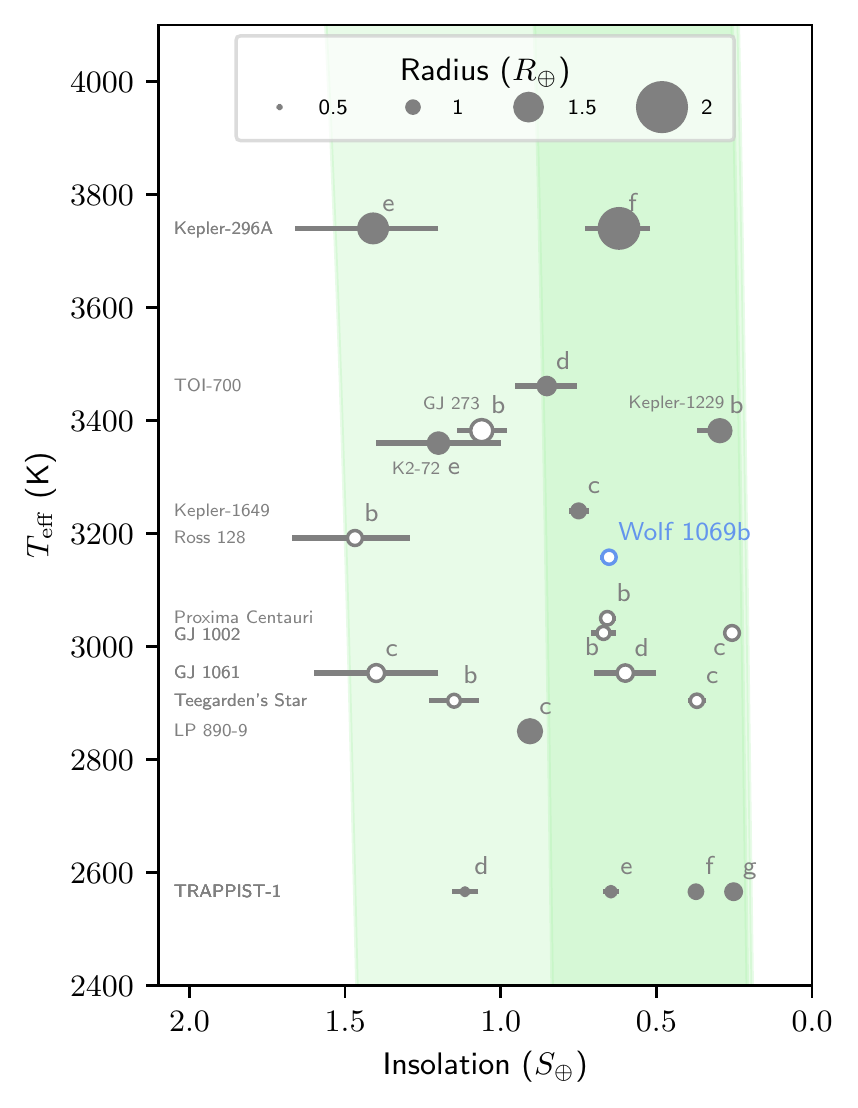}
    \caption{M-dwarf (T$_\textnormal{eff}$ < 4000\,K) planetary systems with at least one detected planet in the conservative sample of potentially habitable exoplanets (i.e., 0.5\,$R_\oplus<R_\textnormal{p}<1.6\,R_\oplus$ or 0.1\,$M_\oplus<M\sin{i}_\textnormal{p}<3\,M_\oplus$) defined by the Habitable Exoplanet Catalog. The optimistic and conservative HZ regions for a one Earth-mass planet following the definition as set out by \cite{Kopparapu2013} are shaded with light and dark green, respectively. Only the planets in either the conservative or optimistic HZ of each planetary system are shown. White-filled and gray-filled points indicate nontransiting and transiting detections, respectively. The size of the circles is proportional to the planetary radius, estimated with the mass–radius relationship of \cite{Zeng2016} for nontransiting RV planets. The data used in this plot is further discussed in Appendix~\ref{appendix:planetshz}. Plot inspired by \cite{Zechmeister2019_Teegarden} and \cite{Dreizler2020_gj1061}.}
    \label{fig:hz}
\end{figure}

\wolf~b is in the slow rotator regime, and possibly in tidal equilibirum rotation \citep[e.g.,][]{Heller2011}, that can lead to unique atmospheric circulation pathways \citep[e.g.,][]{Dole1964,Yang2019,delGenio2019b}.
The impacts of this slow rotation on both the potential habitability and impact on observations have been discussed in detail by several 3D GCMs \citep[see e.g.,][]{Edson2012,Leconte2013}.
Preliminary results from GCMs climate simulations using both the ExoCAM model \citep{Wolf2022} and the ROCKE-3D model \citep{Way2017} suggest that \wolf~b could maintain moderate temperatures and surface liquid water for a large range of atmospheric compositions and surface types. Simulations explore a variety of surface pressures, N$_2$, CO$_2$, CH$_4$, and H$_2$O abundances, along with desert, solid rock, slab ocean, and dynamic ocean surfaces. The comprehensive analysis of these 3D climate results and the observational signals that could be used to differentiate between climate states of \wolf~b show the planet to be durably habitable (Crouse et al. in prep.). Figure~\ref{fig:ExoCAM} shows the surface temperature map produced with the ExoCAM GCM assuming a Modern Earth-like atmospheric composition. The red line delimiting the open ocean shows that a significant fraction of the day side surface could maintain liquid water, therefore day-side habitable conditions.
While the presence and nature of any atmosphere on \wolf~b (and existing M-dwarf planets in general) remains theoretical, the support of habitable conditions over such a wide range of possible atmospheric states puts \wolf~b in the same elevated class as \object{Proxima Centauri}~b \citep{Turbet2016, delGenio2019}, \object{TRAPPIST-1}~e \citep{Wolf2017, Turbet2018, Fauchez2019}, and \object{TOI-700}~d \citep{Suissa2020} as a primary target to search for habitability and biosignature markers. 

Similar to \object{Proxima Centauri}~b, \wolf~b does not transit its host star, meaning that observation and analysis of thermal emission and reflected light phase curves will need to be employed to probe its atmosphere. Given the brightness of the host star and the distance to Earth, and assuming atmosphere models and albedo similar to the ones predicted for the \object{TRAPPIST-1} planets and \object{Proxima~Centauri}~b \citep{Turbet2022}, the atmospheric characterization of \wolf~b might be within the reach of the ELT\footnote{Extremely Large Telescope.} instrumentation. 
ANDES\footnote{ArmazoNes high Dispersion Echelle Spectrograph, \url{https://elt.eso.org/instrument/ANDES/}} \citep[formerly known as HIRES;][]{Maiolino2013} will be the first instrument theoretically capable of detecting the reflected light from HZ rocky planet atmosphere in the early 2030's. However, for \wolf~b, the small angular separation in the sky between the planet and the host stars, 7.01\,milliarcsec, makes these observations very challenging, even with the use of extreme adaptive optics systems. Further instrumental advances, such as the proposed PCS\footnote{Planetary Camera and Spectrograph.} instrument for the ELT \citep{Kasper2021} or space-based coronographic/interFerometric missions, might be needed. While such observations are very challenging, many of the nearest planets found in the conservative HZ around M dwarfs are nontransiting, RV detections, indicating that perhaps more time and investment into the development of such observations should be considered if we want to establish ground statistics using all of the thus-far detected, potentially habitable worlds. 

\subsection{The case of \wolf~b as a lone, short-period planet} \label{sec:wolfplanetarysimulations}

\begin{figure}
    \centering
    \includegraphics[width=0.5\textwidth]{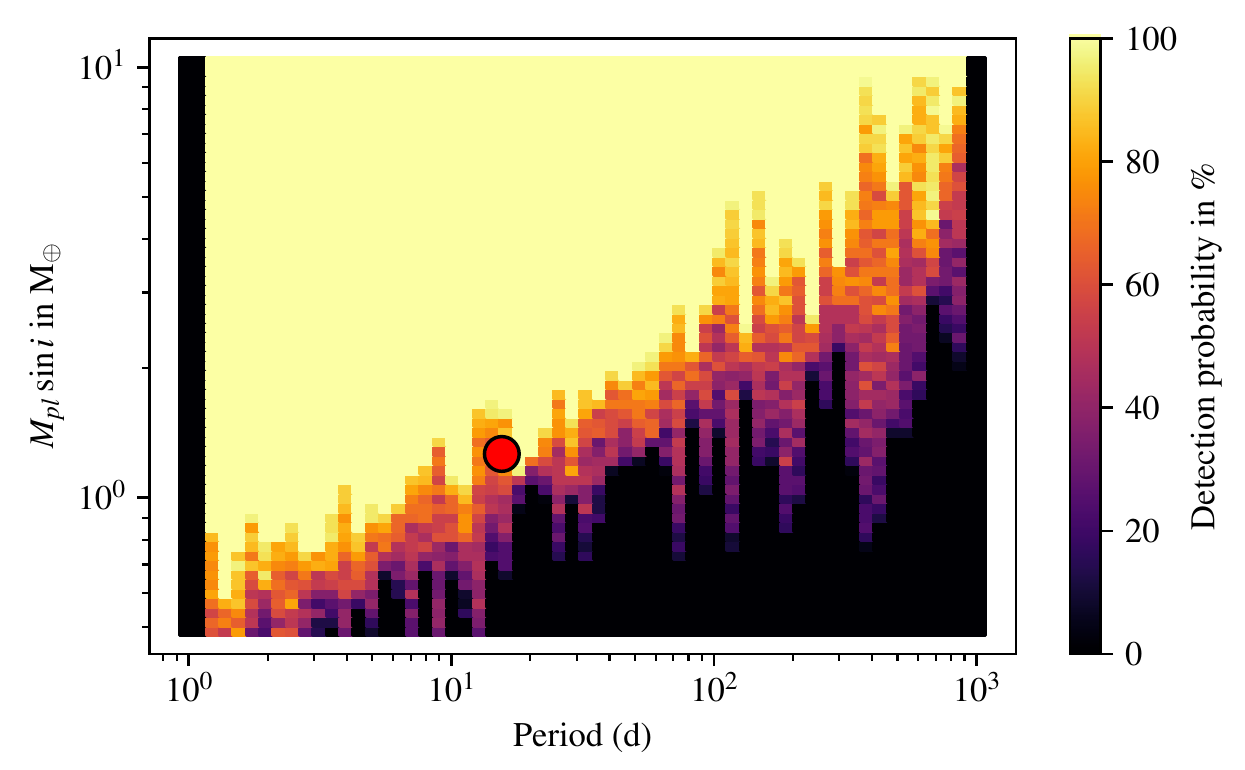}
    \caption{
    RV detection map of \wolf\ from an injection-and-retrieval experiment after subtracting the 15.6\,d, 90.3\,d and 165\,d signals. The red circle indicates the planet \wolf~b.}
    \label{fig:detectionmap}
\end{figure}

Our comprehensive analysis of the RV and photometric data suggests that \wolf~b is the only bonafide planet in the sensitive domain of the planetary parameter space.
We characterized this domain with injection-and-retrieval tests by taking the residuals of the winning model without a GP (Sect.~\ref{sec:modelcomparison}) and creating simulated RV time series using Eqn.~2 in \cite{Sabotta2021}. This was repeated 50 times on 30 log-uniformly distributed grid points in mass and 60 in period, allowing us to rule out additional planets with at least one Earth mass and periods of less than 10~days (Fig.~\ref{fig:detectionmap}).
\wolf~b joins a sample of currently two RV-detected, single terrestrial planets ($\lesssim 2\ M_\oplus$), which have all been detected around M dwarfs less massive than $0.5\,M_\odot$. 
These objects are \object{GJ~393}~b~\citep{Amado2021} and \object{Ross~128}~b~\citep{Bonfils2018}, where the former resides in the HZ region of its host star (see also Sect.~\ref{sec:hz}), whereas the latter receives too much flux. 
Likewise, transiting planets following the same criteria include \object{GJ~367}~b~\citep{Lam2021} and \object{GJ~1252}~b~\citep{Shporer2020}, where the latter has rather a tentative measured mass of $2.09 \pm 0.56\,M_{\oplus}$ and both are not found in the HZ of their parent stars.
Nonetheless, the small sample size raises the question how frequent the solitary occurrence of such a planet is.
The overall occurrence rate of small, rocky planets on close orbits has been shown to be larger around host stars of later spectral type~\citep[e.g.,][]{Mulders2015,Hardegree-Ullman2019,Hsu2020}.
However, there is some indication that this rule might not apply for the latest M~dwarfs~\citep{Gibbs2020,Sebastian2021,Sestovic2020,Brady2021,Mulders2021} and that systems, such as the one presented here, could be in fact rare.

Planet formation models following the core accretion paradigm~\citep{Pollack1996} generally suggest a high multiplicity of Earth-mass planets around mid-M-dwarf host stars~\citep{Burn2021}.
However, these models produce many planets beyond current detection limits, and bias-corrected synthetic populations show sufficiently reduced rates to be compatible with the observations~\citep{Schlecker2022}.
The discovery of a single planet comparable to \wolf~b is consistent with this picture. 

\begin{figure}
    \centering
    \includegraphics[width=0.5\textwidth]{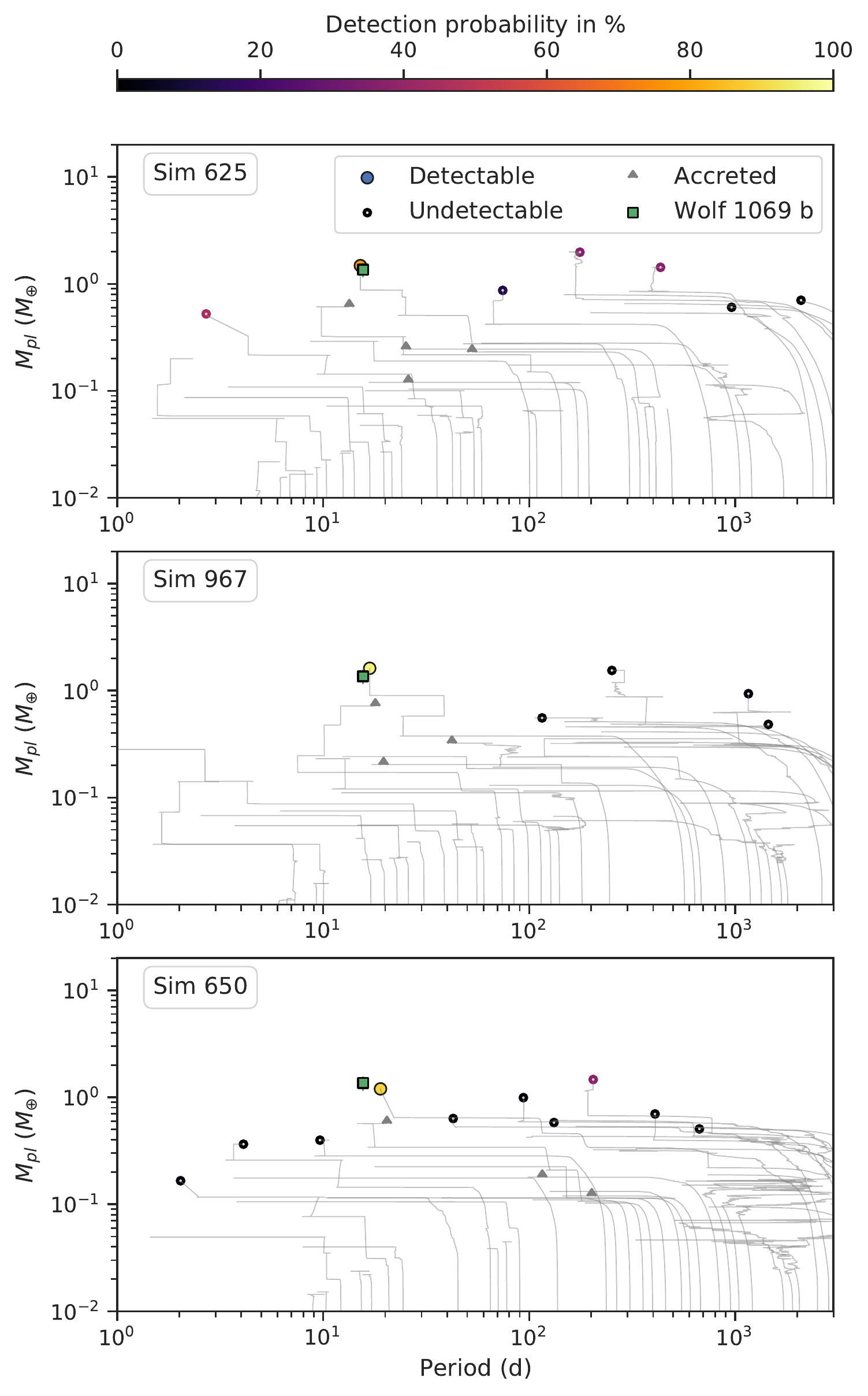}
    \caption{Formation paths and final planets in planet formation simulations taken from the population synthesis work of \citet{Burn2021}. We show the three simulations with a single detectable planet closest to \wolf~b in relative mass and orbital period. A planet is labeled ``undetectable'' if the detection probability is below 50\,\%. Formation tracks are shown as gray lines that can end in either a filled circles (detectable planet), a triangle (accreted by detectable), a ring (undetectable), or without a marker (accreted by other planets or ejected).}
    \label{fig:simulation}
\end{figure}

We tested this scenario by applying the computed detection sensitivity (Fig.~\ref{fig:detectionmap}) to the synthetic planetary population \textit{NGM10} around an 0.1\,M$_{\odot}$ star presented in \citet{Burn2021}.
Using 50\,\% detection probability limits, 48 out of a total of 1000 systems would result in a single detection.
Out of those, we show in Fig.~\ref{fig:simulation} the three simulations that result in planets closest to \wolf~b on the period-versus-mass plane. Simulations leading to a single planet detection went through a stage of giant impacts reducing the number of planets in the inner system and increasing the mass of the detectable planet with respect to the rest of the system. This is exemplified by the three best-fitting simulations which show three to four mergers with embryos more massive than the lunar mass. 

While the scenario of the formation of a single planet cannot be ruled out, those simulations show that it is also possible in $\sim$5\,\% of the cases to form a seemingly lone planet if multiple embryos formed at the same time. However, if future observations extend the detection limits to larger orbits and lower planetary masses, this formation theory will be more severely challenged. While a single, late stage giant impact with a similarly massive body is currently in agreement with observations (e.g., Sim 650), this could be ruled out with better sensitivity. Then, a more dynamic history of the system is required (as in Sim 967 where the complete inner system was ejected or accreted by the detectable planet).
A handicap of particular importance for thorough analyses of planet multiplicity is the omission of early core formation phases in current formation models~\citep[see e.g.,][]{Ormel2017,Schlecker2021b}.
Future planet population synthesis studies have to take into account dust evolution, planetesimal formation, and planetary embryo formation in a self-consistent manner \citep{Voelkel2020,Voelkel2021}.

As for the observational prospects, dedicated measurements with a high-precision spectrograph focused on searching for sub-Earth-mass planets in the \wolf\ system could shed light on a potential inner planet candidate (as was the case with \object{Proxima Centauri}~[d] first identified by \citealt{SuarezMascareno2020} and later announced as a convincing planet candidate by \citealt{Faria2022}, with a periodicity of $\sim$5.12\,d and K$\sim$40\,c\ms), or even further rule out this possibility.

\subsection{Radio emission from star-planet interaction}

Auroral radio emission from stars and planets is due to the electron cyclotron maser (ECM) instability \citep{Melrose1982}, whereby plasma processes within the star (or planet) magnetosphere generate a population of unstable electrons that amplifies the  emission. The characteristic frequency of the ECM emission is given by the electron gyrofrequency, $\nu_G = 2.8 \, B$\,MHz/G, where $B$ is the local magnetic field in the source region in Gauss. ECM emission is a coherent mechanism that yields broadband ($\Delta\,\nu \sim \nu_G/2$), highly polarized (sometimes reaching 100\%), amplified nonthermal radiation.  

For Jupiter-like planets, which have magnetic fields of $B_{\rm pl} \simeq 10$\,G, the direct detection of radio emission from them is plausible, as the associated gyrosynchrotron frequency falls above the $\simeq$10\,MHz ionosphere cutoff. However, the detection of radio emission from Earth-sized exoplanets, which are also the type of planets comprising a large majority of the CARMENES sample, is doomed to fail, as the associated frequency falls below the ionosphere cutoff. 

Fortunately, if the velocity, $v_{\rm rel}$, of the plasma relative to the planetary body is less than the Alfv\'en speed, $v_A$, in other words $M_A = v_{\rm rel}/v_A < 1$, where $M_A$ is the Alfv\'en Mach number, then energy and momentum can be transported upstream of the flow along Alfv\'en wings.  Jupiter’s interaction with its Galilean satellites is a well-known example of sub-Alfvénic interaction, producing auroral radio emission \citep{Zarka2007}.  In the case of star-planet  interaction, the radio emission arises from the magnetosphere of the host star, induced by the exoplanet crossing the star magnetosphere, and the relevant magnetic field is that of the star, $B_\star$, not the exoplanet magnetic field. Since M-dwarf stars have magnetic fields ranging from about 100\,G and up to above 2-3\,kG, their auroral emission falls in the range from a few hundred MHz up to a few GHz. This interaction is expected to yield detectable auroral radio emission via the cyclotron emission mechanism (e.g., \citealt{Turnpenney2018,PerezTorres2021}). 

\begin{figure}[]
\centering
\includegraphics[width=0.5\textwidth]{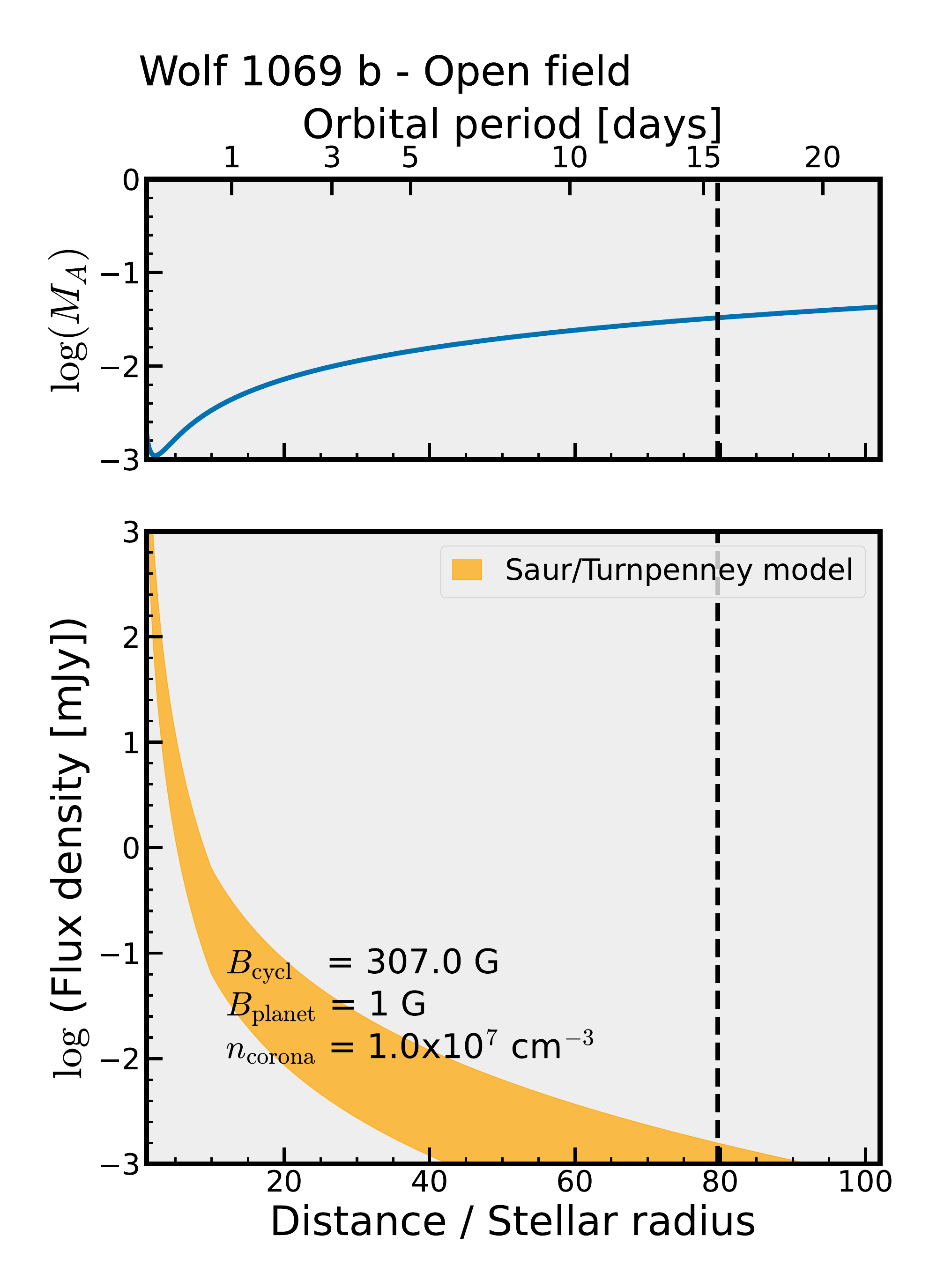}
\caption{\label{fig:wolf1069-spi-radio} 
 Expected flux density for auroral radio emission arising from star-planet interaction in the system \wolf, as a function of orbital distance. 
 The interaction is expected to be in the sub-Alfv\'enic regime (i.e., $M_A = v_{\rm rel}/v_{\rm Alfv} \leq 1$; top panel) at the location of each planet (vertical dashed line). 
}
\end{figure}

We followed the prescriptions in Appendix B of \citet{PerezTorres2021} to estimate the flux density expected to arise from the interaction between the planet \wolf~b and its host star at a frequency of 860\,MHz, which corresponds to the cyclotron frequency of the star magnetic field of 307\,G, from \cite{Reiners2022}. We computed the radio emission arising from star-planet interaction for two different magnetic field geometries: a closed dipolar geometry, and an open Parker spiral geometry. For the dipolar case, the motion of the plasma relative to \wolf~b happens in the supra-Alfv\'enic regime. Therefore no energy or momentum can be transferred to the star through Alfv\'en waves. In the open Parker spiral case, however, the plasma motion proceeds in the sub-Alfv\'enic regime.  We show in Fig.~\ref{fig:wolf1069-spi-radio} the predicted flux density as a function of orbital distance arising from the interaction of a magnetized exoplanet (1\,G) with its host star. The yellow shaded areas encompass the range of values from 0.01 to 0.1 for the efficiency factor, $\epsilon$, in converting Poynting flux into ECM radio emission. The expected flux density is less than 2\,$\mu$\,Jy. This is an extremely low value, which is not within the reach of even the most sensitive radio interferometers. 
The reasons behind this extremely faint signal are mainly two: First, the relatively large distance to the system (9.6\,pc away); and second, the large separation of the planet from its host star (about 80 stellar radii).
Therefore, the chances of detecting radio emission from star-planet interaction in \wolf\ are essentially null.

\section{Conclusion} \label{sec:conclusion}

Using CARMENES spectroscopic measurements, we presented the discovery of a nontransiting exoplanet, \wolf~b, with a period of $15.564 \pm 0.015$\,d, minimum mass of $1.26 \pm 0.21\,M_{\oplus}$, and insolation of $0.652^{+0.029}_{-0.027}\,S_{\oplus}$, putting it safely in the conservative HZ around a low-mass M-dwarf star. This makes \wolf~b the sixth closest ($d \sim 9.6$\,pc), Earth-mass planet in the conservative HZ from us, following \object{Proxima Centauri}~b, \object{GJ~1061}~d, \object{Teegarden's Star}~c, and \object{GJ~1002}~b and~c. Preliminary investigations of the potential habitability of the planet using GCM climate simulations suggest the planet to be a promising addition to the group of current targets to search for biosignature markers, such as \object{Proxima Centauri}~b, TRAPPIST-1~e, and TOI-700~d. \wolf~b unfortunately does not transit its stellar host, though future observations with thermal emission and reflected light phase curves could shed light on the properties of its atmosphere. We additionally investigated whether star-planet interactions in \wolf\ would be feasible to observe with radio emissions, but found these potential observations unfruitful.

The \wolf\ system becomes more intriguing as there is no significant evidence of closer-in planets ($P<10$\,d) greater than one Earth mass, based on our detectability limits. This configuration is a plausible outcome based on a select few synthetic planetary population simulations, and even suggestive of a planet formation history including a late giant impact phase. The detection of potential inner sub-Earth-mass planets with further sub-\ms\ RV observations could then confirm or reject this formation theory.  

The stellar host itself is a relatively inactive, low-mass M5.0 dwarf, though exhibits periods of higher activity levels, for which we determine its photometric rotation period to be 150--170\,d. This rotation period was also present in the CARMENES RVs, and thus, modeled with a dSHO-GP in the final fit. 
The RVs showed one more additional significant periodicity at 90.3\,d with a low amplitude (i.e., <1\,\ms), however, we demonstrate that there is currently not enough supporting evidence in favor of a planetary origin and it appears to be an effect of telluric contamination. Further RV investigation could be beneficial.
To conclude, \wolf~b is a noteworthy discovery that will allow further exploration into the habitability of Earth-mass planets around M dwarfs, as well as case study in testing planetary formation theories.


\begin{acknowledgements}
Part of this work was supported by the German Deutsche Forschungsgemeinschaft, DFG project number Ts~17/2--1.
      
CARMENES is an instrument at the Centro Astron\'omico Hispano-Alem\'an (CAHA) at Calar Alto (Almer\'{\i}a, Spain), operated jointly by the Junta de Andaluc\'ia and the Instituto de Astrof\'isica de Andaluc\'ia (CSIC).
  
CARMENES was funded by the Max-Planck-Gesellschaft (MPG), the Consejo Superior de Investigaciones Cient\'{\i}ficas (CSIC),
  the Ministerio de Econom\'ia y Competitividad (MINECO) and the European Regional Development Fund (ERDF) through projects FICTS-2011-02, ICTS-2017-07-CAHA-4, and CAHA16-CE-3978, 
  and the members of the CARMENES Consortium 
  (Max-Planck-Institut f\"ur Astronomie,
  Instituto de Astrof\'{\i}sica de Andaluc\'{\i}a,
  Landessternwarte K\"onigstuhl,
  Institut de Ci\`encies de l'Espai,
  Institut f\"ur Astrophysik G\"ottingen,
  Universidad Complutense de Madrid,
  Th\"uringer Landessternwarte Tautenburg,
  Instituto de Astrof\'{\i}sica de Canarias,
  Hamburger Sternwarte,
  Centro de Astrobiolog\'{\i}a and
  Centro Astron\'omico Hispano-Alem\'an), 
  with additional contributions by the MINECO, 
  the Deutsche Forschungsgemeinschaft through the Major Research Instrumentation Programme and Research Unit FOR2544 ``Blue Planets around Red Stars'', 
  the Klaus Tschira Stiftung, 
  the states of Baden-W\"urttemberg and Niedersachsen, 
  and by the Junta de Andaluc\'{\i}a.
  
This work was based on data from the CARMENES data archive at CAB (CSIC-INTA).
Data were partly collected with the 90\,cm and 150\,cm telescopes at Observatorio de Sierra Nevada (OSN), operated by the Instituto de Astrof\'isica de Andaluc\'i a (IAA, CSIC); we deeply acknowledge the OSN telescope operators for their very appreciable support.
The Telescopi Joan Or\'o (TJO) of the Observatori Astron\'omic del Montsec  is owned by the Generalitat de Catalunya and operated by the Institut d’Estudis Espacials de Catalunya (IEEC). 
  
 We acknowledge financial support from the Agencia Estatal de Investigaci\'on of the Ministerio de Ciencia e Innovaci\'on (AEI/10.13039/501100011033) and the ERDF ``A way of making Europe'' through projects 
  PID2019-109522GB-C5[1:4],    
PID2019-107061GB-C64, and PID2019-110689RB-100,
and the Centre of Excellence ``Severo Ochoa'' and ``Mar\'ia de Maeztu'' awards to the Instituto de Astrof\'isica de Canarias (SEV-2015-0548), Instituto de Astrof\'isica de Andaluc\'ia (SEV-2017-0709), and Centro de Astrobiolog\'ia (MDM-2017-0737);
the European Research Council under the Horizon 2020 Framework Program (ERC Advanced Grant Origins 832428 and under Marie Sk\l{}odowska-Curie grant 895525); 
the Generalitat de Catalunya/CERCA programme;
the DFG through the priority program SPP 1992 ``Exploring the Diversity of Extrasolar Planets (JE 701/5-1)'' and the Research Unit FOR 2544 ``Blue Planets around Red Stars'' (KU~3625/2-1);
the Bulgarian National Science Fund through program ``VIHREN-2021'' (KP-06-DV/5); the SNSF under grant P2BEP2\_195285;
the National Science Foundation under award No.~1753373, and by a Clare Boothe Luce Professorship.
We thank the anonymous referee for the insightful comments that helped improve the quality of this paper.

\\
\textit{Software:} 
\texttt{astropy} \cite{astropy},
\texttt{celerite} \citep{celerite}, 
\exostriker\ \citep{exostriker},
\texttt{dynesty} \citep{dynesty,dynesty2020},
\texttt{george} \citep{george}, 
\juliet\ \citep{juliet},
\texttt{matplotlib} \citep{matplotlib}, 
\texttt{numpy} \citep{numpy},
\texttt{pandas} \citep{pandas}, 
\texttt{PyFITS} \citep{pyfits}, 
\raccoon\ \citep{Lafarga2020},
\texttt{radvel} \citep{radvel}, 
\serval\ \citep{serval}
\texttt{scipy} \citep{scipy}, 
\end{acknowledgements}

\bibliographystyle{aa} 
\bibliography{biblio} 

\begin{thebibliography}{161}
\expandafter\ifx\csname natexlab\endcsname\relax\def\natexlab#1{#1}\fi

\bibitem[{{Amado} {et~al.}(2021){Amado}, {Bauer}, {Rodr{\'\i}guez L{\'o}pez},
  {Rodr{\'\i}guez}, {Cardona Guill{\'e}n}, {Perger}, {Caballero},
  {L{\'o}pez-Gonz{\'a}lez}, {Mu{\~n}oz Rodr{\'\i}guez}, {Pozuelos},
  {S{\'a}nchez-Rivero}, {Schlecker}, {Quirrenbach}, {Ribas}, {Reiners},
  {Almenara}, {Astudillo-Defru}, {Azzaro}, {B{\'e}jar}, {Bohemann}, {Bonfils},
  {Bouchy}, {Cifuentes}, {Cort{\'e}s-Contreras}, {Delfosse}, {Dreizler},
  {Forveille}, {Hatzes}, {Henning}, {Jeffers}, {Kaminski}, {K{\"u}rster},
  {Lafarga}, {Lodieu}, {Lovis}, {Mayor}, {Montes}, {Morales}, {Morales},
  {Murgas}, {Ortiz}, {Pall{\'e}}, {Pepe}, {Perdelwitz}, {Pollaco}, {Santos},
  {Sch{\"o}fer}, {Schweitzer}, {S{\'e}gransan}, {Shan}, {Stock}, {Tal-Or},
  {Udry}, {Zapatero Osorio}, \& {Zechmeister}}]{Amado2021}
{Amado}, P.~J., {Bauer}, F.~F., {Rodr{\'\i}guez L{\'o}pez}, C., {et~al.} 2021,
  \aap, 650, A188

\bibitem[{{Ambikasaran} {et~al.}(2015){Ambikasaran}, {Foreman-Mackey},
  {Greengard}, {Hogg}, \& {O'Neil}}]{george}
{Ambikasaran}, S., {Foreman-Mackey}, D., {Greengard}, L., {Hogg}, D.~W., \&
  {O'Neil}, M. 2015, IEEE Transactions on Pattern Analysis and Machine
  Intelligence, 38, 252

\bibitem[{Anglada-Escud{\'{e}} {et~al.}(2016)Anglada-Escud{\'{e}}, Amado,
  Barnes, Berdi{\~{n}}as, Butler, Coleman, {De La Cueva}, Dreizler, Endl,
  Giesers, Jeffers, Jenkins, Jones, Kiraga, K{\"{u}}rster,
  L{\'{o}}pez-Gonz{\'{a}}lez, Marvin, Morales, Morin, Nelson, Ortiz, Ofir,
  Paardekooper, Reiners, Rodr{\'{i}}guez, Rodr{\'{i}}guez-L{\'{o}}pez,
  Sarmiento, Strachan, Tsapras, Tuomi, \& Zechmeister}]{Anglada-Escude2016a}
Anglada-Escud{\'{e}}, G., Amado, P.~J., Barnes, J., {et~al.} 2016, Nature, 536,
  437

\bibitem[{{Anglada-Escud{\'e}} {et~al.}(2013){Anglada-Escud{\'e}}, {Tuomi},
  {Gerlach}, {Barnes}, {Heller}, {Jenkins}, {Wende}, {Vogt}, {Butler},
  {Reiners}, \& {Jones}}]{Anglada-Escude2013}
{Anglada-Escud{\'e}}, G., {Tuomi}, M., {Gerlach}, E., {et~al.} 2013, \aap, 556,
  A126

\bibitem[{{Astropy Collaboration} {et~al.}(2018){Astropy Collaboration},
  Price-Whelan, Sipocz, Günther, Lim, Crawford, Conseil, Shupe, Craig,
  Dencheva, Ginsburg, {Vand erPlas}, Bradley, Pérez-Suárez, de~Val-Borro,
  Aldcroft, Cruz, Robitaille, Tollerud, Ardelean, Babej, Bach, Bachetti,
  Bakanov, Bamford, Barentsen, Barmby, Baumbach, Berry, Biscani, Boquien,
  Bostroem, Bouma, Brammer, Bray, Breytenbach, Buddelmeijer, Burke, Calderone,
  {Cano RodrCara}, Cardoso, Cheedella, Copin, Corrales, Crichton, D’Avella,
  Deil, Depagne, Dietrich, Donath, Droettboom, Earl, Erben, Fabbro, Ferreira,
  Finethy, Fox, Garrison, Gibbons, Goldstein, Gommers, Greco, Greenfield,
  Groener, Grollier, Hagen, Hirst, Homeier, Horton, Hosseinzadeh, Hu, Hunkeler,
  Ivezić, Jain, Jenness, Kanarek, Kendrew, Kern, Kerzendorf, Khvalko, King,
  Kirkby, Kulkarni, Kumar, Lee, Lenz, Littlefair, Ma, Macleod, Mastropietro,
  McCully, Montagnac, Morris, Mueller, Mumford, Muna, Murphy, Nelson, Nguyen,
  Ninan, Nöthe, Ogaz, Oh, Parejko, Parley, Pascual, Patil, Patil, Plunkett,
  Prochaska, Rastogi, {Reddy Janga}, Sabater, Sakurikar, Seifert, Sherbert,
  Sherwood-Taylor, Shih, Sick, Silbiger, Singanamalla, Singer, Sladen, Sooley,
  Sornarajah, Streicher, Teuben, Thomas, Tremblay, Turner, Terrón, {van
  Kerkwijk}, de~{La Vega}, Watkins, Weaver, Whitmore, Woillez, Zabalza, \&
  {Astropy Contributors}}]{astropy}
{Astropy Collaboration}, Price-Whelan, A.~M., Sipocz, B.~M., {et~al.} 2018,
  \aj, 156, 123

\bibitem[{{Baluev}(2009)}]{Baluev2009}
{Baluev}, R.~V. 2009, \mnras, 393, 969

\bibitem[{Barrett {et~al.}(2012)Barrett, Hsu, Hanley, Taylor, Droettboom, Bray,
  Hack, Greenfield, Wyckoff, Jedrzejewski, de~{La Pena}, \& Hodge}]{pyfits}
Barrett, P., Hsu, J.~C., Hanley, C., {et~al.} 2012, PyFITS: Python FITS Module

\bibitem[{{Bidelman}(1985)}]{Bidelman1985}
{Bidelman}, W.~P. 1985, \apjs, 59, 197

\bibitem[{{Bluhm} {et~al.}(2020){Bluhm}, {Luque}, {Espinoza}, {Pall{\'e}},
  {Caballero}, {Dreizler}, {Livingston}, {Mathur}, {Quirrenbach}, {Stock}, {Van
  Eylen}, {Nowak}, {L{\'o}pez}, {Csizmadia}, {Zapatero Osorio}, {Sch{\"o}fer},
  {Lillo-Box}, {Oshagh}, {Gonz{\'a}lez-{\'A}lvarez}, {Amado}, {Barrado},
  {B{\'e}jar}, {Cale}, {Chaturvedi}, {Cifuentes}, {Cochran}, {Collins},
  {Collins}, {Cort{\'e}s-Contreras}, {D{\'\i}ez Alonso}, {El Mufti},
  {Ercolino}, {Fridlund}, {Gaidos}, {Garc{\'\i}a}, {Georgieva},
  {Gonz{\'a}lez-Cuesta}, {Guerra}, {Hatzes}, {Henning}, {Herrero}, {Hidalgo},
  {Isopi}, {Jeffers}, {Jenkins}, {Jensen}, {K{\'a}bath}, {Kaminski}, {Kemmer},
  {Korth}, {Kossakowski}, {K{\"u}rster}, {Lafarga}, {Mallia}, {Montes},
  {Morales}, {Morales-Calder{\'o}n}, {Murgas}, {Narita}, {Passegger}, {Pedraz},
  {Persson}, {Plavchan}, {Rauer}, {Redfield}, {Reffert}, {Reiners}, {Ribas},
  {Ricker}, {Rodr{\'\i}guez-L{\'o}pez}, {Santos}, {Seager}, {Schlecker},
  {Schweitzer}, {Shan}, {Soto}, {Subjak}, {Tal-Or}, {Trifonov}, {Vanaverbeke},
  {Vanderspek}, {Wittrock}, {Zechmeister}, \& {Zohrabi}}]{Bluhm2020_toi1235}
{Bluhm}, P., {Luque}, R., {Espinoza}, N., {et~al.} 2020, \aap, 639, A132

\bibitem[{{Bonfils} {et~al.}(2018){Bonfils}, {Astudillo-Defru}, {D{\'\i}az},
  {Almenara}, {Forveille}, {Bouchy}, {Delfosse}, {Lovis}, {Mayor}, {Murgas},
  {Pepe}, {Santos}, {S{\'e}gransan}, {Udry}, \& {W{\"u}nsche}}]{Bonfils2018}
{Bonfils}, X., {Astudillo-Defru}, N., {D{\'\i}az}, R., {et~al.} 2018, \aap,
  613, A25

\bibitem[{{Bonfils} {et~al.}(2013){Bonfils}, {Delfosse}, {Udry}, {Forveille},
  {Mayor}, {Perrier}, {Bouchy}, {Gillon}, {Lovis}, {Pepe}, {Queloz}, {Santos},
  {S{\'e}gransan}, \& {Bertaux}}]{Bonfils2013}
{Bonfils}, X., {Delfosse}, X., {Udry}, S., {et~al.} 2013, \aap, 549, A109

\bibitem[{{Brady} \& {Bean}(2021)}]{Brady2021}
{Brady}, M. \& {Bean}, J. 2021, arXiv e-prints, arXiv:2112.08337

\bibitem[{{Burn} {et~al.}(2021){Burn}, {Schlecker}, {Mordasini}, {Emsenhuber},
  {Alibert}, {Henning}, {Klahr}, \& {Benz}}]{Burn2021}
{Burn}, R., {Schlecker}, M., {Mordasini}, C., {et~al.} 2021, \aap, 656, A72

\bibitem[{{Butters} {et~al.}(2010){Butters}, {West}, {Anderson}, {Collier
  Cameron}, {Clarkson}, {Enoch}, {Haswell}, {Hellier}, {Horne}, {Joshi},
  {Kane}, {Lister}, {Maxted}, {Parley}, {Pollacco}, {Smalley}, {Street},
  {Todd}, {Wheatley}, \& {Wilson}}]{Butters10}
{Butters}, O.~W., {West}, R.~G., {Anderson}, D.~R., {et~al.} 2010, \aap, 520,
  L10

\bibitem[{{Caballero} {et~al.}(2016{\natexlab{a}}){Caballero},
  {Cort{\'e}s-Contreras}, {Alonso-Floriano}, {Montes}, {Quirrenbach}, {Amado},
  {Ribas}, {Reiners}, {Abellan}, {B{\'e}jar}, {Brinkm{\"o}ller}, {Czesla},
  {Dorda}, {Gallardo}, {Gonz{\'a}lez-{\'A}lvarez}, {Hidalgo}, {Holgado},
  {Jeffers}, {Kim}, {Klutsch}, {Lamert}, {Llamas}, {L{\'o}pez-Santiago},
  {Mart{\'\i}nez-Rodr{\'\i}guez}, {Morales}, {Mundt}, {Passegger},
  {Sch{\"o}fer}, {Seifert}, \& {Zechmeister}}]{Cab16}
{Caballero}, J.~A., {Cort{\'e}s-Contreras}, M., {Alonso-Floriano}, F.~J.,
  {et~al.} 2016{\natexlab{a}}, in 19th Cambridge Workshop on Cool Stars,
  Stellar Systems, and the Sun (CS19), Cambridge Workshop on Cool Stars,
  Stellar Systems, and the Sun, 148

\bibitem[{{Caballero} {et~al.}(2016{\natexlab{b}}){Caballero}, {Gu{\`a}rdia},
  {L{\'o}pez del Fresno}, {Zechmeister}, {de Juan}, {Alonso-Floriano}, {Amado},
  {Colom{\'e}}, {Cort{\'e}s-Contreras}, {Garc{\'\i}a-Piquer}, {Gesa}, {de
  Guindos}, {Hagen}, {Helmling}, {Hern{\'a}ndez Casta{\~n}o}, {K{\"u}rster},
  {L{\'o}pez-Santiago}, {Montes}, {Morales Mu{\~n}oz}, {Pavlov}, {Quirrenbach},
  {Reiners}, {Ribas}, {Seifert}, \& {Solano}}]{Caballero2016_SPIE}
{Caballero}, J.~A., {Gu{\`a}rdia}, J., {L{\'o}pez del Fresno}, M., {et~al.}
  2016{\natexlab{b}}, in Society of Photo-Optical Instrumentation Engineers
  (SPIE) Conference Series, Vol. 9910, Observatory Operations: Strategies,
  Processes, and Systems VI, ed. A.~B. {Peck}, R.~L. {Seaman}, \& C.~R. {Benn},
  99100E

\bibitem[{{Chadney} {et~al.}(2016){Chadney}, {Galand}, {Koskinen}, {Miller},
  {Sanz-Forcada}, {Unruh}, \& {Yelle}}]{Chadney2016}
{Chadney}, J.~M., {Galand}, M., {Koskinen}, T.~T., {et~al.} 2016, \aap, 587,
  A87

\bibitem[{{Cifuentes} {et~al.}(2020){Cifuentes}, {Caballero},
  {Cort{\'e}s-Contreras}, {Montes}, {Abell{\'a}n}, {Dorda}, {Holgado},
  {Zapatero Osorio}, {Morales}, {Amado}, {Passegger}, {Quirrenbach}, {Reiners},
  {Ribas}, {Sanz-Forcada}, {Schweitzer}, {Seifert}, \&
  {Solano}}]{Cifuentes2020}
{Cifuentes}, C., {Caballero}, J.~A., {Cort{\'e}s-Contreras}, M., {et~al.} 2020,
  \aap, 642, A115

\bibitem[{{Cohen} {et~al.}(2014){Cohen}, {Drake}, {Glocer}, {Garraffo},
  {Poppenhaeger}, {Bell}, {Ridley}, \& {Gombosi}}]{Cohen2014}
{Cohen}, O., {Drake}, J.~J., {Glocer}, A., {et~al.} 2014, \apj, 790, 57

\bibitem[{{Collins} {et~al.}(2017){Collins}, {Kielkopf}, {Stassun}, \&
  {Hessman}}]{Collins2017}
{Collins}, K.~A., {Kielkopf}, J.~F., {Stassun}, K.~G., \& {Hessman}, F.~V.
  2017, \aj, 153, 77

\bibitem[{{Colome} \& {Ribas}(2006)}]{Colome2006}
{Colome}, J. \& {Ribas}, I. 2006, IAU Special Session, 6, 11

\bibitem[{{Cutri} \& {et al.}(2012)}]{Cutri2012}
{Cutri}, R.~M. \& {et al.} 2012, VizieR Online Data Catalog, 2311

\bibitem[{{Cutri} \& {et al.}(2014)}]{Cutri2014}
{Cutri}, R.~M. \& {et al.} 2014, VizieR Online Data Catalog, 2328

\bibitem[{{David} {et~al.}(2019){David}, {Petigura}, {Luger}, {Foreman-Mackey},
  {Livingston}, {Mamajek}, \& {Hillenbrand}}]{David2019}
{David}, T.~J., {Petigura}, E.~A., {Luger}, R., {et~al.} 2019, \apjl, 885, L12

\bibitem[{{Dawson} \& {Fabrycky}(2010)}]{DawsonFabrycky2010}
{Dawson}, R.~I. \& {Fabrycky}, D.~C. 2010, \apj, 722, 937

\bibitem[{{Del Genio} {et~al.}(2019{\natexlab{a}}){Del Genio}, {Way},
  {Amundsen}, {Aleinov}, {Kelley}, {Kiang}, \& {Clune}}]{delGenio2019}
{Del Genio}, A.~D., {Way}, M.~J., {Amundsen}, D.~S., {et~al.}
  2019{\natexlab{a}}, Astrobiology, 19, 99

\bibitem[{{Del Genio} {et~al.}(2019{\natexlab{b}}){Del Genio}, {Way}, {Kiang},
  {Aleinov}, {Puma}, \& {Cook}}]{delGenio2019b}
{Del Genio}, A.~D., {Way}, M.~J., {Kiang}, N.~Y., {et~al.} 2019{\natexlab{b}},
  \apj, 887, 197

\bibitem[{{Delrez} {et~al.}(2022){Delrez}, {Murray}, {Pozuelos}, {Narita},
  {Ducrot}, {Timmermans}, {Watanabe}, {Burgasser}, {Hirano}, {Rackham},
  {Stassun}, {Van Grootel}, {Aganze}, {Cointepas}, {Howell}, {Kaltenegger},
  {Niraula}, {Sebastian}, {Almenara}, {Barkaoui}, {Baycroft}, {Bonfils},
  {Bouchy}, {Burdanov}, {Caldwell}, {Charbonneau}, {Ciardi}, {Collins},
  {Daylan}, {Demory}, {de Wit}, {Dransfield}, {Fajardo-Acosta}, {Fausnaugh},
  {Fukui}, {Furlan}, {Garcia}, {Gnilka}, {G{\'o}mez Maqueo Chew},
  {G{\'o}mez-Mu{\~n}oz}, {G{\"u}nther}, {Harakawa}, {Heng}, {Hooton}, {Hori},
  {Ikoma}, {Jehin}, {Jenkins}, {Kagetani}, {Kawauchi}, {Kimura}, {Kodama},
  {Kotani}, {Krishnamurthy}, {Kudo}, {Kunovac}, {Kusakabe}, {Latham},
  {Littlefield}, {McCormac}, {Melis}, {Mori}, {Murgas}, {Palle}, {Pedersen},
  {Queloz}, {Ricker}, {Sabin}, {Schanche}, {Schroffenegger}, {Seager}, {Shiao},
  {Sohy}, {Standing}, {Tamura}, {Theissen}, {Thompson}, {Triaud}, {Vanderspek},
  {Vievard}, {Wells}, {Winn}, {Zou}, {Z{\'u}{\~n}iga-Fern{\'a}ndez}, \&
  {Gillon}}]{Delrez2022}
{Delrez}, L., {Murray}, C.~A., {Pozuelos}, F.~J., {et~al.} 2022, \aap, 667, A59

\bibitem[{{Dieterich} {et~al.}(2012){Dieterich}, {Henry}, {Golimowski},
  {Krist}, \& {Tanner}}]{Dieterich2012}
{Dieterich}, S.~B., {Henry}, T.~J., {Golimowski}, D.~A., {Krist}, J.~E., \&
  {Tanner}, A.~M. 2012, \aj, 144, 64

\bibitem[{{D{\'\i}ez Alonso} {et~al.}(2019){D{\'\i}ez Alonso}, {Caballero},
  {Montes}, {de Cos Juez}, {Dreizler}, {Dubois}, {Jeffers}, {Lalitha}, {Naves},
  {Reiners}, {Ribas}, {Vanaverbeke}, {Amado}, {B{\'e}jar},
  {Cort{\'e}s-Contreras}, {Herrero}, {Hidalgo}, {K{\"u}rster}, {Logie},
  {Quirrenbach}, {Rau}, {Seifert}, {Sch{\"o}fer}, \& {Tal-Or}}]{DiezAlonso2019}
{D{\'\i}ez Alonso}, E., {Caballero}, J.~A., {Montes}, D., {et~al.} 2019, \aap,
  621, A126

\bibitem[{{Dobos} {et~al.}(2022){Dobos}, {Haris}, {Kamp}, \& {van der
  Tak}}]{Dobos2022}
{Dobos}, V., {Haris}, A., {Kamp}, I. E.~E., \& {van der Tak}, F. F.~S. 2022,
  \mnras, 513, 5290

\bibitem[{{Dole}(1964)}]{Dole1964}
{Dole}, S.~H. 1964, {Habitable planets for man}

\bibitem[{{Dong} {et~al.}(2018){Dong}, {Jin}, {Lingam}, {Airapetian}, {Ma}, \&
  {van der Holst}}]{Dong2018}
{Dong}, C., {Jin}, M., {Lingam}, M., {et~al.} 2018, Proceedings of the National
  Academy of Science, 115, 260

\bibitem[{{Dreizler} {et~al.}(2020){Dreizler}, {Jeffers}, {Rodr{\'\i}guez},
  {Zechmeister}, {Barnes}, {Haswell}, {Coleman}, {Lalitha}, {Hidalgo Soto},
  {Strachan}, {Hambsch}, {L{\'o}pez-Gonz{\'a}lez}, {Morales}, {Rodr{\'\i}guez
  L{\'o}pez}, {Berdi{\~n}as}, {Ribas}, {Pall{\'e}}, {Reiners}, \&
  {Anglada-Escud{\'e}}}]{Dreizler2020_gj1061}
{Dreizler}, S., {Jeffers}, S.~V., {Rodr{\'\i}guez}, E., {et~al.} 2020, \mnras,
  493, 536

\bibitem[{{Dressing} \& {Charbonneau}(2015)}]{DressingCharbonneau2015}
{Dressing}, C.~D. \& {Charbonneau}, D. 2015, \apj, 807, 45

\bibitem[{{Eastman} {et~al.}(2013){Eastman}, {Gaudi}, \& {Agol}}]{Eastman2013}
{Eastman}, J., {Gaudi}, B.~S., \& {Agol}, E. 2013, \pasp, 125, 83

\bibitem[{{Edson} {et~al.}(2012){Edson}, {Kasting}, {Pollard}, {Lee}, \&
  {Bannon}}]{Edson2012}
{Edson}, A.~R., {Kasting}, J.~F., {Pollard}, D., {Lee}, S., \& {Bannon}, P.~R.
  2012, Astrobiology, 12, 562

\bibitem[{{Espinoza} {et~al.}(2019){Espinoza}, {Kossakowski}, \&
  {Brahm}}]{juliet}
{Espinoza}, N., {Kossakowski}, D., \& {Brahm}, R. 2019, \mnras, 490, 2262

\bibitem[{Faria {et~al.}(2022)Faria, Su{\'a}rez~Mascare{\~n}o, Figueira, Silva,
  Damasso, Demangeon, Pepe, Santos, Rebolo, Cristiani, Adibekyan, Alibert,
  Allart, Barros, Cabral, D'Odorico, Di~Marcantonio, Dumusque, Ehrenreich,
  Gonz{\'a}lez~Hern{\'a}ndez, Hara, {Lillo-Box}, Lo~Curto, Lovis, Martins,
  M{\'e}gevand, Mehner, Micela, Molaro, Nunes, Pall{\'e}, Poretti, Sousa,
  Sozzetti, Tabernero, Udry, \& Zapatero~Osorio}]{Faria2022}
Faria, J.~P., Su{\'a}rez~Mascare{\~n}o, A., Figueira, P., {et~al.} 2022, \aap,
  658, A115

\bibitem[{{Fauchez} {et~al.}(2019){Fauchez}, {Turbet}, {Villanueva}, {Wolf},
  {Arney}, {Kopparapu}, {Lincowski}, {Mandell}, {de Wit}, {Pidhorodetska},
  {Domagal-Goldman}, \& {Stevenson}}]{Fauchez2019}
{Fauchez}, T.~J., {Turbet}, M., {Villanueva}, G.~L., {et~al.} 2019, \apj, 887,
  194

\bibitem[{{Feroz} {et~al.}(2011){Feroz}, {Balan}, \& {Hobson}}]{Feroz2011}
{Feroz}, F., {Balan}, S.~T., \& {Hobson}, M.~P. 2011, \mnras, 415, 3462

\bibitem[{{Feroz} \& {Hobson}(2014)}]{Feroz2014}
{Feroz}, F. \& {Hobson}, M.~P. 2014, \mnras, 437, 3540

\bibitem[{{Foreman-Mackey} {et~al.}(2017){Foreman-Mackey}, {Agol},
  {Ambikasaran}, \& {Angus}}]{celerite}
{Foreman-Mackey}, D., {Agol}, E., {Ambikasaran}, S., \& {Angus}, R. 2017, \aj,
  154, 220

\bibitem[{{Fuhrmeister} {et~al.}(2020){Fuhrmeister}, {Czesla}, {Hildebrandt},
  {Nagel}, {Schmitt}, {Jeffers}, {Caballero}, {Hintz}, {Johnson},
  {Sch{\"o}fer}, {Zechmeister}, {Reiners}, {Ribas}, {Amado}, {Quirrenbach},
  {Nortmann}, {Bauer}, {B{\'e}jar}, {Cort{\'e}s-Contreras}, {Dreizler},
  {Galad{\'\i}-Enr{\'\i}quez}, {Hatzes}, {Kaminski}, {K{\"u}rster}, {Lafarga},
  \& {Montes}}]{Fuhrmeister2020}
{Fuhrmeister}, B., {Czesla}, S., {Hildebrandt}, L., {et~al.} 2020, \aap, 640,
  A52

\bibitem[{{Fuhrmeister} {et~al.}(2022){Fuhrmeister}, {Czesla}, {Nagel},
  {Reiners}, {Schmitt}, {Jeffers}, {Caballero}, {Shulyak}, {Johnson},
  {Zechmeister}, {Montes}, {L{\'o}pez-Gallifa}, {Ribas}, {Quirrenbach},
  {Amado}, {Galad{\'\i}-Enr{\'\i}quez}, {Hatzes}, {K{\"u}rster}, {Danielski},
  {B{\'e}jar}, {Kaminski}, {Morales}, \& {Zapatero Osorio}}]{Fuhrmeister2022}
{Fuhrmeister}, B., {Czesla}, S., {Nagel}, E., {et~al.} 2022, \aap, 657, A125

\bibitem[{{Fulton} {et~al.}(2018){Fulton}, {Petigura}, {Blunt}, \&
  {Sinukoff}}]{radvel}
{Fulton}, B.~J., {Petigura}, E.~A., {Blunt}, S., \& {Sinukoff}, E. 2018, \pasp,
  130, 044504

\bibitem[{{Gaia Collaboration} {et~al.}(2022){Gaia Collaboration}, {Vallenari},
  {Brown}, {Prusti}, {de Bruijne}, {Arenou}, {Babusiaux}, {Biermann},
  {Creevey}, {Ducourant}, {Evans}, {Eyer}, {Guerra}, {Hutton}, {Jordi},
  {Klioner}, {Lammers}, {Lindegren}, {Luri}, {Mignard}, {Panem}, {Pourbaix},
  {Randich}, {Sartoretti}, {Soubiran}, {Tanga}, {Walton}, {Bailer-Jones},
  {Bastian}, {Drimmel}, {Jansen}, {Katz}, {Lattanzi}, {van Leeuwen}, {Bakker},
  {Cacciari}, {Casta{\~n}eda}, {De Angeli}, {Fabricius}, {Fouesneau},
  {Fr{\'e}mat}, {Galluccio}, {Guerrier}, {Heiter}, {Masana}, {Messineo},
  {Mowlavi}, {Nicolas}, {Nienartowicz}, {Pailler}, {Panuzzo}, {Riclet}, {Roux},
  {Seabroke}, {Sordo{\o}rcit}, {Th{\'e}venin}, {Gracia-Abril}, {Portell},
  {Teyssier}, {Altmann}, {Andrae}, {Audard}, {Bellas-Velidis}, {Benson},
  {Berthier}, {Blomme}, {Burgess}, {Busonero}, {Busso}, {C{\'a}novas}, {Carry},
  {Cellino}, {Cheek}, {Clementini}, {Damerdji}, {Davidson}, {de Teodoro},
  {Nu{\~n}ez Campos}, {Delchambre}, {Dell'Oro}, {Esquej},
  {Fern{\'a}ndez-Hern{\'a}ndez}, {Fraile}, {Garabato}, {Garc{\'\i}a-Lario},
  {Gosset}, {Haigron}, {Halbwachs}, {Hambly}, {Harrison}, {Hern{\'a}ndez},
  {Hestroffer}, {Hodgkin}, {Holl}, {Jan{\ss}en}, {Jevardat de Fombelle},
  {Jordan}, {Krone-Martins}, {Lanzafame}, {L{\"o}ffler}, {Marchal}, {Marrese},
  {Moitinho}, {Muinonen}, {Osborne}, {Pancino}, {Pauwels}, {Recio-Blanco},
  {Reyl{\'e}}, {Riello}, {Rimoldini}, {Roegiers}, {Rybizki}, {Sarro}, {Siopis},
  {Smith}, {Sozzetti}, {Utrilla}, {van Leeuwen}, {Abbas}, {{\'A}brah{\'a}m},
  {Abreu Aramburu}, {Aerts}, {Aguado}, {Ajaj}, {Aldea-Montero}, {Altavilla},
  {{\'A}lvarez}, {Alves}, {Anders}, {Anderson}, {Anglada Varela}, {Antoja},
  {Baines}, {Baker}, {Balaguer-N{\'u}{\~n}ez}, {Balbinot}, {Balog}, {Barache},
  {Barbato}, {Barros}, {Barstow}, {Bartolom{\'e}}, {Bassilana}, {Bauchet},
  {Becciani}, {Bellazzini}, {Berihuete}, {Bernet}, {Bertone}, {Bianchi},
  {Binnenfeld}, {Blanco-Cuaresma}, {Blazere}, {Boch}, {Bombrun}, {Bossini},
  {Bouquillon}, {Bragaglia}, {Bramante}, {Breedt}, {Bressan}, {Brouillet},
  {Brugaletta}, {Bucciarelli}, {Burlacu}, {Butkevich}, {Buzzi}, {Caffau},
  {Cancelliere}, {Cantat-Gaudin}, {Carballo}, {Carlucci}, {Carnerero},
  {Carrasco}, {Casamiquela}, {Castellani}, {Castro-Ginard}, {Chaoul},
  {Charlot}, {Chemin}, {Chiaramida}, {Chiavassa}, {Chornay}, {Comoretto},
  {Contursi}, {Cooper}, {Cornez}, {Cowell}, {Crifo}, {Cropper}, {Crosta},
  {Crowley}, {Dafonte}, {Dapergolas}, {David}, {David}, {de Laverny}, {De
  Luise}, {De March}, {De Ridder}, {de Souza}, {de Torres}, {del Peloso}, {del
  Pozo}, {Delbo}, {Delgado}, {Delisle}, {Demouchy}, {Dharmawardena}, {Di
  Matteo}, {Diakite}, {Diener}, {Distefano}, {Dolding}, {Edvardsson}, {Enke},
  {Fabre}, {Fabrizio}, {Faigler}, {Fedorets}, {Fernique}, {Fienga}, {Figueras},
  {Fournier}, {Fouron}, {Fragkoudi}, {Gai}, {Garcia-Gutierrez},
  {Garcia-Reinaldos}, {Garc{\'\i}a-Torres}, {Garofalo}, {Gavel}, {Gavras},
  {Gerlach}, {Geyer}, {Giacobbe}, {Gilmore}, {Girona}, {Giuffrida}, {Gomel},
  {Gomez}, {Gonz{\'a}lez-N{\'u}{\~n}ez}, {Gonz{\'a}lez-Santamar{\'\i}a},
  {Gonz{\'a}lez-Vidal}, {Granvik}, {Guillout}, {Guiraud},
  {Guti{\'e}rrez-S{\'a}nchez}, {Guy}, {Hatzidimitriou}, {Hauser}, {Haywood},
  {Helmer}, {Helmi}, {Sarmiento}, {Hidalgo}, {Hilger}, {H{\l}adczuk}, {Hobbs},
  {Holland}, {Huckle}, {Jardine}, {Jasniewicz}, {Jean-Antoine Piccolo},
  {Jim{\'e}nez-Arranz}, {Jorissen}, {Juaristi Campillo}, {Julbe}, {Karbevska},
  {Kervella}, {Khanna}, {Kontizas}, {Kordopatis}, {Korn}, {K{\'o}sp{\'a}l},
  {Kostrzewa-Rutkowska}, {Kruszy{\'n}ska}, {Kun}, {Laizeau}, {Lambert},
  {Lanza}, {Lasne}, {Le Campion}, {Lebreton}, {Lebzelter}, {Leccia}, {Leclerc},
  {Lecoeur-Taibi}, {Liao}, {Licata}, {Lindstr{\o}m}, {Lister}, {Livanou},
  {Lobel}, {Lorca}, {Loup}, {Madrero Pardo}, {Magdaleno Romeo}, {Managau},
  {Mann}, {Manteiga}, {Marchant}, {Marconi}, {Marcos}, {Marcos Santos},
  {Mar{\'\i}n Pina}, {Marinoni}, {Marocco}, {Marshall}, {Polo},
  {Mart{\'\i}n-Fleitas}, {Marton}, {Mary}, {Masip}, {Massari},
  {Mastrobuono-Battisti}, {Mazeh}, {McMillan}, {Messina}, {Michalik}, {Millar},
  {Mints}, {Molina}, {Molinaro}, {Moln{\'a}r}, {Monari}, {Mongui{\'o}},
  {Montegriffo}, {Montero}, {Mor}, {Mora}, {Morbidelli}, {Morel}, {Morris},
  {Muraveva}, {Murphy}, {Musella}, {Nagy}, {Noval}, {Oca{\~n}a}, {Ogden},
  {Ordenovic}, {Osinde}, {Pagani}, {Pagano}, {Palaversa}, {Palicio},
  {Pallas-Quintela}, {Panahi}, {Payne-Wardenaar}, {Pe{\~n}alosa Esteller},
  {Penttil{\"a}}, {Pichon}, {Piersimoni}, {Pineau}, {Plachy}, {Plum}, {Poggio},
  {Pr{\v{s}}a}, {Pulone}, {Racero}, {Ragaini}, {Rainer}, {Raiteri}, {Rambaux},
  {Ramos}, {Ramos-Lerate}, {Re Fiorentin}, {Regibo}, {Richards}, {Rios Diaz},
  {Ripepi}, {Riva}, {Rix}, {Rixon}, {Robichon}, {Robin}, {Robin}, {Roelens},
  {Rogues}, {Rohrbasser}, {Romero-G{\'o}mez}, {Rowell}, {Royer}, {Ruz Mieres},
  {Rybicki}, {Sadowski}, {S{\'a}ez N{\'u}{\~n}ez}, {Sagrist{\`a} Sell{\'e}s},
  {Sahlmann}, {Salguero}, {Samaras}, {Sanchez Gimenez}, {Sanna},
  {Santove{\~n}a}, {Sarasso}, {Schultheis}, {Sciacca}, {Segol}, {Segovia},
  {S{\'e}gransan}, {Semeux}, {Shahaf}, {Siddiqui}, {Siebert}, {Siltala},
  {Silvelo}, {Slezak}, {Slezak}, {Smart}, {Snaith}, {Solano}, {Solitro},
  {Souami}, {Souchay}, {Spagna}, {Spina}, {Spoto}, {Steele},
  {Steidelm{\"u}ller}, {Stephenson}, {S{\"u}veges}, {Surdej}, {Szabados},
  {Szegedi-Elek}, {Taris}, {Taylo}, {Teixeira}, {Tolomei}, {Tonello}, {Torra},
  {Torra}, {Torralba Elipe}, {Trabucchi}, {Tsounis}, {Turon}, {Ulla}, {Unger},
  {Vaillant}, {van Dillen}, {van Reeven}, {Vanel}, {Vecchiato}, {Viala},
  {Vicente}, {Voutsinas}, {Weiler}, {Wevers}, {Wyrzykowski}, {Yoldas}, {Yvard},
  {Zhao}, {Zorec}, {Zucker}, \& {Zwitter}}]{GaiaDR3}
{Gaia Collaboration}, {Vallenari}, A., {Brown}, A.~G.~A., {et~al.} 2022, arXiv
  e-prints, arXiv:2208.00211

\bibitem[{{Gardner} {et~al.}(2009){Gardner}, {Mather}, {Clampin}, {Doyon},
  {Flanagan}, {Franx}, {Greenhouse}, {Hammel}, {Hutchings}, {Jakobsen},
  {Lilly}, {Lunine}, {McCaughrean}, {Mountain}, {Rieke}, {Rieke}, {Sonneborn},
  {Stiavelli}, {Windhorst}, \& {Wright}}]{jwst}
{Gardner}, J.~P., {Mather}, J.~C., {Clampin}, M., {et~al.} 2009, Astrophysics
  and Space Science Proceedings, 10, 1

\bibitem[{{Gibbs} {et~al.}(2020){Gibbs}, {Bixel}, {Rackham}, {Apai},
  {Schlecker}, {Espinoza}, {Mancini}, {Chen}, {Henning}, {Gabor}, {Boyle},
  {Perez Chavez}, {Mousseau}, {Dietrich}, {Jay Socia}, {Ip}, {Ngeow}, {Tsai},
  {Bhandare}, {Marian}, {Baehr}, {Brown}, {H{\"a}berle}, {Keppler},
  {Molaverdikhani}, \& {Sarkis}}]{Gibbs2020}
{Gibbs}, A., {Bixel}, A., {Rackham}, B.~V., {et~al.} 2020, \aj, 159, 169

\bibitem[{{Gillen} {et~al.}(2020){Gillen}, {Briegal}, {Hodgkin},
  {Foreman-Mackey}, {Van Leeuwen}, {Jackman}, {McCormac}, {West}, {Queloz},
  {Bayliss}, {Goad}, {Watson}, {Wheatley}, {Belardi}, {Burleigh}, {Casewell},
  {Jenkins}, {Raynard}, {Smith}, {Tilbrook}, \& {Vines}}]{Gillen2020}
{Gillen}, E., {Briegal}, J.~T., {Hodgkin}, S.~T., {et~al.} 2020, \mnras, 492,
  1008

\bibitem[{{Gliese}(1957)}]{GL57}
{Gliese}, W. 1957, Astron.~Rechen-Institut, Heidelberg, 89 Seiten, 8

\bibitem[{{Gliese} \& {Jahrei{\ss}}(1979)}]{GlieseJahreiss1979}
{Gliese}, W. \& {Jahrei{\ss}}, H. 1979, \aaps, 38, 423

\bibitem[{{Hardegree-Ullman} {et~al.}(2019){Hardegree-Ullman}, {Cushing},
  {Muirhead}, \& {Christiansen}}]{Hardegree-Ullman2019}
{Hardegree-Ullman}, K.~K., {Cushing}, M.~C., {Muirhead}, P.~S., \&
  {Christiansen}, J.~L. 2019, VizieR Online Data Catalog, J/AJ/158/75

\bibitem[{{Heller} {et~al.}(2011){Heller}, {Leconte}, \& {Barnes}}]{Heller2011}
{Heller}, R., {Leconte}, J., \& {Barnes}, R. 2011, \aap, 528, A27

\bibitem[{{Hilton}(2011)}]{Hilton2011}
{Hilton}, E.~J. 2011, PhD thesis, University of Washington, Seattle

\bibitem[{Hippke \& Heller(2019)}]{Hippke2019}
Hippke, M. \& Heller, R. 2019, \aap, 623, A39

\bibitem[{{Hsu} {et~al.}(2020){Hsu}, {Ford}, \& {Terrien}}]{Hsu2020}
{Hsu}, D.~C., {Ford}, E.~B., \& {Terrien}, R. 2020, \mnras, 498, 2249

\bibitem[{{Hunter}(2007)}]{matplotlib}
{Hunter}, J.~D. 2007, Computing in Science and Engineering, 9, 90

\bibitem[{{Irwin} {et~al.}(2015){Irwin}, {Berta-Thompson}, {Charbonneau},
  {Dittmann}, {Falco}, {Newton}, \& {Nutzman}}]{Irwin2015_MEarth}
{Irwin}, J.~M., {Berta-Thompson}, Z.~K., {Charbonneau}, D., {et~al.} 2015, in
  Cambridge Workshop on Cool Stars, Stellar Systems, and the Sun, Vol.~18, 18th
  Cambridge Workshop on Cool Stars, Stellar Systems, and the Sun, 767--772

\bibitem[{{Jacobson} {et~al.}(2017){Jacobson}, {Rubie}, {Hernlund},
  {Morbidelli}, \& {Nakajima}}]{Jacobson2017}
{Jacobson}, S.~A., {Rubie}, D.~C., {Hernlund}, J., {Morbidelli}, A., \&
  {Nakajima}, M. 2017, Earth and Planetary Science Letters, 474, 375

\bibitem[{{Janson} {et~al.}(2014){Janson}, {Bergfors}, {Brandner},
  {Kudryavtseva}, {Hormuth}, {Hippler}, \& {Henning}}]{Janson2014}
{Janson}, M., {Bergfors}, C., {Brandner}, W., {et~al.} 2014, \apj, 789, 102

\bibitem[{{Jeffers} {et~al.}(2022){Jeffers}, {Barnes}, {Sch{\"o}fer},
  {Quirrenbach}, {Zechmeister}, {Amado}, {Caballero}, {Fern{\'a}ndez},
  {Rodr{\'\i}guez}, {Ribas}, {Reiners}, {Cardona Guill{\'e}n}, {Cifuentes},
  {Czesla}, {Hatzes}, {K{\"u}rster}, {Montes}, {Morales}, {Pedraz}, \&
  {Sadegi}}]{Jeffers2022}
{Jeffers}, S.~V., {Barnes}, J.~R., {Sch{\"o}fer}, P., {et~al.} 2022, \aap, 663,
  A27

\bibitem[{{Jenkins} {et~al.}(2016){Jenkins}, {Twicken}, {McCauliff},
  {Campbell}, {Sanderfer}, {Lung}, {Mansouri-Samani}, {Girouard}, {Tenenbaum},
  {Klaus}, {Smith}, {Caldwell}, {Chacon}, {Henze}, {Heiges}, {Latham},
  {Morgan}, {Swade}, {Rinehart}, \& {Vanderspek}}]{Jenkins2016}
{Jenkins}, J.~M., {Twicken}, J.~D., {McCauliff}, S., {et~al.} 2016, in Society
  of Photo-Optical Instrumentation Engineers (SPIE) Conference Series, Vol.
  9913, Software and Cyberinfrastructure for Astronomy IV, ed. G.~{Chiozzi} \&
  J.~C. {Guzman}, 99133E

\bibitem[{{Jenkins} {et~al.}(2009){Jenkins}, {Ramsey}, {Jones}, {Pavlenko},
  {Gallardo}, {Barnes}, \& {Pinfield}}]{Jenkins2009}
{Jenkins}, J.~S., {Ramsey}, L.~W., {Jones}, H.~R.~A., {et~al.} 2009, \apj, 704,
  975

\bibitem[{{J{\'o}dar} {et~al.}(2013){J{\'o}dar}, {P{\'e}rez-Garrido},
  {D{\'\i}az-S{\'a}nchez}, {Vill{\'o}}, {Rebolo}, \&
  {P{\'e}rez-Prieto}}]{Jodar2013}
{J{\'o}dar}, E., {P{\'e}rez-Garrido}, A., {D{\'\i}az-S{\'a}nchez}, A., {et~al.}
  2013, \mnras, 429, 859

\bibitem[{{Kasper} {et~al.}(2021){Kasper}, {Cerpa Urra}, {Pathak}, {Bonse},
  {Nousiainen}, {Engler}, {Heritier}, {Kammerer}, {Leveratto}, {Rajani},
  {Bristow}, {Le Louarn}, {Madec}, {Str{\"o}bele}, {Verinaud}, {Glauser},
  {Quanz}, {Helin}, {Keller}, {Snik}, {Boccaletti}, {Chauvin}, {Mouillet},
  {Kulcs{\'a}r}, \& {Raynaud}}]{Kasper2021}
{Kasper}, M., {Cerpa Urra}, N., {Pathak}, P., {et~al.} 2021, The Messenger,
  182, 38

\bibitem[{{Kasting} {et~al.}(1993){Kasting}, {Whitmire}, \&
  {Reynolds}}]{Kasting1993}
{Kasting}, J.~F., {Whitmire}, D.~P., \& {Reynolds}, R.~T. 1993, \icarus, 101,
  108

\bibitem[{{Kemmer} {et~al.}(2020){Kemmer}, {Stock}, {Kossakowski}, {Kaminski},
  {Molaverdikhani}, {Schlecker}, {Caballero}, {Amado}, {Astudillo-Defru},
  {Bonfils}, {Ciardi}, {Collins}, {Espinoza}, {Fukui}, {Hirano}, {Jenkins},
  {Latham}, {Matthews}, {Narita}, {Pall{\'e}}, {Parviainen}, {Quirrenbach},
  {Reiners}, {Ribas}, {Ricker}, {Schlieder}, {Seager}, {Vanderspek}, {Winn},
  {Almenara}, {B{\'e}jar}, {Bluhm}, {Bouchy}, {Boyd}, {Christiansen},
  {Cifuentes}, {Cloutier}, {Collins}, {Cort{\'e}s-Contreras}, {Crossfield},
  {Crouzet}, {de Leon}, {Della-Rose}, {Delfosse}, {Dreizler}, {Esparza-Borges},
  {Essack}, {Forveille}, {Figueira}, {Galad{\'\i}-Enr{\'\i}quez}, {Gan},
  {Glidden}, {Gonzales}, {Guerra}, {Harakawa}, {Hatzes}, {Henning}, {Herrero},
  {Hodapp}, {Hori}, {Howell}, {Ikoma}, {Isogai}, {Jeffers}, {K{\"u}rster},
  {Kawauchi}, {Kimura}, {Klagyivik}, {Kotani}, {Kurokawa}, {Kusakabe},
  {Kuzuhara}, {Lafarga}, {Livingston}, {Luque}, {Matson}, {Morales}, {Mori},
  {Muirhead}, {Murgas}, {Nishikawa}, {Nishiumi}, {Omiya}, {Reffert},
  {Rodr{\'\i}guez L{\'o}pez}, {Santos}, {Sch{\"o}fer}, {Schwarz}, {Shiao},
  {Tamura}, {Terada}, {Twicken}, {Ueda}, {Vievard}, {Watanabe}, \&
  {Zechmeister}}]{Kemmer2020_toi488}
{Kemmer}, J., {Stock}, S., {Kossakowski}, D., {et~al.} 2020, \aap, 642, A236

\bibitem[{{Kopparapu} {et~al.}(2013){Kopparapu}, {Ramirez}, {Kasting}, {Eymet},
  {Robinson}, {Mahadevan}, {Terrien}, {Domagal-Goldman}, {Meadows}, \&
  {Deshpande}}]{Kopparapu2013}
{Kopparapu}, R.~K., {Ramirez}, R., {Kasting}, J.~F., {et~al.} 2013, \apj, 765,
  131

\bibitem[{{Kopparapu} {et~al.}(2014){Kopparapu}, {Ramirez}, {SchottelKotte},
  {Kasting}, {Domagal-Goldman}, \& {Eymet}}]{Kopparapu2014}
{Kopparapu}, R.~K., {Ramirez}, R.~M., {SchottelKotte}, J., {et~al.} 2014,
  \apjl, 787, L29

\bibitem[{{Kopparapu} {et~al.}(2020){Kopparapu}, {Wolf}, \&
  {Meadows}}]{Kopparapu2019}
{Kopparapu}, R.~K., {Wolf}, E.~T., \& {Meadows}, V.~S. 2020, in Planetary
  Astrobiology, ed. V.~S. {Meadows}, G.~N. {Arney}, B.~E. {Schmidt}, \& D.~J.
  {Des Marais}, 449

\bibitem[{{Kossakowski} {et~al.}(2021){Kossakowski}, {Kemmer}, {Bluhm},
  {Stock}, {Caballero}, {B{\'e}jar}, {Guill{\'e}n}, {Lodieu}, {Collins},
  {Oshagh}, {Schlecker}, {Espinoza}, {Pall{\'e}}, {Henning}, {Kreidberg},
  {K{\"u}rster}, {Amado}, {Anderson}, {Morales}, {Cartwright}, {Charbonneau},
  {Chaturvedi}, {Cifuentes}, {Conti}, {Cort{\'e}s-Contreras}, {Dreizler},
  {Galad{\'\i}-Enr{\'\i}quez}, {Guerra}, {Hart}, {Hellier}, {Henze}, {Herrero},
  {Jeffers}, {Jenkins}, {Jensen}, {Kaminski}, {Kielkopf}, {Kunimoto},
  {Lafarga}, {Latham}, {Lillo-Box}, {Luque}, {Molaverdikhani}, {Montes},
  {Morello}, {Morgan}, {Nowak}, {Pavlov}, {Perger}, {Quintana}, {Quirrenbach},
  {Reffert}, {Reiners}, {Ricker}, {Ribas}, {L{\'o}pez}, {Osorio}, {Seager},
  {Sch{\"o}fer}, {Schweitzer}, {Trifonov}, {Vanaverbeke}, {Vanderspek}, {West},
  {Winn}, \& {Zechmeister}}]{Kossakowski2021_toi1201}
{Kossakowski}, D., {Kemmer}, J., {Bluhm}, P., {et~al.} 2021, \aap, 656, A124

\bibitem[{{Lafarga} {et~al.}(2020){Lafarga}, {Ribas}, {Lovis}, {Perger},
  {Zechmeister}, {Bauer}, {K{\"u}rster}, {Cort{\'e}s-Contreras}, {Morales},
  {Herrero}, {Rosich}, {Baroch}, {Reiners}, {Caballero}, {Quirrenbach},
  {Amado}, {Alacid}, {B{\'e}jar}, {Dreizler}, {Hatzes}, {Henning}, {Jeffers},
  {Kaminski}, {Montes}, {Pedraz}, {Rodr{\'\i}guez-L{\'o}pez}, \&
  {Schmitt}}]{Lafarga2020}
{Lafarga}, M., {Ribas}, I., {Lovis}, C., {et~al.} 2020, \aap, 636, A36

\bibitem[{{Lafarga} {et~al.}(2021){Lafarga}, {Ribas}, {Reiners}, {Quirrenbach},
  {Amado}, {Caballero}, {Azzaro}, {B{\'e}jar}, {Cort{\'e}s-Contreras},
  {Dreizler}, {Hatzes}, {Henning}, {Jeffers}, {Kaminski}, {K{\"u}rster},
  {Montes}, {Morales}, {Oshagh}, {Rodr{\'\i}guez-L{\'o}pez}, {Sch{\"o}fer},
  {Schweitzer}, \& {Zechmeister}}]{Lafarga2021}
{Lafarga}, M., {Ribas}, I., {Reiners}, A., {et~al.} 2021, \aap, 652, A28

\bibitem[{{Lam} {et~al.}(2021){Lam}, {Csizmadia}, {Astudillo-Defru}, {Bonfils},
  {Gandolfi}, {Padovan}, {Esposito}, {Hellier}, {Hirano}, {Livingston},
  {Murgas}, {Smith}, {Collins}, {Mathur}, {Garcia}, {Howell}, {Santos}, {Dai},
  {Ricker}, {Vanderspek}, {Latham}, {Seager}, {Winn}, {Jenkins}, {Albrecht},
  {Almenara}, {Artigau}, {Barrag{\'a}n}, {Bouchy}, {Cabrera}, {Charbonneau},
  {Chaturvedi}, {Chaushev}, {Christiansen}, {Cochran}, {De Meideiros},
  {Delfosse}, {D{\'\i}az}, {Doyon}, {Eigm{\"u}ller}, {Figueira}, {Forveille},
  {Fridlund}, {Gaisn{\'e}}, {Goffo}, {Georgieva}, {Grziwa}, {Guenther},
  {Hatzes}, {Johnson}, {Kab{\'a}th}, {Knudstrup}, {Korth}, {Lewin}, {Lissauer},
  {Lovis}, {Luque}, {Melo}, {Morgan}, {Morris}, {Mayor}, {Narita}, {Osborne},
  {Palle}, {Pepe}, {Persson}, {Quinn}, {Rauer}, {Redfield}, {Schlieder},
  {S{\'e}gransan}, {Serrano}, {Smith}, {{\v{S}}ubjak}, {Twicken}, {Udry}, {Van
  Eylen}, \& {Vezie}}]{Lam2021}
{Lam}, K. W.~F., {Csizmadia}, S., {Astudillo-Defru}, N., {et~al.} 2021,
  Science, 374, 1271

\bibitem[{{Lamman} {et~al.}(2020){Lamman}, {Baranec}, {Berta-Thompson}, {Law},
  {Schonhut-Stasik}, {Ziegler}, {Salama}, {Jensen-Clem}, {Duev}, {Riddle},
  {Kulkarni}, {Winters}, \& {Irwin}}]{Lamman2020}
{Lamman}, C., {Baranec}, C., {Berta-Thompson}, Z.~K., {et~al.} 2020, \aj, 159,
  139

\bibitem[{{Leconte} {et~al.}(2013){Leconte}, {Forget}, {Charnay}, {Wordsworth},
  {Selsis}, {Millour}, \& {Spiga}}]{Leconte2013}
{Leconte}, J., {Forget}, F., {Charnay}, B., {et~al.} 2013, \aap, 554, A69

\bibitem[{{Luque} \& {Pall{\'e}}(2022)}]{LuquePalle2022}
{Luque}, R. \& {Pall{\'e}}, E. 2022, Science, 377, 1211

\bibitem[{{Luyten}(1995)}]{Luyten1955}
{Luyten}, W.~J. 1995, VizieR Online Data Catalog, I/54A

\bibitem[{{Maiolino} {et~al.}(2013){Maiolino}, {Haehnelt}, {Murphy}, {Queloz},
  {Origlia}, {Alcala}, {Alibert}, {Amado}, {Allende Prieto}, {Ammler-von Eiff},
  {Asplund}, {Barstow}, {Becker}, {Bonfils}, {Bouchy}, {Bragaglia}, {Burleigh},
  {Chiavassa}, {Cimatti}, {Cirasuolo}, {Cristiani}, {D'Odorico}, {Dravins},
  {Emsellem}, {Farihi}, {Figueira}, {Fynbo}, {Gansicke}, {Gillon},
  {Gustafsson}, {Hill}, {Israelyan}, {Korn}, {Larsen}, {De Laverny}, {Liske},
  {Lovis}, {Marconi}, {Martins}, {Molaro}, {Nisini}, {Oliva}, {Petitjean},
  {Pettini}, {Recio Blanco}, {Rebolo}, {Reiners}, {Rodriguez-Lopez}, {Ryde},
  {Santos}, {Savaglio}, {Snellen}, {Strassmeier}, {Tanvir}, {Testi}, {Tolstoy},
  {Triaud}, {Vanzi}, {Viel}, \& {Volonteri}}]{Maiolino2013}
{Maiolino}, R., {Haehnelt}, M., {Murphy}, M.~T., {et~al.} 2013, arXiv e-prints,
  arXiv:1310.3163

\bibitem[{{Mann} {et~al.}(2015){Mann}, {Feiden}, {Gaidos}, {Boyajian}, \& {von
  Braun}}]{Mann2015}
{Mann}, A.~W., {Feiden}, G.~A., {Gaidos}, E., {Boyajian}, T., \& {von Braun},
  K. 2015, \apj, 804, 64

\bibitem[{{Marfil} {et~al.}(2021){Marfil}, {Tabernero}, {Montes}, {Caballero},
  {L{\'a}zaro}, {Gonz{\'a}lez Hern{\'a}ndez}, {Nagel}, {Passegger},
  {Schweitzer}, {Ribas}, {Reiners}, {Quirrenbach}, {Amado}, {Cifuentes},
  {Cort{\'e}s-Contreras}, {Dreizler}, {Duque-Arribas},
  {Galad{\'\i}-Enr{\'\i}quez}, {Henning}, {Jeffers}, {Kaminski}, {K{\"u}rster},
  {Lafarga}, {L{\'o}pez-Gallifa}, {Morales}, {Shan}, \&
  {Zechmeister}}]{Marfil2021}
{Marfil}, E., {Tabernero}, H.~M., {Montes}, D., {et~al.} 2021, \aap, 656, A162

\bibitem[{{Mart{\'\i}nez-Rodr{\'\i}guez}
  {et~al.}(2019){Mart{\'\i}nez-Rodr{\'\i}guez}, {Caballero}, {Cifuentes},
  {Piro}, \& {Barnes}}]{MartinezRodiguez2019}
{Mart{\'\i}nez-Rodr{\'\i}guez}, H., {Caballero}, J.~A., {Cifuentes}, C.,
  {Piro}, A.~L., \& {Barnes}, R. 2019, \apj, 887, 261

\bibitem[{{Melrose} \& {Dulk}(1982)}]{Melrose1982}
{Melrose}, D.~B. \& {Dulk}, G.~A. 1982, \apj, 259, 844

\bibitem[{{Montes} {et~al.}(2001){Montes}, {L{\'o}pez-Santiago}, {G{\'a}lvez},
  {Fern{\'a}ndez-Figueroa}, {De Castro}, \& {Cornide}}]{Montes2001}
{Montes}, D., {L{\'o}pez-Santiago}, J., {G{\'a}lvez}, M.~C., {et~al.} 2001,
  \mnras, 328, 45

\bibitem[{{Mulders} {et~al.}(2021){Mulders}, {Dr{\k{a}}{\.z}kowska}, {van der
  Marel}, {Ciesla}, \& {Pascucci}}]{Mulders2021}
{Mulders}, G.~D., {Dr{\k{a}}{\.z}kowska}, J., {van der Marel}, N., {Ciesla},
  F.~J., \& {Pascucci}, I. 2021, \apjl, 920, L1

\bibitem[{{Mulders} {et~al.}(2015){Mulders}, {Pascucci}, \&
  {Apai}}]{Mulders2015}
{Mulders}, G.~D., {Pascucci}, I., \& {Apai}, D. 2015, \apj, 814, 130

\bibitem[{{Nagel} {et~al.}(2022){Nagel}, {Czesla}, {Kaminski}, {Zechmeister},
  {Tal-Or}, {Schmitt}, \& {Reiners}}]{Nagel2022}
{Nagel}, E., {Czesla}, S., {Kaminski}, A., {et~al.} 2022, \aap, submitted

\bibitem[{{Newton} {et~al.}(2017){Newton}, {Irwin}, {Charbonneau}, {Berlind},
  {Calkins}, \& {Mink}}]{Newton2017}
{Newton}, E.~R., {Irwin}, J., {Charbonneau}, D., {et~al.} 2017, \apj, 834, 85

\bibitem[{{Newton} {et~al.}(2016){Newton}, {Irwin}, {Charbonneau},
  {Berta-Thompson}, {Dittmann}, \& {West}}]{Newton2016}
{Newton}, E.~R., {Irwin}, J., {Charbonneau}, D., {et~al.} 2016, \apj, 821, 93

\bibitem[{Oliphant(2006)}]{numpy}
Oliphant, T.~E. 2006, A guide to NumPy, Vol.~1 ({Trelgol Publishing USA})

\bibitem[{{Ormel}(2017)}]{Ormel2017}
{Ormel}, C.~W. 2017, in Astrophysics and Space Science Library, Vol. 445,
  Formation, Evolution, and Dynamics of Young Solar Systems, ed. M.~{Pessah} \&
  O.~{Gressel}, 197

\bibitem[{{Passegger} {et~al.}(2019){Passegger}, {Schweitzer}, {Shulyak},
  {Nagel}, {Hauschildt}, {Reiners}, {Amado}, {Caballero},
  {Cort{\'e}s-Contreras}, {Dom{\'\i}nguez-Fern{\'a}ndez}, {Quirrenbach},
  {Ribas}, {Azzaro}, {Anglada-Escud{\'e}}, {Bauer}, {B{\'e}jar}, {Dreizler},
  {Guenther}, {Henning}, {Jeffers}, {Kaminski}, {K{\"u}rster}, {Lafarga},
  {Mart{\'\i}n}, {Montes}, {Morales}, {Schmitt}, \&
  {Zechmeister}}]{Passegger2019}
{Passegger}, V.~M., {Schweitzer}, A., {Shulyak}, D., {et~al.} 2019, \aap, 627,
  A161

\bibitem[{{P{\'e}rez-Torres} {et~al.}(2021){P{\'e}rez-Torres}, {G{\'o}mez},
  {Ortiz}, {Leto}, {Anglada}, {G{\'o}mez}, {Rodr{\'\i}guez}, {Trigilio},
  {Amado}, {Alberdi}, {Anglada-Escud{\'e}}, {Osorio}, {Umana}, {Berdi{\~n}as},
  {L{\'o}pez-Gonz{\'a}lez}, {Morales}, {Rodr{\'\i}guez-L{\'o}pez}, \&
  {Chibueze}}]{PerezTorres2021}
{P{\'e}rez-Torres}, M., {G{\'o}mez}, J.~F., {Ortiz}, J.~L., {et~al.} 2021,
  \aap, 645, A77

\bibitem[{{Perryman}(2018)}]{Perryman2018}
{Perryman}, M. 2018, {The Exoplanet Handbook}

\bibitem[{{Pollacco} {et~al.}(2006){Pollacco}, {Skillen}, {Collier Cameron},
  {Christian}, {Hellier}, {Irwin}, {Lister}, {Street}, {West}, {Anderson},
  {Clarkson}, {Deeg}, {Enoch}, {Evans}, {Fitzsimmons}, {Haswell}, {Hodgkin},
  {Horne}, {Kane}, {Keenan}, {Maxted}, {Norton}, {Osborne}, {Parley}, {Ryans},
  {Smalley}, {Wheatley}, \& {Wilson}}]{Pollacco2006}
{Pollacco}, D.~L., {Skillen}, I., {Collier Cameron}, A., {et~al.} 2006, \pasp,
  118, 1407

\bibitem[{Pollack {et~al.}(1996)Pollack, Hubickyj, Bodenheimer, Lissauer,
  Podolak, \& Greenzweig}]{Pollack1996}
Pollack, J.~B., Hubickyj, O., Bodenheimer, P., {et~al.} 1996, Icarus, 124, 62

\bibitem[{{Probst}(1983)}]{Probst1983}
{Probst}, R.~G. 1983, \apjs, 53, 335

\bibitem[{{Quirrenbach} {et~al.}(2014){Quirrenbach}, {Amado}, {Caballero},
  {Mundt}, {Reiners}, {Ribas}, {Seifert}, {Abril}, {Aceituno},
  {Alonso-Floriano}, {Ammler-von Eiff}, {Antona Jim{\'e}nez},
  {Anwand-Heerwart}, {Azzaro}, {Bauer}, {Barrado}, {Becerril}, {B{\'e}jar},
  {Ben{\'{\i}}tez}, {Berdi{\~n}as}, {C{\'a}rdenas}, {Casal}, {Claret},
  {Colom{\'e}}, {Cort{\'e}s-Contreras}, {Czesla}, {Doellinger}, {Dreizler},
  {Feiz}, {Fern{\'a}ndez}, {Galad{\'{\i}}}, {G{\'a}lvez-Ortiz},
  {Garc{\'{\i}}a-Piquer}, {Garc{\'{\i}}a-Vargas}, {Garrido}, {Gesa}, {G{\'o}mez
  Galera}, {Gonz{\'a}lez {\'A}lvarez}, {Gonz{\'a}lez Hern{\'a}ndez},
  {Gr{\"o}zinger}, {Gu{\`a}rdia}, {Guenther}, {de Guindos},
  {Guti{\'e}rrez-Soto}, {Hagen}, {Hatzes}, {Hauschildt}, {Helmling}, {Henning},
  {Hermann}, {Hern{\'a}ndez Casta{\~n}o}, {Herrero}, {Hidalgo}, {Holgado},
  {Huber}, {Huber}, {Jeffers}, {Joergens}, {de Juan}, {Kehr}, {Klein},
  {K{\"u}rster}, {Lamert}, {Lalitha}, {Laun}, {Lemke}, {Lenzen}, {L{\'o}pez del
  Fresno}, {L{\'o}pez Mart{\'{\i}}}, {L{\'o}pez-Santiago}, {Mall}, {Mandel},
  {Mart{\'{\i}}n}, {Mart{\'{\i}}n-Ruiz}, {Mart{\'{\i}}nez-Rodr{\'{\i}}guez},
  {Marvin}, {Mathar}, {Mirabet}, {Montes}, {Morales Mu{\~n}oz}, {Moya},
  {Naranjo}, {Ofir}, {Oreiro}, {Pall{\'e}}, {Panduro}, {Passegger},
  {P{\'e}rez-Calpena}, {P{\'e}rez Medialdea}, {Perger}, {Pluto}, {Ram{\'o}n},
  {Rebolo}, {Redondo}, {Reffert}, {Reinhardt}, {Rhode}, {Rix}, {Rodler},
  {Rodr{\'{\i}}guez}, {Rodr{\'{\i}}guez-L{\'o}pez},
  {Rodr{\'{\i}}guez-P{\'e}rez}, {Rohloff}, {Rosich}, {S{\'a}nchez-Blanco},
  {S{\'a}nchez Carrasco}, {Sanz-Forcada}, {Sarmiento}, {Sch{\"a}fer},
  {Schiller}, {Schmidt}, {Schmitt}, {Solano}, {Stahl}, {Storz}, {St{\"u}rmer},
  {Su{\'a}rez}, {Ulbrich}, {Veredas}, {Wagner}, {Winkler}, {Zapatero Osorio},
  {Zechmeister}, {Abell{\'a}n de Paco}, {Anglada-Escud{\'e}}, {del Burgo},
  {Klutsch}, {Lizon}, {L{\'o}pez-Morales}, {Morales}, {Perryman}, {Tulloch}, \&
  {Xu}}]{CARMENES}
{Quirrenbach}, A., {Amado}, P.~J., {Caballero}, J.~A., {et~al.} 2014, in
  \procspie, Vol. 9147, Ground-based and Airborne Instrumentation for Astronomy
  V, 91471F

\bibitem[{{Quirrenbach} {et~al.}(2018){Quirrenbach}, {Amado}, {Ribas},
  {Reiners}, {Caballero}, {Seifert}, {Aceituno}, {Azzaro}, {Baroch}, {Barrado},
  \& et~al.}]{CARMENES18}
{Quirrenbach}, A., {Amado}, P.~J., {Ribas}, I., {et~al.} 2018, in Society of
  Photo-Optical Instrumentation Engineers (SPIE) Conference Series, Vol. 10702,
  Ground-based and Airborne Instrumentation for Astronomy VII, 107020W

\bibitem[{{Rajpurohit} {et~al.}(2018){Rajpurohit}, {Allard}, {Rajpurohit},
  {Sharma}, {Teixeira}, {Mousis}, \& {Rajpurohit}}]{Rajpurohit2018}
{Rajpurohit}, A.~S., {Allard}, F., {Rajpurohit}, S., {et~al.} 2018, \aap, 620,
  A180

\bibitem[{{Reid} {et~al.}(1995){Reid}, {Hawley}, \& {Gizis}}]{Reid1995}
{Reid}, I.~N., {Hawley}, S.~L., \& {Gizis}, J.~E. 1995, \aj, 110, 1838

\bibitem[{{Reiners} {et~al.}(2022){Reiners}, {Shulyak}, {K{\"a}pyl{\"a}},
  {Ribas}, {Nagel}, {Zechmeister}, {Caballero}, {Shan}, {Fuhrmeister},
  {Quirrenbach}, {Amado}, {Montes}, {Jeffers}, {Azzaro}, {B{\'e}jar},
  {Chaturvedi}, {Henning}, {K{\"u}rster}, \& {Pall{\'e}}}]{Reiners2022}
{Reiners}, A., {Shulyak}, D., {K{\"a}pyl{\"a}}, P.~J., {et~al.} 2022, \aap,
  662, A41

\bibitem[{{Reiners} {et~al.}(2018){Reiners}, {Zechmeister}, {Caballero},
  {Ribas}, {Morales}, {Jeffers}, {Sch{\"o}fer}, {Tal-Or}, {Quirrenbach},
  {Amado}, {Kaminski}, {Seifert}, {Abril}, {Aceituno}, {Alonso-Floriano},
  {Ammler-von Eiff}, {Antona}, {Anglada-Escud{\'e}}, {Anwand-Heerwart},
  {Arroyo-Torres}, {Azzaro}, {Baroch}, {Barrado}, {Bauer}, {Becerril},
  {B{\'e}jar}, {Ben{\'\i}tez}, {Berdinas}, {Bergond}, {Bl{\"u}mcke},
  {Brinkm{\"o}ller}, {del Burgo}, {Cano}, {C{\'a}rdenas V{\'a}zquez}, {Casal},
  {Cifuentes}, {Claret}, {Colom{\'e}}, {Cort{\'e}s-Contreras}, {Czesla},
  {D{\'\i}ez-Alonso}, {Dreizler}, {Feiz}, {Fern{\'a}ndez}, {Ferro},
  {Fuhrmeister}, {Galad{\'\i}-Enr{\'\i}quez}, {Garcia-Piquer}, {Garc{\'\i}a
  Vargas}, {Gesa}, {G{\'o}mez Galera}, {Gonz{\'a}lez Hern{\'a}ndez},
  {Gonz{\'a}lez-Peinado}, {Gr{\"o}zinger}, {Grohnert}, {Gu{\`a}rdia},
  {Guenther}, {Guijarro}, {de Guindos}, {Guti{\'e}rrez-Soto}, {Hagen},
  {Hatzes}, {Hauschildt}, {Hedrosa}, {Helmling}, {Henning}, {Hermelo},
  {Hern{\'a}ndez Arab{\'\i}}, {Hern{\'a}ndez Casta{\~n}o}, {Hern{\'a}ndez
  Hernando}, {Herrero}, {Huber}, {Huke}, {Johnson}, {de Juan}, {Kim}, {Klein},
  {Kl{\"u}ter}, {Klutsch}, {K{\"u}rster}, {Lafarga}, {Lamert}, {Lamp{\'o}n},
  {Lara}, {Laun}, {Lemke}, {Lenzen}, {Launhardt}, {L{\'o}pez del Fresno},
  {L{\'o}pez-Gonz{\'a}lez}, {L{\'o}pez-Puertas}, {L{\'o}pez Salas},
  {L{\'o}pez-Santiago}, {Luque}, {Mag{\'a}n Madinabeitia}, {Mall}, {Mancini},
  {Mandel}, {Marfil}, {Mar{\'\i}n Molina}, {Maroto Fern{\'a}ndez},
  {Mart{\'\i}n}, {Mart{\'\i}n-Ruiz}, {Marvin}, {Mathar}, {Mirabet}, {Montes},
  {Moreno-Raya}, {Moya}, {Mundt}, {Nagel}, {Naranjo}, {Nortmann}, {Nowak},
  {Ofir}, {Oreiro}, {Pall{\'e}}, {Panduro}, {Pascual}, {Passegger}, {Pavlov},
  {Pedraz}, {P{\'e}rez-Calpena}, {P{\'e}rez Medialdea}, {Perger}, {Perryman},
  {Pluto}, {Rabaza}, {Ram{\'o}n}, {Rebolo}, {Redondo}, {Reffert}, {Reinhart},
  {Rhode}, {Rix}, {Rodler}, {Rodr{\'\i}guez}, {Rodr{\'\i}guez-L{\'o}pez},
  {Rodr{\'\i}guez Trinidad}, {Rohloff}, {Rosich}, {Sadegi},
  {S{\'a}nchez-Blanco}, {S{\'a}nchez Carrasco}, {S{\'a}nchez-L{\'o}pez},
  {Sanz-Forcada}, {Sarkis}, {Sarmiento}, {Sch{\"a}fer}, {Schmitt}, {Schiller},
  {Schweitzer}, {Solano}, {Stahl}, {Strachan}, {St{\"u}rmer}, {Su{\'a}rez},
  {Tabernero}, {Tala}, {Trifonov}, {Tulloch}, {Ulbrich}, {Veredas}, {Vico
  Linares}, {Vilardell}, {Wagner}, {Winkler}, {Wolthoff}, {Xu}, {Yan}, \&
  {Zapatero Osorio}}]{Reiners2018}
{Reiners}, A., {Zechmeister}, M., {Caballero}, J.~A., {et~al.} 2018, \aap, 612,
  A49

\bibitem[{{Ribas} {et~al.}(2022){Ribas}, {Reiners}, {Zechmeister}, {Caballero},
  {Morales}, {Sabotta}, \& {Baroch}}]{Ribas2022}
{Ribas}, I., {Reiners}, A., {Zechmeister}, M., {et~al.} 2022, \aap, submitted

\bibitem[{{Ricker} {et~al.}(2015){Ricker}, {Winn}, {Vanderspek}, {Latham},
  {Bakos}, {Bean}, {Berta-Thompson}, {Brown}, {Buchhave}, {Butler}, {Butler},
  {Chaplin}, {Charbonneau}, {Christensen-Dalsgaard}, {Clampin}, {Deming},
  {Doty}, {De Lee}, {Dressing}, {Dunham}, {Endl}, {Fressin}, {Ge}, {Henning},
  {Holman}, {Howard}, {Ida}, {Jenkins}, {Jernigan}, {Johnson}, {Kaltenegger},
  {Kawai}, {Kjeldsen}, {Laughlin}, {Levine}, {Lin}, {Lissauer}, {MacQueen},
  {Marcy}, {McCullough}, {Morton}, {Narita}, {Paegert}, {Palle}, {Pepe},
  {Pepper}, {Quirrenbach}, {Rinehart}, {Sasselov}, {Sato}, {Seager},
  {Sozzetti}, {Stassun}, {Sullivan}, {Szentgyorgyi}, {Torres}, {Udry}, \&
  {Villasenor}}]{Ricker2015}
{Ricker}, G.~R., {Winn}, J.~N., {Vanderspek}, R., {et~al.} 2015, JATIS, 1,
  014003

\bibitem[{{Robertson} \& {Mahadevan}(2014)}]{Robertson2014}
{Robertson}, P. \& {Mahadevan}, S. 2014, \apjl, 793, L24

\bibitem[{{Rodr{\'\i}guez} {et~al.}(2010){Rodr{\'\i}guez}, {Garc{\'\i}a},
  {Costa}, {Lampens}, {van Cauteren}, {Mkrtichian}, {Olson}, {Amado},
  {Daszy{\'n}ska-Daszkiewicz}, {Turcu}, {Kim}, {Zhou},
  {L{\'o}pez-Gonz{\'a}lez}, {Rolland}, {D{\'\i}az-Fraile}, {Wood}, {Hintz},
  {Pop}, {Moldovan}, {Etzel}, {Casanova}, {Sota}, {Aceituno}, \&
  {Lee}}]{Rodriguez2010}
{Rodr{\'\i}guez}, E., {Garc{\'\i}a}, J.~M., {Costa}, V., {et~al.} 2010, \mnras,
  408, 2149

\bibitem[{{Rojas-Ayala} {et~al.}(2012){Rojas-Ayala}, {Covey}, {Muirhead}, \&
  {Lloyd}}]{RojasAyala2012}
{Rojas-Ayala}, B., {Covey}, K.~R., {Muirhead}, P.~S., \& {Lloyd}, J.~P. 2012,
  \apj, 748, 93

\bibitem[{{Sabotta} {et~al.}(2021){Sabotta}, {Schlecker}, {Chaturvedi},
  {Guenther}, {Mu{\~n}oz Rodr{\'\i}guez}, {Mu{\~n}oz S{\'a}nchez}, {Caballero},
  {Shan}, {Reffert}, {Ribas}, {Reiners}, {Hatzes}, {Amado}, {Klahr}, {Morales},
  {Quirrenbach}, {Henning}, {Dreizler}, {Pall{\'e}}, {Perger}, {Azzaro},
  {Jeffers}, {Kaminski}, {K{\"u}rster}, {Lafarga}, {Montes}, {Passegger}, \&
  {Zechmeister}}]{Sabotta2021}
{Sabotta}, S., {Schlecker}, M., {Chaturvedi}, P., {et~al.} 2021, \aap, 653,
  A114

\bibitem[{{Schlecker} {et~al.}(2022){Schlecker}, {Burn}, {Sabotta}, {Seifert},
  {Henning}, {Emsenhuber}, {Mordasini}, {Reffert}, {Shan}, \&
  {Klahr}}]{Schlecker2022}
{Schlecker}, M., {Burn}, R., {Sabotta}, S., {et~al.} 2022, \aap, 664, A180

\bibitem[{{Schlecker} {et~al.}(2021){Schlecker}, {Pham}, {Burn}, {Alibert},
  {Mordasini}, {Emsenhuber}, {Klahr}, {Henning}, \& {Mishra}}]{Schlecker2021b}
{Schlecker}, M., {Pham}, D., {Burn}, R., {et~al.} 2021, \aap, 656, A73

\bibitem[{{Sch{\"o}fer} {et~al.}(2019){Sch{\"o}fer}, {Jeffers}, {Reiners},
  {Shulyak}, {Fuhrmeister}, {Johnson}, {Zechmeister}, {Ribas}, {Quirrenbach},
  {Amado}, {Caballero}, {Anglada-Escud{\'e}}, {Bauer}, {B{\'e}jar},
  {Cort{\'e}s-Contreras}, {Dreizler}, {Guenther}, {Kaminski}, {K{\"u}rster},
  {Lafarga}, {Montes}, {Morales}, {Pedraz}, \& {Tal-Or}}]{Schoefer2019}
{Sch{\"o}fer}, P., {Jeffers}, S.~V., {Reiners}, A., {et~al.} 2019, \aap, 623,
  A44

\bibitem[{{Schweitzer} {et~al.}(2019){Schweitzer}, {Passegger}, {Cifuentes},
  {B{\'e}jar}, {Cort{\'e}s-Contreras}, {Caballero}, {del Burgo}, {Czesla},
  {K{\"u}rster}, {Montes}, {Zapatero Osorio}, {Ribas}, {Reiners},
  {Quirrenbach}, {Amado}, {Aceituno}, {Anglada-Escud{\'e}}, {Bauer},
  {Dreizler}, {Jeffers}, {Guenther}, {Henning}, {Kaminski}, {Lafarga},
  {Marfil}, {Morales}, {Schmitt}, {Seifert}, {Solano}, {Tabernero}, \&
  {Zechmeister}}]{Schweitzer2019}
{Schweitzer}, A., {Passegger}, V.~M., {Cifuentes}, C., {et~al.} 2019, \aap,
  625, A68

\bibitem[{{Seager}(2010)}]{Seager2010}
{Seager}, S. 2010, {Exoplanets}

\bibitem[{{Sebastian} {et~al.}(2021){Sebastian}, {Gillon}, {Ducrot},
  {Pozuelos}, {Garcia}, {G{\"u}nther}, {Delrez}, {Queloz}, {Demory}, {Triaud},
  {Burgasser}, {de Wit}, {Burdanov}, {Dransfield}, {Jehin}, {McCormac},
  {Murray}, {Niraula}, {Pedersen}, {Rackham}, {Sohy}, {Thompson}, \& {Van
  Grootel}}]{Sebastian2021}
{Sebastian}, D., {Gillon}, M., {Ducrot}, E., {et~al.} 2021, \aap, 645, A100

\bibitem[{{Segura} {et~al.}(2003){Segura}, {Krelove}, {Kasting}, {Sommerlatt},
  {Meadows}, {Crisp}, {Cohen}, \& {Mlawer}}]{Segura2003}
{Segura}, A., {Krelove}, K., {Kasting}, J.~F., {et~al.} 2003, Astrobiology, 3,
  689

\bibitem[{{Segura} {et~al.}(2010){Segura}, {Walkowicz}, {Meadows}, {Kasting},
  \& {Hawley}}]{Segura2010}
{Segura}, A., {Walkowicz}, L.~M., {Meadows}, V., {Kasting}, J., \& {Hawley}, S.
  2010, Astrobiology, 10, 751

\bibitem[{{Sestovic} \& {Demory}(2020)}]{Sestovic2020}
{Sestovic}, M. \& {Demory}, B.-O. 2020, \aap, 641, A170

\bibitem[{{Shields} {et~al.}(2016){Shields}, {Ballard}, \&
  {Johnson}}]{Shields2016}
{Shields}, A.~L., {Ballard}, S., \& {Johnson}, J.~A. 2016, \physrep, 663, 1

\bibitem[{{Shporer} {et~al.}(2020){Shporer}, {Collins}, {Astudillo-Defru},
  {Irwin}, {Bonfils}, {Collins}, {Matthews}, {Winters}, {Anderson},
  {Armstrong}, {Charbonneau}, {Cloutier}, {Daylan}, {Gan}, {G{\"u}nther},
  {Hellier}, {Horne}, {Huang}, {Jensen}, {Kielkopf}, {Palle}, {Sefako},
  {Stassun}, {Tan}, {Vanderburg}, {Ricker}, {Latham}, {Vanderspek}, {Seager},
  {Winn}, {Jenkins}, {Colon}, {Dressing}, {L{\'e}epine}, {Muirhead}, {Rose},
  {Twicken}, \& {Villasenor}}]{Shporer2020}
{Shporer}, A., {Collins}, K.~A., {Astudillo-Defru}, N., {et~al.} 2020, \apjl,
  890, L7

\bibitem[{{Skrutskie} {et~al.}(2006){Skrutskie}, {Cutri}, {Stiening},
  {Weinberg}, {Schneider}, {Carpenter}, {Beichman}, {Capps}, {Chester},
  {Elias}, {Huchra}, {Liebert}, {Lonsdale}, {Monet}, {Price}, {Seitzer},
  {Jarrett}, {Kirkpatrick}, {Gizis}, {Howard}, {Evans}, {Fowler}, {Fullmer},
  {Hurt}, {Light}, {Kopan}, {Marsh}, {McCallon}, {Tam}, {Van Dyk}, \&
  {Wheelock}}]{Skrutskie2006_2MASS}
{Skrutskie}, M.~F., {Cutri}, R.~M., {Stiening}, R., {et~al.} 2006, \aj, 131,
  1163

\bibitem[{{Smette} {et~al.}(2015){Smette}, {Sana}, {Noll}, {Horst}, {Kausch},
  {Kimeswenger}, {Barden}, {Szyszka}, {Jones}, {Gallenne}, {Vinther},
  {Ballester}, \& {Taylor}}]{Smette2015}
{Smette}, A., {Sana}, H., {Noll}, S., {et~al.} 2015, \aap, 576, A77

\bibitem[{{Smith} {et~al.}(2012){Smith}, {Stumpe}, {Van Cleve}, {Jenkins},
  {Barclay}, {Fanelli}, {Girouard}, {Kolodziejczak}, {McCauliff}, {Morris}, \&
  {Twicken}}]{Smith2012}
{Smith}, J.~C., {Stumpe}, M.~C., {Van Cleve}, J.~E., {et~al.} 2012, \pasp, 124,
  1000

\bibitem[{{Speagle} \& {Barbary}(2018)}]{dynesty}
{Speagle}, J. \& {Barbary}, K. 2018, {dynesty: Dynamic Nested Sampling
  package}, Astrophysics Source Code Library

\bibitem[{{Speagle}(2020)}]{dynesty2020}
{Speagle}, J.~S. 2020, \mnras, 493, 3132

\bibitem[{{Stelzer} {et~al.}(2013){Stelzer}, {Marino}, {Micela},
  {L{\'o}pez-Santiago}, \& {Liefke}}]{Stelzer2013}
{Stelzer}, B., {Marino}, A., {Micela}, G., {L{\'o}pez-Santiago}, J., \&
  {Liefke}, C. 2013, \mnras, 431, 2063

\bibitem[{{Stock} \& {Kemmer}(2020)}]{Stock2020_aliasfinder}
{Stock}, S. \& {Kemmer}, J. 2020, Journal of Open Source Software, 5(45), 1771

\bibitem[{{Stock} {et~al.}(2020){Stock}, {Nagel}, {Kemmer}, {Passegger},
  {Reffert}, {Quirrenbach}, {Caballero}, {Czesla}, {B{\'e}jar}, {Cardona},
  {D{\'\i}ez-Alonso}, {Herrero}, {Lalitha}, {Schlecker}, {Tal-Or},
  {Rodr{\'\i}guez}, {Rodr{\'\i}guez-L{\'o}pez}, {Ribas}, {Reiners}, {Amado},
  {Bauer}, {Bluhm}, {Cort{\'e}s-Contreras}, {Gonz{\'a}lez-Cuesta}, {Dreizler},
  {Hatzes}, {Henning}, {Jeffers}, {Kaminski}, {K{\"u}rster}, {Lafarga},
  {L{\'o}pez-Gonz{\'a}lez}, {Montes}, {Morales}, {Pedraz}, {Sch{\"o}fer},
  {Schweitzer}, {Trifonov}, {Zapatero Osorio}, \&
  {Zechmeister}}]{Stock2020_threesuperearths}
{Stock}, S., {Nagel}, E., {Kemmer}, J., {et~al.} 2020, \aap, 643, A112

\bibitem[{{Stumpe} {et~al.}(2014){Stumpe}, {Smith}, {Catanzarite}, {Van Cleve},
  {Jenkins}, {Twicken}, \& {Girouard}}]{Stumpe2014}
{Stumpe}, M.~C., {Smith}, J.~C., {Catanzarite}, J.~H., {et~al.} 2014, \pasp,
  126, 100

\bibitem[{{Stumpe} {et~al.}(2012){Stumpe}, {Smith}, {Van Cleve}, {Twicken},
  {Barclay}, {Fanelli}, {Girouard}, {Jenkins}, {Kolodziejczak}, {McCauliff}, \&
  {Morris}}]{Stumpe2012}
{Stumpe}, M.~C., {Smith}, J.~C., {Van Cleve}, J.~E., {et~al.} 2012, \pasp, 124,
  985

\bibitem[{{Su{\'a}rez Mascare{\~n}o} {et~al.}(2020){Su{\'a}rez Mascare{\~n}o},
  {Faria}, {Figueira}, {Lovis}, {Damasso}, {Gonz{\'a}lez Hern{\'a}ndez},
  {Rebolo}, {Cristiani}, {Pepe}, {Santos}, {Zapatero Osorio}, {Adibekyan},
  {Hojjatpanah}, {Sozzetti}, {Murgas}, {Abreu}, {Affolter}, {Alibert},
  {Aliverti}, {Allart}, {Allende Prieto}, {Alves}, {Amate}, {Avila}, {Baldini},
  {Bandi}, {Barros}, {Bianco}, {Benz}, {Bouchy}, {Broeng}, {Cabral},
  {Calderone}, {Cirami}, {Coelho}, {Conconi}, {Coretti}, {Cumani}, {Cupani},
  {D'Odorico}, {Deiries}, {Delabre}, {Di Marcantonio}, {Dumusque},
  {Ehrenreich}, {Fragoso}, {Genolet}, {Genoni}, {G{\'e}nova Santos}, {Hughes},
  {Iwert}, {Kerber}, {Knusdstrup}, {Landoni}, {Lavie}, {Lillo-Box}, {Lizon},
  {Lo Curto}, {Maire}, {Manescau}, {Martins}, {M{\'e}gevand}, {Mehner},
  {Micela}, {Modigliani}, {Molaro}, {Monteiro}, {Monteiro}, {Moschetti},
  {Mueller}, {Nunes}, {Oggioni}, {Oliveira}, {Pall{\'e}}, {Pariani},
  {Pasquini}, {Poretti}, {Rasilla}, {Redaelli}, {Riva}, {Santana Tschudi},
  {Santin}, {Santos}, {Segovia}, {Sosnowska}, {Sousa}, {Span{\`o}}, {Tenegi},
  {Udry}, {Zanutta}, \& {Zerbi}}]{SuarezMascareno2020}
{Su{\'a}rez Mascare{\~n}o}, A., {Faria}, J.~P., {Figueira}, P., {et~al.} 2020,
  \aap, 639, A77

\bibitem[{{Su{\'a}rez Mascare{\~n}o} {et~al.}(2022){Su{\'a}rez Mascare{\~n}o},
  {Gonz{\'a}lez-{\'A}lvarez}, {Zapatero Osorio}, {Lillo-Box}, {Faria}, \&
  {Passegger}}]{Mascareno2022}
{Su{\'a}rez Mascare{\~n}o}, A., {Gonz{\'a}lez-{\'A}lvarez}, E., {Zapatero
  Osorio}, M.~R., {et~al.} 2022, \aap, Forthcoming article

\bibitem[{{Suissa} {et~al.}(2020){Suissa}, {Wolf}, {Kopparapu}, {Villanueva},
  {Fauchez}, {Mandell}, {Arney}, {Gilbert}, {Schlieder}, {Barclay}, {Quintana},
  {Lopez}, {Rodriguez}, \& {Vanderburg}}]{Suissa2020}
{Suissa}, G., {Wolf}, E.~T., {Kopparapu}, R.~k., {et~al.} 2020, \aj, 160, 118

\bibitem[{{Tamuz} {et~al.}(2005){Tamuz}, {Mazeh}, \& {Zucker}}]{Tamuz05}
{Tamuz}, O., {Mazeh}, T., \& {Zucker}, S. 2005, \mnras, 356, 1466

\bibitem[{{Terrien} {et~al.}(2015){Terrien}, {Mahadevan}, {Deshpande}, \&
  {Bender}}]{Terrien2015}
{Terrien}, R.~C., {Mahadevan}, S., {Deshpande}, R., \& {Bender}, C.~F. 2015,
  \apjs, 220, 16

\bibitem[{{The pandas development team}(2020)}]{pandas}
{The pandas development team}. 2020, pandas-dev/pandas: Pandas

\bibitem[{{Trifonov}(2019)}]{exostriker}
{Trifonov}, T. 2019, {The Exo-Striker: Transit and radial velocity interactive
  fitting tool for orbital analysis and N-body simulations}

\bibitem[{{Trifonov} {et~al.}(2020){Trifonov}, {Tal-Or}, {Zechmeister},
  {Kaminski}, {Zucker}, \& {Mazeh}}]{Trifonov2020}
{Trifonov}, T., {Tal-Or}, L., {Zechmeister}, M., {et~al.} 2020, \aap, 636, A74

\bibitem[{{Trotta}(2008)}]{Trotta2008}
{Trotta}, R. 2008, Contemporary Physics, 49, 71

\bibitem[{{Turbet} {et~al.}(2018){Turbet}, {Bolmont}, {Leconte}, {Forget},
  {Selsis}, {Tobie}, {Caldas}, {Naar}, \& {Gillon}}]{Turbet2018}
{Turbet}, M., {Bolmont}, E., {Leconte}, J., {et~al.} 2018, \aap, 612, A86

\bibitem[{{Turbet} {et~al.}(2022){Turbet}, {Fauchez}, {Sergeev}, {Boutle},
  {Tsigaridis}, {Way}, {Wolf}, {Domagal-Goldman}, {Forget}, {Haqq-Misra},
  {Kopparapu}, {Lambert}, {Manners}, {Mayne}, \& {Sohl}}]{Turbet2022}
{Turbet}, M., {Fauchez}, T.~J., {Sergeev}, D.~E., {et~al.} 2022, The Planetary
  Science Journal, 3, 211

\bibitem[{{Turbet} {et~al.}(2016){Turbet}, {Leconte}, {Selsis}, {Bolmont},
  {Forget}, {Ribas}, {Raymond}, \& {Anglada-Escud{\'e}}}]{Turbet2016}
{Turbet}, M., {Leconte}, J., {Selsis}, F., {et~al.} 2016, \aap, 596, A112

\bibitem[{{Turnpenney} {et~al.}(2018){Turnpenney}, {Nichols}, {Wynn}, \&
  {Burleigh}}]{Turnpenney2018}
{Turnpenney}, S., {Nichols}, J.~D., {Wynn}, G.~A., \& {Burleigh}, M.~R. 2018,
  \apj, 854, 72

\bibitem[{Virtanen {et~al.}(2020)Virtanen, Gommers, {Oliphant Travis E.},
  Haberland, Reddy, Cournapeau, Burovski, Peterson, Weckesser, Bright, {van der
  Walt Stéfan J.}, Brett, Wilson, {Jarrod Millman}, Mayorov, Nelson, Jones,
  Kern, Larson, Carey, {Polat, Feng, Yu}, {Moore Eric W.}, {Vand erPlas},
  Laxalde, Perktold, Cimrman, Henriksen, Quintero, Harris, Archibald, Ribeiro,
  Pedregosa, {van Mulbregt}, \& Contributors}]{scipy}
Virtanen, P., Gommers, R., {Oliphant Travis E.}, {et~al.} 2020, Nature Methods,
  17, 261

\bibitem[{{Voelkel} {et~al.}(2021){Voelkel}, {Deienno}, {Kretke}, \&
  {Klahr}}]{Voelkel2021}
{Voelkel}, O., {Deienno}, R., {Kretke}, K., \& {Klahr}, H. 2021, \aap, 645,
  A131

\bibitem[{{Voelkel} {et~al.}(2020){Voelkel}, {Klahr}, {Mordasini},
  {Emsenhuber}, \& {Lenz}}]{Voelkel2020}
{Voelkel}, O., {Klahr}, H., {Mordasini}, C., {Emsenhuber}, A., \& {Lenz}, C.
  2020, \aap, 642, A75

\bibitem[{{Way} {et~al.}(2017){Way}, {Aleinov}, {Amundsen}, {Chandler},
  {Clune}, {Del Genio}, {Fujii}, {Kelley}, {Kiang}, {Sohl}, \&
  {Tsigaridis}}]{Way2017}
{Way}, M.~J., {Aleinov}, I., {Amundsen}, D.~S., {et~al.} 2017, \apjs, 231, 12

\bibitem[{{Weis}(1984)}]{Weis1984}
{Weis}, E.~W. 1984, \apjs, 55, 289

\bibitem[{{Wolf}(2017)}]{Wolf2017}
{Wolf}, E.~T. 2017, \apjl, 839, L1

\bibitem[{{Wolf} {et~al.}(2022){Wolf}, {Kopparapu}, {Haqq-Misra}, \&
  {Fauchez}}]{Wolf2022}
{Wolf}, E.~T., {Kopparapu}, R., {Haqq-Misra}, J., \& {Fauchez}, T.~J. 2022, The
  Planetary Science Journal, 3, 7

\bibitem[{{Wolf}(1920)}]{Wolf1920}
{Wolf}, M. 1920, Astronomische Nachrichten, 212, 303

\bibitem[{{Wood} {et~al.}(1994){Wood}, {Brown}, {Linsky}, {Kellett}, {Bromage},
  {Hodgkin}, \& {Pye}}]{Wood1994}
{Wood}, B.~E., {Brown}, A., {Linsky}, J.~L., {et~al.} 1994, \apjs, 93, 287

\bibitem[{{Yang} {et~al.}(2019){Yang}, {Abbot}, {Koll}, {Hu}, \&
  {Showman}}]{Yang2019}
{Yang}, J., {Abbot}, D.~S., {Koll}, D. D.~B., {Hu}, Y., \& {Showman}, A.~P.
  2019, \apj, 871, 29

\bibitem[{{Zacharias} {et~al.}(2012){Zacharias}, {Finch}, {Girard}, {Henden},
  {Bartlett}, {Monet}, \& {Zacharias}}]{Zacharias2013}
{Zacharias}, N., {Finch}, C.~T., {Girard}, T.~M., {et~al.} 2012, VizieR Online
  Data Catalog, I/322A

\bibitem[{{Zarka}(2007)}]{Zarka2007}
{Zarka}, P. 2007, \planss, 55, 598

\bibitem[{{Zechmeister} {et~al.}(2019){Zechmeister}, {Dreizler}, {Ribas},
  {Reiners}, {Caballero}, {Bauer}, {B{\'e}jar}, {Gonz{\'a}lez-Cuesta},
  {Herrero}, {Lalitha}, {L{\'o}pez-Gonz{\'a}lez}, {Luque}, {Morales},
  {Pall{\'e}}, {Rodr{\'\i}guez}, {Rodr{\'\i}guez L{\'o}pez}, {Tal-Or},
  {Anglada-Escud{\'e}}, {Quirrenbach}, {Amado}, {Abril}, {Aceituno},
  {Aceituno}, {Alonso-Floriano}, {Ammler-von Eiff}, {Antona Jim{\'e}nez},
  {Anwand-Heerwart}, {Arroyo-Torres}, {Azzaro}, {Baroch}, {Barrado},
  {Becerril}, {Ben{\'\i}tez}, {Berdi{\~n}as}, {Bergond}, {Bluhm},
  {Brinkm{\"o}ller}, {del Burgo}, {Calvo Ortega}, {Cano}, {Cardona
  Guill{\'e}n}, {Carro}, {C{\'a}rdenas V{\'a}zquez}, {Casal},
  {Casasayas-Barris}, {Casanova}, {Chaturvedi}, {Cifuentes}, {Claret},
  {Colom{\'e}}, {Cort{\'e}s-Contreras}, {Czesla}, {D{\'\i}ez-Alonso}, {Dorda},
  {Fern{\'a}ndez}, {Fern{\'a}ndez-Mart{\'\i}n}, {Fuhrmeister}, {Fukui},
  {Galad{\'\i}-Enr{\'\i}quez}, {Gallardo Cava}, {Garcia de la Fuente},
  {Garcia-Piquer}, {Garc{\'\i}a Vargas}, {Gesa}, {G{\'o}ngora Rueda},
  {Gonz{\'a}lez-{\'A}lvarez}, {Gonz{\'a}lez Hern{\'a}ndez},
  {Gonz{\'a}lez-Peinado}, {Gr{\"o}zinger}, {Gu{\`a}rdia}, {Guijarro}, {de
  Guindos}, {Hatzes}, {Hauschildt}, {Hedrosa}, {Helmling}, {Henning},
  {Hermelo}, {Hern{\'a}ndez Arabi}, {Hern{\'a}ndez Casta{\~n}o}, {Hern{\'a}ndez
  Otero}, {Hintz}, {Huke}, {Huber}, {Jeffers}, {Johnson}, {de Juan},
  {Kaminski}, {Kemmer}, {Kim}, {Klahr}, {Klein}, {Kl{\"u}ter}, {Klutsch},
  {Kossakowski}, {K{\"u}rster}, {Labarga}, {Lafarga}, {Llamas}, {Lamp{\'o}n},
  {Lara}, {Launhardt}, {L{\'a}zaro}, {Lodieu}, {L{\'o}pez del Fresno},
  {L{\'o}pez-Puertas}, {L{\'o}pez Salas}, {L{\'o}pez-Santiago}, {Mag{\'a}n
  Madinabeitia}, {Mall}, {Mancini}, {Mandel}, {Marfil}, {Mar{\'\i}n Molina},
  {Maroto Fern{\'a}ndez}, {Mart{\'\i}n}, {Mart{\'\i}n-Fern{\'a}ndez},
  {Mart{\'\i}n-Ruiz}, {Marvin}, {Mirabet}, {Monta{\~n}{\'e}s-Rodr{\'\i}guez},
  {Montes}, {Moreno-Raya}, {Nagel}, {Naranjo}, {Narita}, {Nortmann}, {Nowak},
  {Ofir}, {Oshagh}, {Panduro}, {Parviainen}, {Pascual}, {Passegger}, {Pavlov},
  {Pedraz}, {P{\'e}rez-Calpena}, {P{\'e}rez Medialdea}, {Perger}, {Perryman},
  {Rabaza}, {Ram{\'o}n Ballesta}, {Rebolo}, {Redondo}, {Reffert}, {Reinhardt},
  {Rhode}, {Rix}, {Rodler}, {Rodr{\'\i}guez Trinidad}, {Rosich}, {Sadegi},
  {S{\'a}nchez-Blanco}, {S{\'a}nchez Carrasco}, {S{\'a}nchez-L{\'o}pez},
  {Sanz-Forcada}, {Sarkis}, {Sarmiento}, {Sch{\"a}fer}, {Schmitt},
  {Sch{\"o}fer}, {Schweitzer}, {Seifert}, {Shulyak}, {Solano}, {Sota}, {Stahl},
  {Stock}, {Strachan}, {Stuber}, {St{\"u}rmer}, {Su{\'a}rez}, {Tabernero},
  {Tala Pinto}, {Trifonov}, {Veredas}, {Vico Linares}, {Vilardell}, {Wagner},
  {Wolthoff}, {Xu}, {Yan}, \& {Zapatero Osorio}}]{Zechmeister2019_Teegarden}
{Zechmeister}, M., {Dreizler}, S., {Ribas}, I., {et~al.} 2019, \aap, 627, A49

\bibitem[{{Zechmeister} \& {K{\"u}rster}(2009)}]{Zechmeister2009_GLS}
{Zechmeister}, M. \& {K{\"u}rster}, M. 2009, \aap, 496, 577

\bibitem[{{Zechmeister} {et~al.}(2009){Zechmeister}, {K{\"u}rster}, \&
  {Endl}}]{ZechmeisterKuerster2009}
{Zechmeister}, M., {K{\"u}rster}, M., \& {Endl}, M. 2009, \aap, 505, 859

\bibitem[{{Zechmeister} {et~al.}(2018){Zechmeister}, {Reiners}, {Amado},
  {Azzaro}, {Bauer}, {B{\'e}jar}, {Caballero}, {Guenther}, {Hagen}, \&
  {Jeffers}}]{serval}
{Zechmeister}, M., {Reiners}, A., {Amado}, P.~J., {et~al.} 2018, \aap, 609, A12

\bibitem[{{Zeng} {et~al.}(2016){Zeng}, {Sasselov}, \& {Jacobsen}}]{Zeng2016}
{Zeng}, L., {Sasselov}, D.~D., \& {Jacobsen}, S.~B. 2016, \apj, 819, 127

\end{thebibliography}

\newpage

\begin{appendix} 

\section{\aliasfinder\ figures} 

\begin{figure*}
    \centering
    \includegraphics{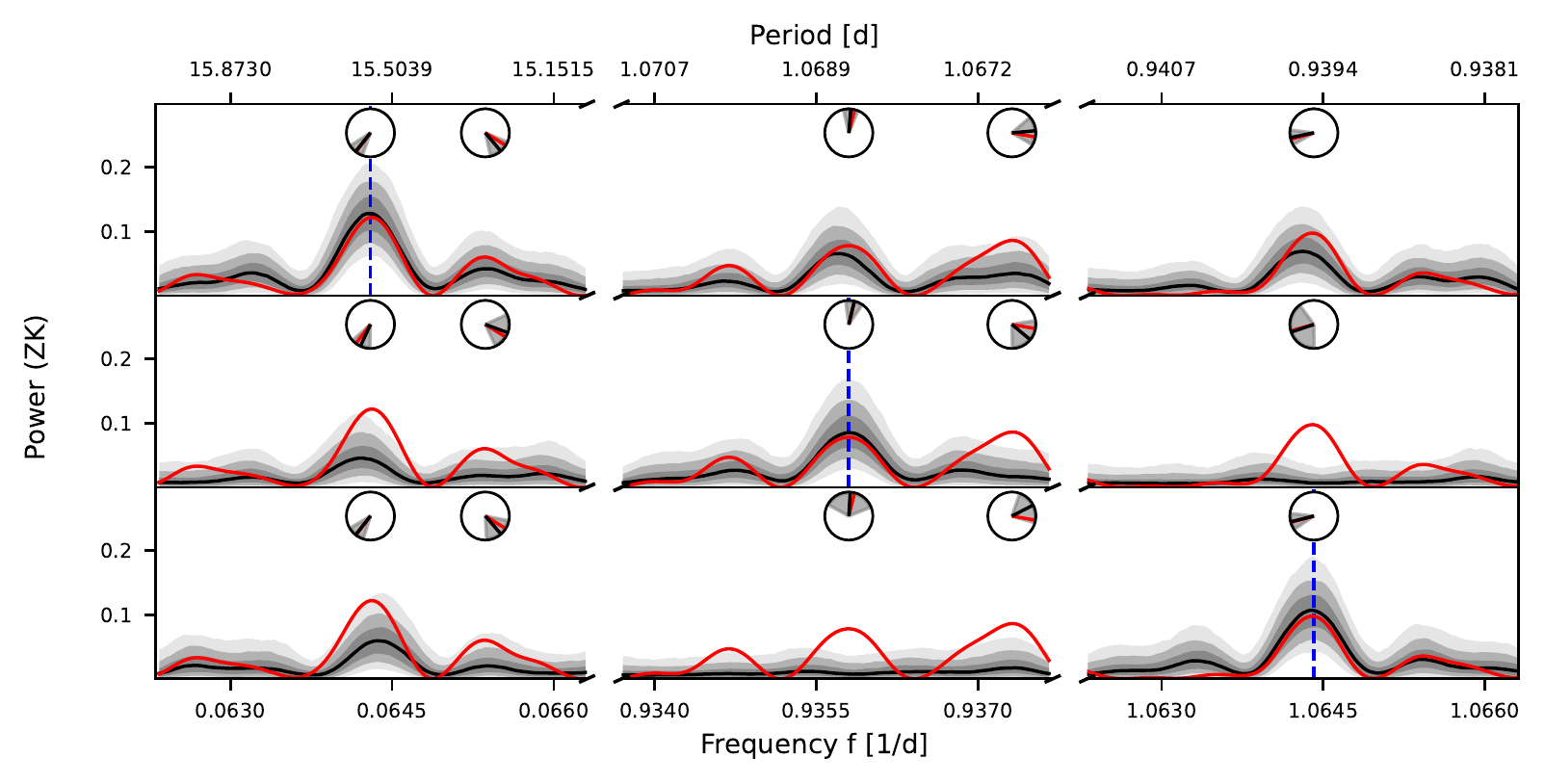}
    \includegraphics{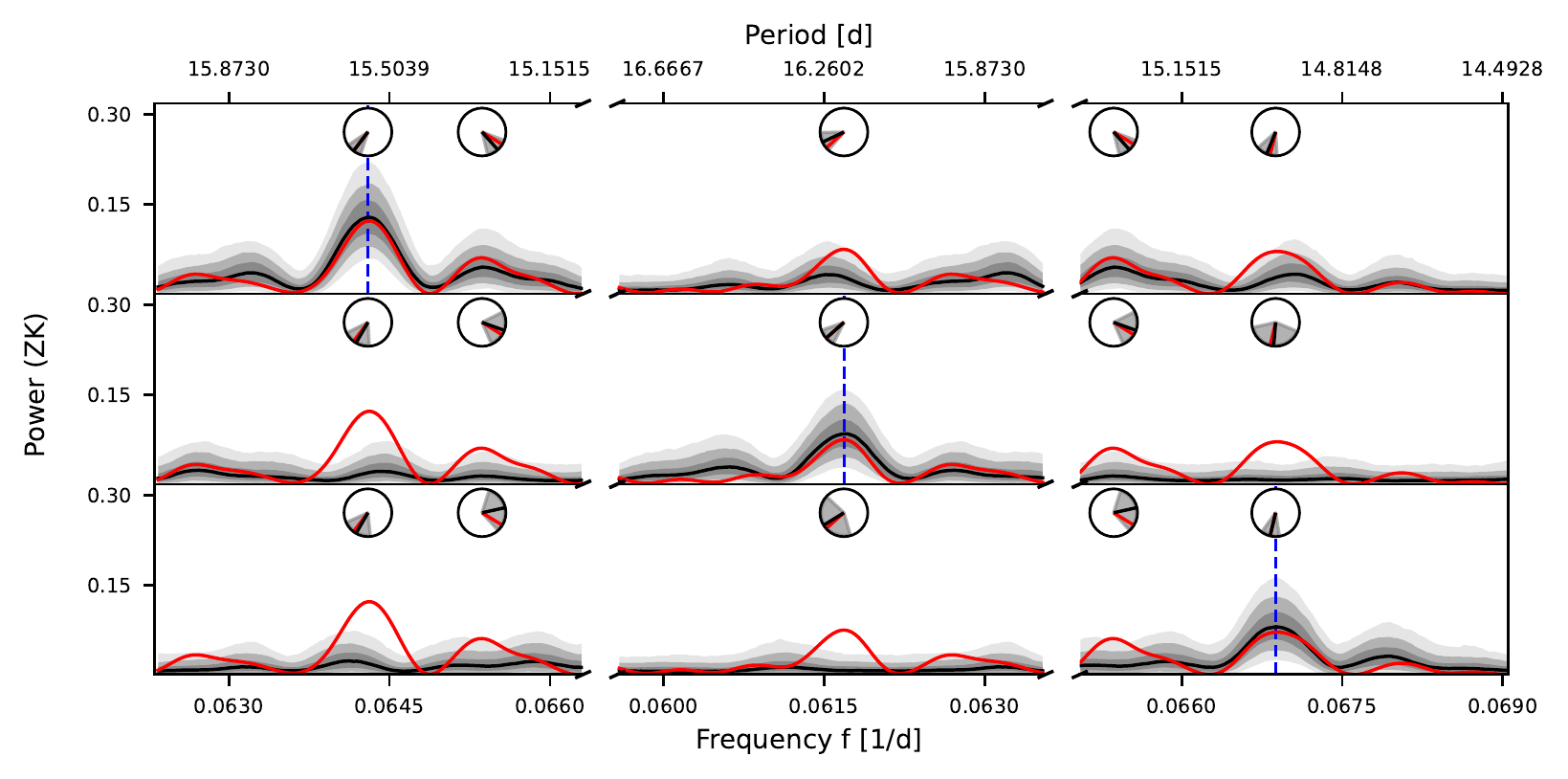}
    \caption{Plots generated by \aliasfinder\ for the daily (\textit{top}) and yearly (\textit{bottom}) aliases for the 15.6\,d signal.
    Each row illustrates the results for one simulated frequency, as indicated by the dashed blue vertical line. Each column is centered on a frequency window corresponding to the simulated frequencies.
    The red line represents the periodogram of the original data set, whereas the black line is the median of the simulations, and the gray shaded regions depict the interquartile range and the confidence range of 90\% and 99\% of the simulations. The clock diagrams indicate the phase. 
    }
    \label{fig:15daliasfinder}
\end{figure*}

The RV data show a variety of aliases related to the 15.6\,d signal. 
In order to establish that 15.6\,d is indeed the true periodicity, we tested the aliasing using \aliasfinder\ \citep{Stock2020_aliasfinder}, which follows the methodology from \cite{DawsonFabrycky2010}. 
The essence behind the algorithm is to examine the GLS periodograms of simulated data sets, in which either of the two aliasing signals are injected, to the GLS periodogram provided by the original data. The injected signal of whichever periodogram matches best to the original one is defined to be the true periodicity apparent in the data. 
The results of this method by simulating 1000 time series for both the daily and yearly sampling frequencies is shown in Fig.~\ref{fig:15daliasfinder}, confirming the true signal to be 15.6\,d.

\section{Known planets in the habitable zone around M dwarfs} \label{appendix:planetshz}

We started from the list collected by PHL at UPR, which obtains its parameters from the NASA exoplanet archive. The list, as last updated on 06 December 2021, comprises 21 planets that are most probable to have a rocky composition and maintain surface liquid water. We individually vetted each system using the most up-to-date literature and updated the planetary and stellar parameters. Most of the planets stayed consistent since the update, though there are some modifications:

\begin{itemize}
    \item \textbf{\object{GJ~667~C}}: We omit the planets d, e, f, and g proposed by \cite{Anglada-Escude2013} in which the second two would reside in the HZ. They emerged as controversial when stellar activity was also modeled within the RVs as a red-noise component \citep{Robertson2014,Feroz2014}.
    Unfortunately, that leaves only planet c in the optimistic HZ in this planetary system. Nonetheless, there is still an inner planet to that of the HZ one, planet b, with a minimum mass of 5.6\,M$_\oplus$. 

    \item \textbf{LP~890-9}: We add the planet b recently unveiled by \cite{Delrez2022} around \object{LP~890-9}, which is next coolest star found to host a HZ planet, after TRAPPIST-1.
    
   \item \textbf{\object{GJ~1002}}: We add planets b and c recently discovered around the M5.5\,V star, both of which reside in the conservative HZ \citep{Mascareno2022}.
    
    


\end{itemize}



\section{Priors and posteriors}

\begin{table*}
\centering
\small
\caption{Prior parameters for the photometric rotation period determination in Sect.~\ref{sec:rotperiod}.}
\label{tab:protphotpriors}
\begin{tabular}{lccl}
\hline
\hline
\noalign{\smallskip}
Parameter name & Prior & Unit & Description \\
\noalign{\smallskip}
\hline
\noalign{\smallskip}
\multicolumn{4}{c}{\textit{Photometric instrumental parameters}}\\[0.1 cm] 
\noalign{\smallskip}
~~~$\mu_{\textnormal{OSN-V-T150}}$  & $\mathcal{U}(0.9,1.1) $ & ppm & Photometric normalization for OSN-V-T150\\[0.1 cm]
~~~$\sigma_{\textnormal{OSN-V-T150}}$  & $\mathcal{J}( 10^{-8},10^{-1}) $ & ppm & Extra jitter term for OSN-V-T150\\[0.1 cm]
~~~$\mu_{\textnormal{OSN-R-T150}}$  & $\mathcal{U}(0.9,1.1) $ & ppm & Photometric normalization for OSN-R-T150\\[0.1 cm]
~~~$\sigma_{\textnormal{OSN-R-T150}}$  & $\mathcal{J}( 10^{-8},10^{-1}) $ & ppm & Extra jitter term for OSN-R-T150\\[0.1 cm]
~~~$\mu_{\textnormal{OSN-I-T150}}$  & $\mathcal{U}(0.9,1.1) $ & ppm & Photometric normalization for OSN-I-T150\\[0.1 cm]
~~~$\sigma_{\textnormal{OSN-I-T150}}$  & $\mathcal{J}( 10^{-8},10^{-1}) $ & ppm & Extra jitter term for OSN-I-T150\\[0.1 cm]
~~~$\mu_{\textnormal{OSN-V-T90}}$  & $\mathcal{U}(0.9,1.1) $ & ppm & Photometric normalization for OSN-V-T90\\[0.1 cm]
~~~$\sigma_{\textnormal{OSN-V-T90}}$  & $\mathcal{J}( 10^{-8},10^{-1}) $ & ppm & Extra jitter term for OSN-V-T90\\[0.1 cm]
~~~$\mu_{\textnormal{OSN-R-T90}}$  & $\mathcal{U}(0.9,1.1) $ & ppm & Photometric normalization for OSN-R-T90\\[0.1 cm]
~~~$\sigma_{\textnormal{OSN-R-T90}}$  & $\mathcal{J}( 10^{-8},10^{-1}) $ & ppm & Extra jitter term for OSN-R-T90\\[0.1 cm]
~~~$\mu_{\textnormal{TJO-R}}$  & $\mathcal{U}(0.9,1.1) $ & ppm & Photometric normalization for TJO-R\\[0.1 cm]
~~~$\sigma_{\textnormal{TJO-R}}$  & $\mathcal{J}( 10^{-8},10^{-1}) $ & ppm & Extra jitter term for TJO-R\\[0.1 cm]
~~~$\mu_{\textnormal{SuperWASP}}$  & $\mathcal{U}(0.9,1.1) $ & ppm & Photometric normalization for SuperWASP\\[0.1 cm]
~~~$\sigma_{\textnormal{SuperWASP}}$  & $\mathcal{J}( 10^{-8},10^{-1}) $ & ppm & Extra jitter term for SuperWASP\\[0.1 cm]
~~~$\mu_{\textnormal{MEarth-tel01-s1}}$  & $\mathcal{U}(0.9,1.1) $ & ppm & Photometric normalization for MEarth-tel01-s1\\[0.1 cm]
~~~$\sigma_{\textnormal{MEarth-tel01-s1}}$  & $\mathcal{J}( 10^{-8},10^{-1}) $ & ppm & Extra jitter term for MEarth-tel01-s1\\[0.1 cm]
~~~$\mu_{\textnormal{MEarth-tel01-s2}}$  & $\mathcal{U}(0.9,1.1) $ & ppm & Photometric normalization for MEarth-tel01-s2\\[0.1 cm]
~~~$\sigma_{\textnormal{MEarth-tel01-s2}}$  & $\mathcal{J}( 10^{-8},10^{-1}) $ & ppm & Extra jitter term for MEarth-tel01-s2\\[0.1 cm]
~~~$\mu_{\textnormal{MEarth-tel05-s2}}$  & $\mathcal{U}(0.9,1.1) $ & ppm & Photometric normalization for MEarth-tel05-s2\\[0.1 cm]
~~~$\sigma_{\textnormal{MEarth-tel05-s2}}$  & $\mathcal{J}( 10^{-8},10^{-1}) $ & ppm & Extra jitter term for MEarth-tel05-s2\\[0.1 cm]
\noalign{\smallskip}
\multicolumn{4}{c}{\textit{dSHO-GP parameters}} \\
\noalign{\smallskip}
~~~$P_\textnormal{rot,\ GP,\ all\tablefootmark{(a)}}$   & $\mathcal{J}(10,200)$ & d & Primary period of the dSHO-GP\\[0.1 cm]
~~~$\delta Q_{\textnormal{GP,\ all\tablefootmark{(a)}}}$  & $\mathcal{J}(10^2,10^5)$ & $\dots$ & Quality factor difference between the first \\[0.1 cm]
& & & and second oscillations of the dSHO-GP\\[0.1 cm]
~~~$Q_{0\ \textnormal{GP,\ all\tablefootmark{(a)}}}$  & $\mathcal{J}(10^{-8},10^3)$ & $\dots$ & Quality factor for the secondary oscillation of the dSHO-GP\\[0.1 cm]

~~~$\sigma_\textnormal{GP,\ OSN-V-T150,OSN-V-T90}$  & $\mathcal{U}(0.0,0.2)$ & $\dots$ & \rdelim\}{7}{17.5mm}[]\\[0.1 cm]
~~~$\sigma_\textnormal{GP,\ OSN-R-T150,OSN-R-T90,TJO-R}$  & $\mathcal{U}(0.0,0.2)$ & $\dots$ &\\[0.1 cm]
~~~$\sigma_\textnormal{GP,\ OSN-I-T150}$  & $\mathcal{U}(0.0,0.2)$ & $\dots$ & \ \ \ \ Amplitude of the dSHO-GP\\[0.1 cm]
~~~$\sigma_\textnormal{GP,\ MEarth-tel01-s1}$  & $\mathcal{U}(0.0,0.2)$ & $\dots$ &  \\[0.1 cm]
~~~$\sigma_\textnormal{GP,\ MEarth-tel01-s2,MEarth-tel05-s2}$  & $\mathcal{U}(0.0,0.2)$ & $\dots$ & \\[0.1 cm]
~~~$\sigma_\textnormal{GP,\ SuperWASP}$  & $\mathcal{U}(0.0,0.2)$ & $\dots$ &  \\[0.1 cm]

~~~$f_\textnormal{GP,\ OSN-V-T150,OSN-V-T90}$  & $\mathcal{U}(0.1,1.0)$ & $\dots$ & \rdelim\}{6}{17.5mm}[]\\[0.1 cm]
~~~$f_\textnormal{GP,\ OSN-R-T150,OSN-R-T90,TJO-R}$  & $\mathcal{U}(0.1,1.0)$ & $\dots$ &\ \ \ \ Fractional amplitude of the \\[0.1 cm]
~~~$f_\textnormal{GP,\ OSN-I-T150}$  & $\mathcal{U}(0.1,1.0)$ & $\dots$ & \ \ \ \ secondary oscillation of the dSHO-GP\\[0.1 cm]
~~~$f_\textnormal{GP,\ MEarth-tel01-s1,MEarth-tel01-s2,MEarth-tel05-s2}$  & $\mathcal{U}(0.1,1.0)$ & $\dots$ &\\[0.1 cm]
~~~$f_\textnormal{GP,\ SuperWASP}$  & $\mathcal{U}(0.1,1.0)$ & $\dots$ & \\[0.1 cm]

\noalign{\smallskip}
\hline
\end{tabular}
\tablefoot{
\tablefoottext{a}{``all'' comprises the following instruments:  OSN-V-T150, OSN-R-T150,OSN-I-T150, OSN-V-T90, OSN-R-T90, TJO-R, SuperWASP, MEarth-tel01-s1, MEarth-tel01-s2, MEarth-tel05-s2}
}
\end{table*}

\begin{table*}[!h]
\centering
\caption[Priors for the RV fits for \wolf]{Priors for the RV fits for \wolf\ with \juliet\ in Sect.~\ref{sec:modelcomparison}. The 90.3\,d signal is not speculated to have planetary origins (Sect.~\ref{sec:90dsignal}), so we denote it as a signal ``2''.}
\label{tab:priors}
\begin{tabular}{lccl}
\hline
\hline
\noalign{\smallskip}
Parameter name & Prior & Units & Description \\
\noalign{\smallskip}
\hline
\noalign{\smallskip}
\multicolumn{4}{c}{\textit{Parameters for planet b}}\\[0.1 cm] 
\noalign{\smallskip}

~~~$P_{b}$  & $\mathcal{U}(15.4,15.7) $ & d & Period.\\[0.1 cm]
~~~$t_{0,b}$  & $\mathcal{U}(2458502.0,2458515.0) $ & d & Time-of-transit center.\\[0.1 cm]
~~~$K_{b}$  & $\mathcal{U}(0.0,5.0) $ & m s$^{-1}$ & Radial velocity semi-amplitude.\\[0.1 cm]
~~~$S_{1,b} = \sqrt{e_\textnormal{b}}\sin \omega_\textnormal{b}$  & $\mathcal{F}{(0.0)}$ (circular) & $\dots$ & Parametrization for $e$ and $\omega$\\
 &  $\mathcal{U}{(-1,1)}$ (eccentric) & $\dots$ & Parametrization for $e$ and $\omega$\\
~~~$S_{2,b} = \sqrt{e_\textnormal{b}}\cos \omega_\textnormal{b}$  & $\mathcal{F}{(0.0)}$ (circular) & $\dots$ & Parametrization for $e$ and $\omega$\\
 &  $\mathcal{U}{(-1,1)}$ (eccentric) & $\dots$ & Parametrization for $e$ and $\omega$\\
\noalign{\smallskip}
\multicolumn{4}{c}{\textit{Parameters for the 90.3\,d signal}}\\[0.1 cm] 
\noalign{\smallskip}
~~~$P_{2}$  & $\mathcal{U}(85.0,95.0) $ & d & Period.\\[0.1 cm]
~~~$t_{0,2}$  & $\mathcal{U}(2458500.0,2458590.0) $ & d & Time-of-transit center.\\[0.1 cm]
~~~$K_{2}$  & $\mathcal{U}(0.0,5.0) $ & m s$^{-1}$ & Radial velocity semi-amplitude.\\[0.1 cm]
~~~$S_{1,2} = \sqrt{e_\textnormal{b}}\sin \omega_\textnormal{b}$  & $\mathcal{F}{(0.0)}$ (circular) & $\dots$ & Parametrization for $e$ and $\omega$\\
~~~$S_{2,2} = \sqrt{e_\textnormal{b}}\cos \omega_\textnormal{b}$  & $\mathcal{F}{(0.0)}$ (circular) & $\dots$ & Parametrization for $e$ and $\omega$\\
\noalign{\smallskip}
\multicolumn{4}{c}{\textit{RV instrumental parameters}}\\[0.1 cm] 
\noalign{\smallskip}
~~~$\gamma_{\textnormal{CARMENES-VIS}}$  & $\mathcal{U}(-20.0,20.0) $ & \ms & Systemic velocity for CARMENES\\[0.1 cm]
~~~$\sigma_{\textnormal{CARMENES-VIS}}$  & $\mathcal{J}(0.01,50.0) $ & \ms & Extra jitter term for CARMENES\\[0.1 cm]
\noalign{\smallskip}
\multicolumn{4}{c}{\textit{dSHO-GP parameters}} \\
\noalign{\smallskip}
~~~$\sigma_\textnormal{GP,\ CARMENES-VIS}$   & $\mathcal{U}(0.0,15.0)$ & \ms & Amplitude of the dSHO-GP\\[0.1 cm]
~~~$Q_{0\ \textnormal{GP,\ CARMENES-VIS}}$  & $\mathcal{J}(10^{-8},10^5)$ & $\dots$ & Quality factor for the secondary oscillation of the dSHO-GP\\[0.1 cm]
~~~$f_\textnormal{GP,\ CARMENES-VIS}$  & $\mathcal{U}(0.1,1.0)$ & $\dots$ & Fractional amplitude of the secondary oscillation of the dSHO-GP\\[0.1 cm]
~~~$\delta Q_\textnormal{GP,\ CARMENES-VIS}$  & $\mathcal{J}(10^2,10^8)$ & $\dots$ & Quality factor difference between the first \\[0.1 cm]
& & & and second oscillations of the dSHO-GP\\[0.1 cm]
~~~$P_\textnormal{rot,\ GP,\ CARMENES-VIS}$  & $\mathcal{U}(155.0,175.0)$ & d & Primary period of the dSHO-GP\\[0.1 cm]
\noalign{\smallskip}
\hline
\end{tabular}
\end{table*}

\begin{table}[!h]
\centering
\caption[Full set of posterior parameters used in the model for \wolf]{Full set of posterior parameters used in the final model choice for \wolf\ and described in Sect.~\ref{sec:modelcomparison}.}
\label{tab:posteriors_all}
\begin{tabular}{lc}
\hline
\hline
\noalign{\smallskip}
Parameter & Posterior  \\
\noalign{\smallskip}
\hline
\noalign{\smallskip}
\multicolumn{2}{c}{\textit{Posterior parameters for planet b}} \\ 
\noalign{\smallskip}
~~~$P_{b}$  & $15.564^{+0.015}_{-0.015}$\\[0.1 cm]
~~~$t_{0,b}$  & $2458511.63^{+0.45}_{-0.46}$\\[0.1 cm]
~~~$K_{b}$  & $1.07^{+0.17}_{-0.17}$\\[0.1 cm]
\noalign{\smallskip}
\multicolumn{2}{c}{\textit{RV instrumental parameters}} \\ 
\noalign{\smallskip}
~~~$\gamma_{\textnormal{CARMENES}}$ (m\,s$^{-1}$)  & $-10.64^{+0.3}_{-0.39}$\\[0.1 cm]
~~~$\sigma_{\textnormal{CARMENES}}$ (m\,s$^{-1}$)  & $0.47^{+0.36}_{-0.41}$\\[0.1 cm]
\noalign{\smallskip}
\multicolumn{2}{c}{\textit{dSHO-GP parameters}} \\ 
\noalign{\smallskip}
~~~$\sigma_\textnormal{GP,\ CARMENES-VIS}$  & $2.80^{+0.80}_{-0.53}$\\[0.1 cm]
~~~$Q_{0\ \textnormal{GP,\ CARMENES-VIS}}$  & $0.00015^{+0.05076}_{-0.00015}$\\[0.1 cm]
~~~$f_\textnormal{GP,\ CARMENES-VIS}$  & $0.61^{+0.25}_{-0.28}$\\[0.1 cm]
~~~$\delta Q_\textnormal{GP,\ CARMENES-VIS}$ & $80000^{+7300000}_{-78000}$\\[0.1 cm]
~~~$P_\textnormal{rot,\ GP,\ CARMENES-VIS}$  & $165.6^{+3.3}_{-3.4}$\\[0.1 cm]
\noalign{\smallskip}
\hline
\end{tabular}
\end{table}

\begin{figure*}[!h]
    \centering
    \includegraphics[width=1\linewidth]{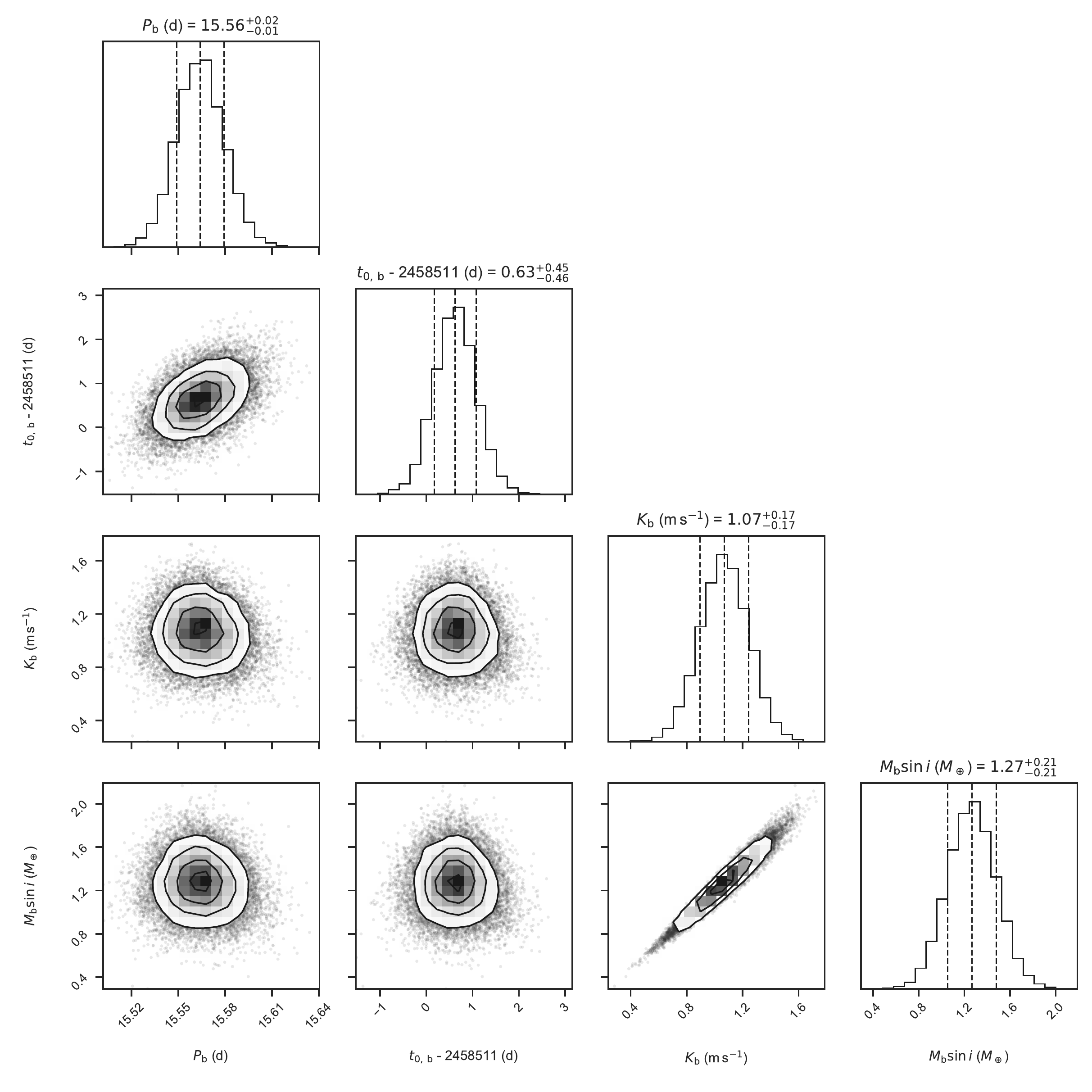}
    \caption[Posterior distributions for the inner-most planet \wolf~b]{Posterior distributions for the inner-most planet \wolf~b from the final RV fit described in Sect.~\ref{sec:modelcomparison}.  
    }
    \label{fig:cornerplot_p1}
\end{figure*}

\begin{figure*}[!h]
    \centering
    \includegraphics[width=1\linewidth]{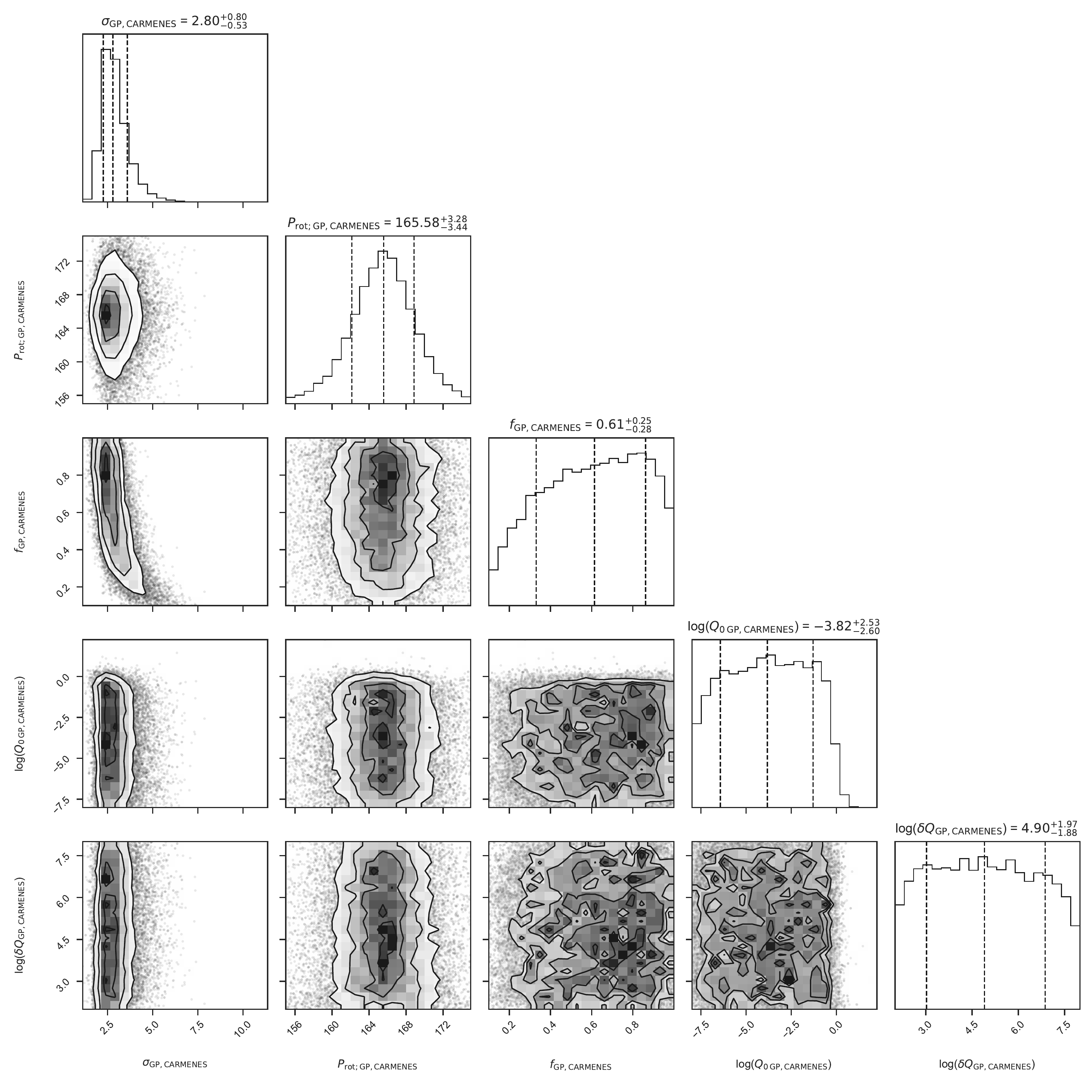}
    \caption[Posterior distributions for the dSHO-GP on the CARMENES RVs]{Posterior distributions for the stellar rotation period using the dSHO-GP from the final RV fit described in Sect.~\ref{sec:modelcomparison}.
    }
    \label{fig:cornerplot_gp}
\end{figure*}

\section{Short tables and data tables}
\begin{table}
\centering 
\small
\begin{center}
\caption{Multiband photometry of \wolf \tablefootmark{a}.}
\label{tab:multibandphotometry}
\centering
\begin{tabular}{lcl}
\hline \hline
\noalign{\smallskip}
Band & Magnitude & Reference \\
 & (mag) &  \\
\noalign{\smallskip}
\hline
\noalign{\smallskip}
~~~$B$        & $15.82 \pm 0.10$      & UCAC4\\
~~~$g'$       &  $14.78 \pm 0.13$     & UCAC4\\
~~~$G_{BP}$   & $14.368 \pm 0.004$     & \gaia\ DR3\\
~~~$V$        & $13.99 \pm 0.05$    & UCAC4\\
~~~$r'$       & $13.41\pm0.05$      & UCAC4\\
~~~$i'$       & $11.58\pm0.09$      & UCAC4\\
~~~$G_{RP}$   & $11.027 \pm 0.004$     & \gaia\ DR3\\
~~~$J$        & $9.029 \pm 0.039$     & 2MASS\\
~~~$H$        & $8.483 \pm 0.073$     & 2MASS\\
~~~$K_S$       & $8.095 \pm 0.021$     & 2MASS\\
~~~$W1$       & $7.877 \pm 0.023$     & AllWISE\\
~~~$W2$       & $7.717 \pm 0.020$     & AllWISE\\
~~~$W3$       & $7.545 \pm 0.016$     & AllWISE\\
~~~$W4$       & $7.445 \pm 0.084$     & AllWISE\\

\noalign{\smallskip}
\hline
\end{tabular}
\tablefoot{
\tablefoottext{a}{\gaia\ EDR3 $G$ magnitude in Table~\ref{tab:stellarparams}.}
}
\tablebib{
    2MASS: \cite{Skrutskie2006_2MASS}; 
    \gaia\ DR3: \cite{GaiaDR3}; 
    UCAC4: \cite{Zacharias2013};
    WISE/AllWISE: \cite{Cutri2012,Cutri2014}.
} 

\end{center}
\end{table}

\begin{table}
\centering
\caption{Telluric-corrected RV data used in this work for \wolf. Data will be available online in machine-readible format.}
\label{tab:rvdata}
\begin{tabular}{c 
S[table-format=2] 
S[table-format=2] 
c}
\hline
\hline
\noalign{\smallskip}
BJD (TDB\tablefootmark{*}) & \multicolumn{1}{c}{RV (m\,s$^{-1}$)} & \multicolumn{1}{c}{$\sigma_\textnormal{RV}$} (m\,s$^{-1}$) & Instrument \\
\noalign{\smallskip}
\hline
\noalign{\smallskip}
2457563.66099 & -11.183 & 1.547 & CARMENES\\[0.1 cm]
2457569.58800 & -10.331 & 3.441 & CARMENES\\[0.1 cm]
2457575.61747 & -12.551 & 1.584 & CARMENES\\[0.1 cm]
2457584.59445 & -15.557 & 2.250 & CARMENES\\[0.1 cm]
2457591.51688 & -11.028 & 1.915 & CARMENES\\[0.1 cm]
2457594.58057 & -14.606 & 2.152 & CARMENES\\[0.1 cm]
2457596.52948 & -15.083 & 1.468 & CARMENES\\[0.1 cm]
2457597.44527 & -16.279 & 1.138 & CARMENES\\[0.1 cm]
2457610.51532 & -13.451 & 1.577 & CARMENES\\[0.1 cm]
2457612.45812 & -12.912 & 1.810 & CARMENES\\[0.1 cm]
2457613.43402 & -12.103 & 1.540 & CARMENES\\[0.1 cm]
\vdots & \vdots & \vdots & \vdots \\[0.1 cm]
2458978.65464 & -12.167 & 2.433 & CARMENES\\[0.1 cm]
2458988.61439 & -9.675 & 1.770 & CARMENES\\[0.1 cm]
2458994.61685 & -9.450 & 1.428 & CARMENES\\[0.1 cm]
2458999.63468 & -6.863 & 1.591 & CARMENES\\[0.1 cm]
2459000.64327 & -7.105 & 1.278 & CARMENES\\[0.1 cm]
2459001.64411 & -7.717 & 1.318 & CARMENES\\[0.1 cm]
2459006.64529 & -5.812 & 1.641 & CARMENES\\[0.1 cm]
2459010.59488 & -5.312 & 1.664 & CARMENES\\[0.1 cm]
2459015.61493 & -7.113 & 1.513 & CARMENES\\[0.1 cm]
2459017.64665 & -9.772 & 3.050 & CARMENES\\[0.1 cm]
\noalign{\smallskip}
\hline
\end{tabular}
\tablefoot{\tablefoottext{*}{Barycentric dynamical time.}}
\end{table}

\end{appendix}

\end{document}